\documentclass[ALICE,manyauthors]{cernphprep}
\usepackage[comma,square,numbers,sort&compress]{natbib}
\usepackage{hyperref}
\usepackage{lineno}
\usepackage{epsfig} 
\usepackage{subfigure}
\usepackage{verbatim}
\usepackage{color}
\usepackage{soul}
\usepackage{caption}  
\usepackage{array}    
\usepackage{multirow} 

\begin{document}%

\begin{titlepage}
\PHyear{2014}
\PHnumber{254}      
\PHdate{08 October}  
%

\title{Charged jet cross sections and properties\\ in
  proton-proton collisions at $\sqrt{\mathbf{s}}$~=~7~TeV}
\ShortTitle{Charged jet properties in pp at 7 TeV}   

\Collaboration{ALICE Collaboration\thanks{See Appendix~\ref{app:collab} for the list of collaboration members}}
\ShortAuthor{ALICE Collaboration} 

\begin{abstract}
  The differential charged jet cross sections, jet fragmentation distributions, 
and jet shapes are measured in minimum bias proton-proton collisions at 
centre-of-mass energy $\sqrt{s}=$~7~TeV using the ALICE detector at the 
LHC. Jets are reconstructed from charged particle momenta in the mid-rapidity 
region using the sequential recombination $k_{\rm T}$ and anti-$k_{\rm T}$ as 
well as the SISCone jet finding algorithms with several resolution parameters 
in the range $R$~=~0.2~--~0.6. Differential jet production cross sections 
measured with the three jet finders are in agreement in the transverse momentum 
($p_{\rm T}$) interval 20 $< p_{\rm T}^{\rm jet,ch} <$ 100~GeV/$c$. They are also 
consistent with prior measurements carried out at the LHC by the ATLAS collaboration. 
The jet charged particle multiplicity rises monotonically with increasing jet $p_{\rm T}$, 
in qualitative agreement with prior observations at lower energies. The transverse profiles 
of leading jets are investigated using radial momentum density distributions as well as 
distributions of the average radius containing 80\% ($\langle R_{\rm 80} \rangle$) of the 
reconstructed jet $p_{\rm T}$. The fragmentation of leading jets with $R$~=~0.4 using 
scaled $p_{\rm T}$ spectra of the jet constituents is studied. The measurements are 
compared to model calculations from event generators (PYTHIA, PHOJET, HERWIG). 
The measured radial density distributions and $\langle R_{\rm 80} \rangle$ distributions 
are well described by the PYTHIA model (tune Perugia-2011). The fragmentation 
distributions are better described by HERWIG. 
\end{abstract}
\end{titlepage}
\setcounter{page}{2}

\section{Introduction}
\label{sec:introduction}
Jets consist of collimated showers of particles resulting from the fragmentation of
hard (high momentum transfer $Q$) 
partons (quarks and gluons)
produced in high energy 
collisions. The production cross sections of jets 
were measured in detail in
proton-antiproton (p$\bar{\rm p}$) collisions at the
Tevatron ($\sqrt{s}$~=~540~GeV, 630~GeV, 1.8~TeV and 1.96~TeV)~\cite{A1_D0-jets-630-1.8-1.96,A1_D0-jets-630-1.8-1.96_1,A1_D0-jets-630-1.8-1.96_2,A2_CDF-jets-540-1.8-1.96,A2_CDF-jets-540-1.8-1.96_1,A2_CDF-jets-540-1.8-1.96_2,A2_CDF-jets-540-1.8-1.96_3,A2_CDF-jets-540-1.8-1.96_4,A2_Abulencia:2005yg,A2_Abulencia:2007ez,A2_Aaltonen:2008eq}. Measurements were also carried out recently at the CERN LHC at higher
energies ($\sqrt{s}$~=~2.76, 7 and 8~TeV) in proton-proton (pp)
collisions~\cite{A3_ATLASchJets, A4_ATLASJets, A5_CMSJets,A5_CMSJets_1,A15_FullJetPaper}. 
Jet shape observables 
were previously measured by the CDF~\cite{A6_cdfprl70, cdfprd65, A8_cdfprd71}, and
D0~\cite{A9_d0plb357} collaborations in $\rm p\bar{p}$ collisions and
more recently by the ATLAS and CMS collaborations in pp
collisions~\cite{A10_atlasprd83, A11_cmsanalysis, A11_1_cms2014}.
The fragmentation functions of jets produced in p$\bar{\rm p}$
collisions were reported by the CDF collaboration~\cite{A12_CDF_FF}. Jet fragmentation
in pp and Pb--Pb collisions at the LHC were reported by
the ATLAS~\cite{A3_ATLASchJets, A13_ATLAS_FF, ATLAS_PbPb_FF} and CMS~\cite{A14_CMS_PbPb_FF} collaborations.
Jet production in e$^{\rm +}$e$^{\rm -}$, ep, p$\bar {\rm p}$, and pp collisions is well described
by perturbative Quantum Chromodynamics (pQCD) calculations.
The measured jet properties
are typically well reproduced by
Monte Carlo (MC) generators such as PYTHIA~\cite{A30_Pythia}, HERWIG~\cite{A36_Herwig,A36_Herwig_1}, and PHOJET~\cite{A35_Phojet}.
The unprecedented beam energy achieved at the Large
Hadron Collider (LHC) in pp collisions
enables an extension of jet production cross section and property measurements
carried out at lower energies. Such measurements  enable further tests of QCD and
help in tuning of MC event generators.\par
In this paper, we present measurements of the jet production cross sections, jet fragmentation
distributions, and transverse jet shape observables 
in pp collisions at $\sqrt{s}$ = 7~TeV. The analysis is restricted 
to charged particle jets, i.e.\  jets reconstructed solely from
charged particle momenta, hereafter called charged jets. 
ALICE has already reported measurements of charged jet production in Pb--Pb
collisions at 2.76~TeV~\cite{A15_1_ChargedJet_PbPb}.
Charged jets are reconstructed with particles having $p_{\rm T}$ down
to values as low as 0.15~GeV/$c$, 
thereby allowing to test perturbative and non-perturbative aspects of
jet production and fragmentation as implemented in 
MC generators.
The measured particle spectra in jets 
reflect the jet fragmentation function, as summarized in~\cite{A16_PDG} (Sec. 19).
The jet shape distributions are related to the details of the parton shower process.\par
Jets also constitute an important probe for the study of the hot and dense QCD matter
created in high energy collisions of heavy nuclei. 
In such collisions, high $p_{\rm T}$ partons penetrate the colored medium and 
lose energy via induced gluon radiation and elastic scattering 
(see~\cite{A17_QuenchingPaper} and references therein). The measurements in pp 
collisions thus provide a baseline for similar measurements in 
nucleus--nucleus (A--A) and proton-nucleus (p--A) collisions.\par
Medium modifications of the parton shower may change the fragmentation pattern relative
to the vacuum~\cite{A18_Wiedemann_Jet_Quenching}. There are empirical indications~\cite{A19_Renk_QM_proceedings} that 
the scale relevant to these effects is given by the medium temperature
of the order of 
few hundred MeV rather than the 
hard scattering scale. At such small particle momenta, the jets measured experimentally 
in pp and A--A collisions also contain contributions from the underlying event (UE). In pp collisions~\cite{cdfprd65}, 
the UE includes gluon radiation in the initial state, the fragmentation 
of beam remnants and multiple parton interactions. In this study, we subtract the UE from the distributions
measured in pp collisions, to allow for a meaningful comparison to models,
because theoretical modeling of the underlying event is very
complex. To disentangle UE and hard parton fragmentation into low momentum particles, 
we correct our measurements using a technique as described in Sec.~\ref{sec:underlyingEventSub}.
This approach will also help to make eventually a comparison with data
from A--A collisions, where the UE in addition includes hadrons from
an expanding fireball.\par 
This paper is organized as follows. Section~\ref{sec:expmethod} describes
the experiment and detectors used for the measurements reported in this work. 
Details of the jet reconstruction
algorithms and parameters are presented in
Sec.~\ref{sec:jetReconstruction}, while 
jet observables are defined and discussed in Sec.~\ref{sec:jetobs}.
Section~\ref{secMC} discusses the MC simulations carried out  for comparisons of measured data to models, data corrections for instrumental effects, 
and systematic error studies.  The procedures applied to correct for
instrumental and UE effects are presented in Sec.~\ref{sec:corrections}. The methods used to evaluate systematic
uncertainties of the measurements are discussed
in Sec.~\ref{sec5syserrors}. Results are presented  and discussed in
comparison with MC Event Generator 
simulations in Sec.~\ref{sec6results}. Section~\ref{sec7summary}
summarizes the results and conclusions of this work.  

\section{Experimental setup and data sample}
\label{sec:expmethod}
The data used in this analysis were collected during the 2010 LHC run
with the ALICE detector~\cite{A20_AliceExpt, A20_AliceExpt1}. This analysis relies primarily on the Time Projection Chamber
(TPC)~\cite{A21_RefTPC}, the Inner Tracking System (ITS)~\cite{A22_RefITS}, and the V0~\cite{A23_RefVZERO} sub-detectors. The V0
and ITS are used for event selection. 
A minimum bias trigger is achieved by requiring at least one hit in
either the V0 forward scintillators or in the two innermost Silicon
Pixel Detector layers (SPD) of the ITS, in coincidence with an LHC bunch crossing. 
The efficiency for detecting inelastic events is about 85\%~\cite{A24_AliceSigmaPaper}.  
The TPC and ITS are used for primary vertex and track reconstruction.
Only events with a primary 
vertex within $\pm$10~cm along the beam direction from the nominal interaction point are
analyzed to minimize dependencies of the TPC acceptance on the vertex position. 
The results reported in this paper are based on 177~$\times$~$\rm 10^6$ minimum bias
events corresponding to an integrated luminosity~\cite{A24_AliceSigmaPaper} of (2.9$\pm$0.1)~nb$^{\rm -1}$.\par 
The ALICE solenoidal magnet is operated with a magnetic field of 0.5~T that provides a good compromise between momentum resolution at high $p_{\rm T}$ and
detection of low $p_{\rm T}$ particles. Charged tracks are reconstructed using the combined information from the
TPC and the ITS utilizing a hybrid reconstruction technique described
in~\cite{A15_FullJetPaper} to assure uniform $\varphi$ distribution.  
The acceptance for charged tracks is $|\eta|<$0.9 over the full azimuth. 
This hybrid technique combines two distinct track classes: (i) tracks
containing at least three hits (of up to six) in the ITS, including at 
least one hit in the SPD, and (ii) tracks containing fewer than three hits 
in the ITS, or no hit in the SPD. 
The momentum of tracks of class (i) is determined without a vertex
constraint. The vertex constraint is however added for tracks of class
(ii) to improve the determination of their transverse momentum.
The track momentum resolution ${\rm \delta} p_{\rm T}/p_{\rm T}$ is approximately 1\% 
at $p_{\rm T}$ = 1~GeV/$c$ for all reconstructed tracks, and  
4\% at $p_{\rm T}$ = 40~GeV/$c$ for 95\% of all tracks. 
For tracks without a hit in the ITS (5\% of the track sample) the resolution 
is 7\% at $p_{\rm T}$ = 40~GeV/$c$. The analysis is restricted to tracks 
with a Distance of Closest Approach (DCA) to the primary vertex 
smaller than 2.4~cm and 3.2~cm in the plane transverse to the beam and the beam 
direction, respectively, in order to suppress contributions from secondary particles produced by weak decays 
and interactions of primary particles with detector materials and beam
pipe. 

Tracks in the TPC are selected by
requiring a $p_{\rm T}$ dependent minimum number of space points ranging from 70 (of up to 159) 
for $p_{\rm T}$ = 0.15~GeV/c to 100 at $p_{\rm T} >$ 20~GeV/c. A $\chi^2$ cut on the track fit is applied. 
Secondary particles which are not produced at the primary vertex may acquire a wrong momentum 
when constrained to the vertex. Therefore, a $\chi^2$ cut on the difference between the parameters 
of the track fit using all the space points 
in the ITS and TPC and using only the TPC space points with the primary vertex position as 
an additional constraint is applied. The track reconstruction efficiency for primary charged particles is 
approximately 60\% at $p_{\rm T}$ = 0.15~GeV/$c$ and 
rises to a value of about 87\% at 1~GeV/$c$ and is approximately uniform up to 10~GeV/$c$ beyond which it decreases slightly. 
The efficiency is uniform in azimuth and within the
pseudorapidity range $|\eta|~\textless$~0.9.
Further details on the track selection procedure and tracking performance can be found in~\cite{A15_FullJetPaper}. 

\section{Jet reconstruction}
\label{sec:jetReconstruction}
The charged jet reconstruction is carried out using the infrared-safe and collinear-safe sequential
recombination algorithms anti-$k_{\rm T}$~\cite{A25_RefAntikt} and $k_{\rm T}$~\cite{A26_RefKt,A26_RefKt_1} from the FastJet
package~\cite{A27_RefFastjet} and  a seedless infrared safe iterative cone
based algorithm, named SISCone~\cite{A28_RefSiscone} to obtain the jet
cross sections. The three jet finders are found to be in
good agreement within the uncertainties as discussed in
Sec.~\ref{algocomp}. All other observables (as discussed in
Sec.~\ref{sec:jetobs}) are analyzed with anti-$k_{\rm T}$ only.
Charged tracks with 
$p_{\rm T}~\textgreater$ 0.15~GeV/$c$ and within $|\eta|~\textless$ 0.9 are the inputs to the jet reconstruction algorithms. 
A boost invariant $p_{\rm T}$ recombination scheme is used to determine the transverse momenta of jets 
by adding the charged particle transverse momenta. 
Jets are reconstructed with resolution parameters $R$ = 0.2, 0.3, 0.4, and 0.6 to enable a systematic study of the production 
cross section and shape properties, as well as to provide a suite of references for measurements performed in p--A and A--A collisions. 
The analyses reported in this work are 
restricted to jets detected within the range $|\eta|~\textless$~(0.9 - $R$) in order to
minimize edge effects in the reconstruction of jets and biases on jet
transverse profile and fragmentation functions. The inclusive jet cross sections are reported as a function of $p_{\rm
  T}$ in the interval 20 $< p_{\rm T}^{\rm jet,ch} <$ 100~GeV/$c$. The
properties of the charged jet with the highest $p_{\rm T}$ in the
event, the so called {\it leading jet}, are presented in the same
$p_{\rm T}$ interval.

\section{ Jet observables}
\label{sec:jetobs}
The results are reported for a suite of charged jet properties including inclusive differential jet cross section, 
charged particle multiplicity in leading jets ($\langle N_{\rm ch} \rangle$), leading jet size ($\langle R_{\rm 80} \rangle$), radial 
distribution of $p_{\rm T}$ within the leading jet ($\langle
\rm{d}{\it p}_{\rm T}^{\rm sum}/\rm{d}{\it r} \rangle$),
and jet 
fragmentation distributions ($F^{p_{\rm T}}$, $F^{z}$, $F^{\xi}$). The definition of these observables and 
the methods used to measure them are presented in this section. Correction techniques applied to measured raw distributions to account for instrumental 
effects (including the detector acceptance and resolution), as well as the UE, 
are discussed in Sec.~\ref{sec:corrections}. 
All observables reported in this work are corrected to particle level
as defined in Sec.~\ref{secMC}.\par
The differential jet cross section is evaluated using the following relation: 
\begin{equation}
  \frac{{\rm {d^{2}}}\sigma^{\rm jet, ch}}{{\rm d} p_{\rm
      T}\rm{d}\eta} (p_{\rm T}^{\rm jet,ch}) =
  \frac{1}{\mathcal{L}^{\rm int}}\frac{ \rm{\Delta}{\it N}_{\rm
      jets}}{\rm {\Delta} {\it p}_{\rm T} \rm{\Delta} \eta} (p_{\rm
    T}^{\rm jet,ch}), 
  \label{xsec-equation}
\end{equation}
where $\mathcal{L}^{\rm int}$ is the integrated luminosity and  $\rm{\Delta}{\it N}_{\rm jets}$ 
the number of jets in the selected intervals of $\rm{\Delta} {\it p}_{\rm T}$ and $\rm{\Delta} \eta$. \par
The charged particle multiplicity in leading jets, $N_{\rm ch}$, is defined as the number of charged particles found within the leading 
jet cone. Results for the mean charged particle multiplicity, $\langle N_{\rm ch} \rangle$, computed in bins of jet 
$p_{\rm T}$ are presented for resolution parameter values $R$~=~0.2, 0.4, and 0.6.\par   
The size of the leading 
jet, $R_{\rm 80}$, is defined
as the radius in the ${\rm\Delta}\eta$~--~${\rm \Delta}\varphi$ 
space that contains 80\% of the total $p_{\rm T}$ found in the jet cone. Results for the mean value, 
$\langle R_{\rm 80} \rangle$, are presented as a function of jet $p_{\rm T}$ for resolution parameter values $R$~=~0.2, 0.4, and 0.6.\par
The distribution of $p_{\rm T}$ density, $\rm{d}{\it p}_{\rm T}^{\rm
  sum}/\rm{d}{\it r}$, within a leading 
jet is measured as a function of the distance $r = \sqrt{(\rm{\Delta}\eta)^{\rm 2}+(\rm{\Delta}\varphi)^{\rm 2}}$ from the jet
direction. The momentum density is calculated jet by jet as a scalar sum of the transverse momenta, $p_{\rm T}^{\rm sum}$, 
of all charged particles produced in annular regions of width $\rm{\Delta} {\it r}$ at radius $r$ centered on the jet direction. 
The mean value  of the momentum density, $\langle \rm{d}{\it p}_{\rm
  T}^{\rm sum}/\rm{d}{\it r} \rangle$, is evaluated as a function of
$r$ using the following relation:
\begin{equation}\label{eq1}
  \langle \frac{\rm {d}{\it p}_{\rm T}^{\rm sum}}{\rm {d}{\it r}}
  \rangle (r) = \frac{1}{\Delta
    r}\frac{1}{N_{\rm jets}}\sum_{\rm i=1}^{N_{\rm jets}} {p}_{\rm T}^{\rm i}(r-\Delta r/{\rm 2},
  r+\Delta r/{\rm 2})
\end{equation}
where $p_{\rm T}^{\rm i} (r-\rm{\Delta} {\it r}/{\rm 2}, r+\rm{\Delta} {\it r}/{\rm 2})$ denotes the summed $p_{\rm T}$ 
of all tracks of jet i, inside the annular ring between $r-\rm{\Delta} {\it r}/{\rm 2}$ and $r+\rm{\Delta} {\it r}/{\rm 2}$. 
The mean value is reported in bins of jet $p_{\rm T}$ for resolution
parameter values $R$ = 0.2, 0.4, and 0.6. $N_{\rm jets}$ denotes the number of jets per bin. 

The fragmentation of the leading jet is reported based on the distributions
\begin{equation}
  F^{p_{\rm T}}( p_{\rm T},p_{\rm T}^{\rm jet, ch}) = \frac{1}{N_{\rm
      jets}} \frac{{\rm d}N}{{\rm d}p_{\rm T}}, \\ 
\end{equation}
\begin{equation}
  F^{z}(z^{\rm ch},p_{\rm T}^{\rm jet, ch}) = \frac{1}{N_{\rm jets}} \frac{{\rm
      d}N}{{\rm d}z^{\rm ch}}, 
\end{equation}
\begin{equation}
  F^{\xi}(\xi^{\rm ch},p_{\rm T}^{\rm jet, ch}) = \frac{1}{N_{\rm jets}}
  \frac{{\rm d}N}{{\rm d}\xi^{\rm ch}},
\end{equation}
where N is the number of charged particles. The scaled $p_{\rm T}$ variables $z^{\rm ch} = p_{\rm T}^{\rm particle} / p_{\rm T}^{\rm jet, ch}$ and 
$\xi^{\rm ch} = \log (1/z^{\rm ch})$ are calculated jet by jet for each track.  In contrast to the definition in~\cite{A16_PDG}, 
the energy carried by neutral particles is not contained in the jet
momentum.
The (scaled) $p_{\rm T}$ spectra of the jet constituents 
are normalized per jet and presented in bins of jet $p_{\rm T}$.  $F^{p_{\rm T}}$, $F^{z}$ and $F^{\xi}$ are 
complementary representations: the particle $p_{\rm T}$ spectra $F^{p_{\rm T}}$ are less sensitive to uncertainties 
in the jet energy scale and may be more suitable as a reference for future measurements in nuclear collisions than the 
standard representation $F^{z}$, whereas the $F^{\xi}$ distributions emphasize fragmentation into low momentum constituents 
and are particularly suited to demonstrate QCD coherence
effects~\cite{A29_QCD_coherence,A29_QCD_coherence_1}. \par
In this work, the averages $\langle N_{\rm ch} \rangle$, $\langle
R_{\rm 80} \rangle$, and $\langle \rm{d}{\it p}_{\rm T}^{\rm
  sum}/\rm{d}{\it r} \rangle$ are referred to as jet shape 
observables (jet shapes) and $F^{p_{\rm T}}$, $F^{z}$ and $F^{\xi}$ as fragmentation distributions. 

\section{Monte Carlo simulations}
\label{secMC}
Instrumental effects, such as the limited particle detection efficiency and the finite track momentum resolution, induce momentum 
dependent particle losses and impact the jet energy scale and structures of the observables reported in 
this work. The effect of the detector response is studied 
using the simulation of the ALICE detector performance for particle
detection and jet reconstruction. Simulated events are 
generated with PYTHIA 6.425~\cite{A30_Pythia} (tune Perugia-0~\cite{A31_PerugiaTunes}) and 
the produced particles are transported with GEANT3~\cite{A32_RefGeant3}. The simulated and real
data are analyzed with the same reconstruction algorithms and using
the same kinematic cuts ($p_{\rm T}~\textgreater$~0.15~GeV/$c$,
$|\eta|~\textless$~0.9) on produced particles. Jets reconstructed
based directly on momenta of
charged particles ($p_{\rm T}~\textgreater$~0.15~GeV/$c$,
$|\eta|~\textless$~0.9) produced by MC generators are hereafter
referred to as {\it particle level} jets whereas those obtained after 
processing the generator outputs  through GEANT and the ALICE
reconstruction software  are referred to as {\it detector level}
jets. As the data are corrected for instrumental effects, their
comparison with simulation is done at particle level only.\par
\begin{figure}[ht]
  \centering
  \includegraphics[width=0.5\textwidth]{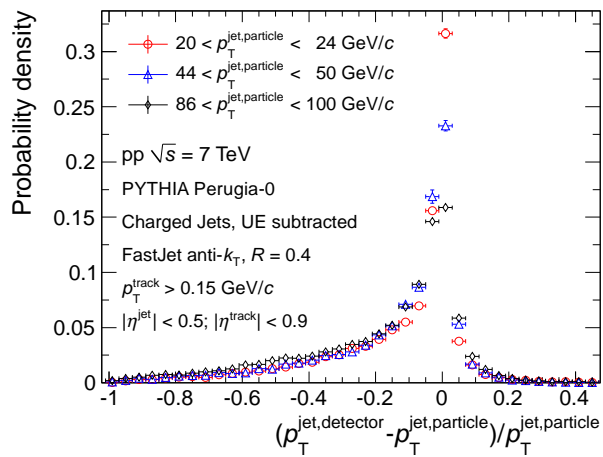}
  \caption{(Color online) Probability distribution of the relative momentum difference
    of simulated ALICE detector response to charged jets in pp
    collisions at $\sqrt{s}$~=~7~TeV for three different $p_{\rm
      T}^{\rm jet,particle}$ intervals. Charged jets are simulated
    using PYTHIA Perugia-0 and reconstructed with the anti-$k_{\rm T}$
    jet finding algorithm with $R$~=~0.4.}
  \label{JetEnergyResolution}
\end{figure}

 
The detector response to simulated charged jets  with $R$~=~0.4 is illustrated in Fig.~\ref{JetEnergyResolution}, showing on a jet-by-jet 
basis the probability distribution of the relative difference between the
charged jet $p_{\rm T}$ at the particle level ($p_{\rm T}^{\rm jet, particle}$) and at the detector level
($p_{\rm T}^{\rm jet, detector}$). 
The probability distribution is shown for three different $p_{\rm T}^{\rm jet, particle}$ intervals. 
The distributions have a pronounced maximum at zero ($p_{\rm T}^{\rm jet, detector}$ = $p_{\rm T}^{\rm jet, particle}$). The 
tracking $p_{\rm T}$ resolution induces upward and downward fluctuations with equal probability, whereas 
the finite detection efficiency of charged particles results in an
asymmetric response. As a function of $p_{\rm T}^{\rm jet, particle}$, 
the probability that $p_{\rm T}^{\rm jet, detector}$ is smaller than $p_{\rm T}^{\rm jet, particle}$ varies between 88 and 92\% and the mean value of the distribution varies between -14\% to -24\%.\par 
The event generators PHOJET~1.12.1.35~\cite{A35_Phojet}, HERWIG~6.510~\cite{A36_Herwig,A36_Herwig_1}, and several PYTHIA tunes
are used for 
comparisons to data and for systematic investigations of the sensitivity of the MC correction factors to variations of the detector 
response as well as to jet fragmentation and hadronization patterns. PYTHIA, PHOJET, and HERWIG 
utilize different approaches to describe the parton shower and
hadronization process. HERWIG makes of angular 
ordering a direct part of the evolution process and thereby takes correctly into account coherence effects in the emission of soft gluons. 
PYTHIA~6.4 is instead based on transverse-momentum-ordered showers~\cite{A37_pt_ordered_showers} in which angular ordering is imposed by an 
additional veto. PHOJET generates angular ordered initial-state radiation, whereas for final 
state radiation the mass-ordered PYTHIA shower algorithm is used. Hadronization in PYTHIA and PHOJET proceeds 
via string breaking as described by the Lund model~\cite{A38_Lund_model}, whereas
HERWIG uses cluster fragmentation. The PYTHIA Perugia tune variations, beginning  with the central tune Perugia-0~\cite{A31_PerugiaTunes}, 
are based on LEP, Tevatron, and SPS data. The Perugia-2011 family of tunes~\cite{A31_PerugiaTunes} and the ATLAS Minimum Bias tune AMBT1~\cite{A39_AMBT1}
belong to the first generation of tunes that also use LHC pp data at $\sqrt{s}$ = 0.9 and 7~TeV with slight variations of the 
parameters controlling the modeling of the UE and fragmentation. Compared to the central Perugia-2011 tune, 
AMBT1 uses a lower value of the infrared regularization scale for
multiple partonic interactions resulting in higher UE activity. 
It also uses a probability density of sum of two Gaussians for the matter distribution inside the proton and a higher non-perturbative 
color-reconnection strength for string fragmentation. 
The HERWIG generator version and PYTHIA tunes used in this work utilize the CTEQ5L parton 
distributions~\cite{A40_CTEQ5L}, except for PYTHIA tune AMBT1 which uses MRST~2007LO*~\cite{A41_MRST2007}. PHOJET uses GRV94~\cite{A42_GRV94}. 

\section{Corrections }
\label{sec:corrections}
Two classes of correction techniques are used to account for instrumental effects in the measurements 
reported in this work. The techniques are known as bin-by-bin
correction and Bayesian unfolding~\cite{A43_unfold-bayes}.
A third technique based on Singular Value Decomposition
(SVD)~\cite{A44_unfold-svd} is also used as a cross check. The techniques
and their comparative merits are presented in the following
subsections. Corrections for contamination from secondary particles and UE are discussed in Secs.~\ref{sec:secondaries} 
and ~\ref{sec:underlyingEventSub} respectively.\par
The jet shapes and fragmentation distributions are corrected using
the bin-by-bin method, while the cross sections are corrected
with the Bayesian unfolding technique. All observables are corrected
for secondaries contamination. All observables, except $\langle R_{\rm 80} \rangle$,
are also corrected for UE contamination. 
\subsection{Bin-by-bin correction method}
\label{binBybin}
The bin-by-bin correction method is used to correct the jet shape observables and fragmentation functions. To validate 
the method, it is also applied to the jet cross sections. It utilizes MC simulations as described in Sec.~\ref{secMC} and is based on ratios of 
values for observables obtained at particle (generator) level and detector 
level as a function of 
variable $\mathbf{x}$. In this work, $\mathbf{x}$ can be 
1-dimensional (e.g. jet $p_{\rm T}$ in case of the jet spectra) or
2-dimensional (e.g. jet $p_{\rm T}$ and particle $p_{\rm T}$ in case of the fragmentation distributions). 
Let $O_{\rm mc}^{\rm part}(\mathbf{x})$ be the observable value 
at the particle level, and $O_{\rm  mc}^{\rm det}(\mathbf{x})$ the value obtained 
at the detector level. The correction factors are defined as the ratio of the particle and detector level 
values of $O_{\rm mc}^{\rm part}(\mathbf{x})$ and $O_{\rm mc}^{\rm det}(\mathbf{x})$ in bins 
of $\mathbf{x}$. The corrected measurements, $O_{\rm data}^{\rm corrected}$, are obtained bin-by-bin by multiplying the 
raw (uncorrected) values, $O_{\rm data}^{\rm uncorrected}$, as follows,
\begin{equation}
  O_{\rm data}^{\rm corrected} (\mathbf{x}) = O_{\rm
    data}^{\rm uncorrected} (\mathbf{x}) \frac{O_{\rm
      mc}^{\rm part}(\mathbf{x})}{O_{\rm mc}^{\rm det} (\mathbf{x})}.
\end{equation}\par
The correction factors depend on the 
shape of the simulated jet spectrum and fragmentation distributions. Systematic uncertainties 
related to the accuracy with which data are reproduced by the simulations are discussed in Sec.~\ref{sec:sysEvGen}. \par 
Correction factors obtained for the jet $p_{\rm T}$ spectra range from 25\% to 50\% and reach a maximum at 
100~GeV/$c$. The bin-by-bin corrections applied to jet shape
observables include subtraction of contamination associated with the
production of secondary particles within the detector. 
Correction factors
obtained for $\langle N_{\rm ch} \rangle$ at $R$ = 0.2 (0.4, 0.6) are of the order of 2-6\% (3-5\%, 4-6\%) while for 
$\langle R_{\rm 80} \rangle$ at $R$~=~0.2 (0.4, 0.6) they are found in the range 5-7\% (2-10\%,
4-9\%). Correction factors applied on radial momentum densities have a maximum value of 12\%(15\%, 19\%) at $R$~=~0.2 (0.4,
0.6).
In contrast, for the fragmentation distributions, 
the bin-by-bin correction and the correction for the contamination from secondaries, discussed 
in Sec.~\ref{sec:secondaries}, are carried out in separate steps. The typical value of the corrections  at the maximum of the $F^{\xi}$ distribution is of the 
order of few percent only. The correction factors for $F^{p_{\rm T}}$ and $F^{z}$
are largest at low particle $p_{\rm T}$ (up to 50\%), where the tracking efficiency is smallest, 
and at the highest $z^{\rm ch}$ (up to 40\%) where the impact of the track momentum resolution is 
strong and detector effects at the track level strongly influence the reconstructed jet momentum.
\subsection{Unfolding using response matrix inversion techniques}
\label{sec:unfolding}
Instrumental effects associated with acceptance, particle losses due to limited efficiency, and finite momentum resolution 
are modeled using a detection response matrix, which is used to correct observables for these effects. The jet $p_{\rm T}$ response matrix is determined by processing 
MC events through a full ALICE detector simulation as described in Sec.~\ref{secMC}. 
The particle level (true), $T(t)$, and detector level (measured), $M(m)$, $p_{\rm T}$
spectra of the leading jet are both subdivided in 11 bins in the
interval 20~$<~p_{\rm T}^{\rm jet,ch}~<~100$~GeV/$c$. 
The matrix elements $R_{mt}$ express the
conditional probability of measuring a jet $p_{\rm T}$ in bin, $
m$ given a true value in bin, $t$.
The measured distribution, $M$, can thus be estimated by multiplying the true distribution, $T$, by the response matrix, 
\begin{equation}
  M = RT.
\end{equation}
Experimentally, the unfolding problem involves the determination of $T$ given $M$. This is symbolically written as
\begin{equation}
  T = R^{-1}M.
\end{equation}
However 
the matrix $\rm  R$ may be singular and 
can not always be inverted analytically. 
Consequently, other numerical techniques are needed to obtain the 
true, physically meaningful, distribution $T$ given a measured distribution $M$. 
Furthermore, the exact solution, even if it exists, is usually unstable 
against small variations in the initial estimates of the measured distribution, and oscillating due to
finite statistics in the measured distribution. 
This problem can be overcome using a regularization condition based on
a priori information about the solution.\par
The Bayesian unfolding technique~\cite{A43_unfold-bayes} is an iterative method based on Bayes' theorem. 
Given an initial hypothesis (a prior), $P_t$, with $t=1, ..., n$, 
for the true momentum and reconstruction efficiency, $\epsilon_t$,
Bayes' theorem provides an estimator of the inverse response matrix elements, $\tilde{R}_{tm}$, 
\begin{equation}
  \tilde{R}_{tm} = \frac{R_{mt}P_t}{\epsilon_t \sum_{t^{\prime}}
    R_{mt^{\prime}}P_{t^{\prime}}}.
\end{equation}
The measured distribution, $M_m$, is thus unfolded as follows
\begin{equation}
 P^{\prime}_t = {\sum_{m} \tilde{R}_{tm} M_m},
\end{equation}
to obtain a posterior estimator, $P^{\prime}_t$, of the true distribution. The inversion is improved iteratively 
by recursively using posterior estimators to update and recalculate the inversion matrix. The number of iterations serves as a 
regularization parameter in the unfolding procedure.  For jet spectra studies, the measured spectra are used as prior 
and convergence is 
obtained typically after three iterations.\par
As an additional cross check, the analysis of charged jet cross
sections is also carried out with the RooUnfold implementation of the Singular Value
Decomposition (SVD) unfolding technique ~\cite{A44_unfold-svd, A45_RooUnfoldHtml} using raw measured spectra as  prior distributions. 
The performance of the Bayesian unfolding, SVD unfolding,  and bin-by-bin correction methods are compared based on
PYTHIA Perugia-0 simulated jets. The three methods produce results that are 
found to be within 4\% of the truth distribution. The 
cross sections reported in this work are obtained with the Bayesian unfolding method.
\subsection{Contamination from secondary particles}
\label{sec:secondaries}
Charged secondary particles are predominantly produced by weak decays of strange particles (e.g. $K^{\mathrm 0}_{\mathrm S}$ and $\Lambda$), decays of charged pions,  
conversions of photons from neutral pion decays and hadronic interactions in the detector material. The charged 
jet transverse momentum, jet shapes and fragmentation distributions include by definition only primary charged particles 
(prompt particles produced in the collisions and all decay products, except products from weak decays of strange particles such as $K^{\mathrm 0}_{\mathrm S}$ and $\Lambda$). Secondary particles introduce ambiguities in the jet energy scale and contribute to the raw reconstructed multiplicity, momentum density, 
and fragmentation distributions. Although their contribution is minimized by the analysis 
cuts described in Sec.~\ref{sec:expmethod}, 
the measured distributions nonetheless must be corrected for a small
residual contamination. The subtraction of the 
secondary particle contamination is implicitly included in the bin-by-bin correction applied for 
measurements of jet shape observables. 
It is however carried out separately and explicitly in the
measurements of the fragmentation function. 
The contribution of secondaries is estimated from MC simulations, separately for 
each bin in jet $p_{\rm T}$ and particle $p_{\rm T}$, $z^{\rm ch}$ 
and $\xi^{\rm ch}$. The correction applied to the
measured fragmentation functions is highest, up to 35\%, at 
small $p_{\rm T}$ and large $\xi^{\rm ch}$. It amounts to few percent only when averaged 
over all jet constituents. To enhance the low strangeness yield in the PYTHIA Perugia-0 simulations 
to the level observed in data, the contamination estimate is multiplied by a data-driven 
correction factor based on measurements~\cite{A46_CMSStrangeness} of strange particle 
production in non-single-diffractive events by the CMS collaboration and simulations from~\cite{A47_mcplots}. The contamination of secondaries from strange particle decays is small, and the effect of the 
strangeness scaling on the final result is less than 1\%. No scaling is 
applied on the correction to the jet spectrum and jet shape observables.

\subsection{Underlying event subtraction}
\label{sec:underlyingEventSub} 
There is no strict definition of the Underlying Event. Operationally, it corresponds to all particles produced in an event that are not an integral part of a jet or produced directly by hard scattering of
partons. The ATLAS~\cite{A48_ATLAS_UE, A49_ATLAS_UE_charged+neutral}, CMS~\cite{A50_CMS_UE} and ALICE~\cite{A51_ALICE_UE} collaborations have already published studies of UE in
pp collisions at $\sqrt{s}$~=~7~TeV. In this work, a similar
method is adopted to determine the UE yield and correct the measured jet observables 
for this source of contamination.\par
The UE particle yield is estimated event-by-event based on circular regions perpendicular to the measured jet cones. 
The circular regions have the same size as the jet resolution parameter and are placed at the same pseudorapidity as the leading jet but 
offset at an azimuthal angle ${\rm \Delta} \varphi = \pi$/2 relative to the jet axis.\par
For the jet cross section measurements, the UE is subtracted 
on a jet-by-jet basis prior to unfolding and the same treatment is applied to jets obtained 
from simulations before jet response matrix is created.\par
In the case of the fragmentation and jet shape observables, no correction for the UE contribution to the reconstructed 
jet energy is applied, but the UE contribution to the measured distributions in each bin of jet $p_{\rm T}$ is subtracted. The 
$p_{\rm T}$ spectra of particles in the perpendicular cone are accumulated and averaged over many 
events. To account for variations of the cone size of the anti-$k_{\rm T}$ jets, the spectra are weighted jet by jet with the ratio of the 
cone size, determined by FastJet, to the nominal aperture of $\pi R^2$ for a jet with resolution parameter $R$. The difference between the 
weighted and unweighted UE distributions is at the level of 1\%. The $\xi^{\rm ch}$ variable is computed jet-by-jet for 
each particle using the transverse momentum of the leading jet. The radial $p_{\rm T}$ sum 
distributions are obtained relative to the axis of the perpendicular cone.\par
The algorithms used for jet reconstruction are sensitive to statistical fluctuations of the particle density 
which are possibly enhanced by local variations of the detection efficiency and secondary particle production. This 
reconstruction bias may differ for the jet region and the UE
region. 
Hence, the UE distributions are 
corrected first for tracking efficiency, resolution and contamination from secondary particles. 
The fully corrected distributions are then subtracted in bins of
the leading jet transverse momentum. The correction is smaller than 2.5\% 
of the charged jet energy, but it is considerable for the fragmentation distributions at the lowest track momentum and 
highest $\xi^{\rm ch}$, where the ratio of UE background to fragmentation signal takes values
up to 2.5. No self-consistent technique exists to subtract the UE in
the $\langle R_{\rm 80} \rangle$
measurements, these measurements are therefore reported  without
correction for UE contamination. However, comparing the radial 
$\langle \rm{d}{\it p}_{\rm T}^{\rm sum}/\rm{d}{\it r} \rangle$ distributions before
and after UE subtraction, the increase in jet size $\langle R_{\rm 80} \rangle$ due
to the UE is estimated to be of the order of few percent only. 
The systematic
uncertainties for not performing the UE subtraction
are
thus found negligible 
compared to other sources of errors in the measurements of $\langle
R_{\rm 80} \rangle$.

\newcommand{\specialcell}[2][l]{\begin{tabular}[#1]{@{}l@{}}#2\end{tabular}}  
\renewcommand{\arraystretch}{1.5}

\begin{table}[bht]

\small 

\centering  

\begin{tabular}{ 
   m{2.9cm}             | 
   r                       
   m{1.45cm}<{\centering}   
   m{1.45cm}<{\centering}   
   m{1.45cm}<{\centering}   
   m{1.65cm}<{\centering}   
   m{1.1cm}<{\centering}   
   m{1.2cm}<{\centering} } 
 
 \hline \hline 
 
Distribution &  \rule{0.2cm}{0cm} \specialcell[c]{ \rule{0.30cm}{0cm} Bin\\
   [-0.2cm] (GeV/$c$) } & Track eff. (\%) & Track $p_{\rm T}$ res. (\%)& Unfolding (\%) &  Normalization (\%) & Sec. (\%) & Total (\%) \\ 
\hline 

\multirow {3}{*}{   $ \displaystyle \frac{ { \rm {d^{2}}}\sigma^{\rm jet, ch} } { {\rm d} p_{\rm T}\rm{d}\eta }$  ($R$ = 0.2)   }
&  20-24 &   \scriptsize \specialcell[c]{ $+$4.6     \\ [-0.2cm] $-$4.2  }   &4.0& 3.0   & 3.5   &1.9& \scriptsize \specialcell[c]{ $+$7.8   \\ [-0.2cm] $-$7.6   } \\ \cline{2-8}
&  50-58 &   \scriptsize \specialcell[c]{ $+$22.1   \\ [-0.2cm] $-$10.5  } & 4.0& 1.6   & 3.5   &2.5& \scriptsize \specialcell[c]{ $+$23.0   \\ [-0.2cm] $-$12.2   } \\ \cline{2-8} 
& 86-100&   \scriptsize \specialcell[c]{ $+$26.0   \\ [-0.2cm] $-$15.3  } & 4.0& 5.2   & 3.5   &2.8& \scriptsize \specialcell[c]{ $+$27.1   \\ [-0.2cm] $-$17.2   } \\ \cline{2-8}

\hline 

\multirow {3}{*}{ $ \displaystyle \frac{ {\rm {d^{2}}}\sigma^{\rm jet, ch} } { {\rm d} p_{\rm T}\rm{d}\eta  }$  ($R$ = 0.4)   } 
&  20-24 &   \scriptsize \specialcell[c]{ $+$7.5   \\ [-0.2cm] $-$4.5  } & 4.0   & 3.0   & 3.5 &2.1  & \scriptsize \specialcell[c]{ $+$9.9   \\ [-0.2cm] $-$7.9   } \\ \cline{2-8}
&  50-58 &   \scriptsize \specialcell[c]{ $+$23.2   \\ [-0.2cm] $-$10.6  } & 4.0   & 1.4   & 3.5 &2.5  & \scriptsize \specialcell[c]{ $+$24.0   \\ [-0.2cm] $-$12.2   } \\ \cline{2-8} 
& 86-100 &   \scriptsize \specialcell[c]{ $+$24.9   \\ [-0.2cm] $-$15.0  } & 4.0   & 5.6   & 3.5 &2.7  & \scriptsize \specialcell[c]{ $+$26.2   \\ [-0.2cm] $-$17.2   } \\ \cline{2-8}
\hline

\multirow {3}{*}{ $ \displaystyle \frac{ {\rm {d^{2}}}\sigma^{\rm jet, ch} } { {\rm d} p_{\rm T}\rm{d}\eta  }$  ($R$ = 0.6)   } 
&  20-24 &   \scriptsize \specialcell[c]{ $+$11.1   \\ [-0.2cm] $-$5.3  } & 4.0   & 6.6   & 3.5 &2.3  & \scriptsize \specialcell[c]{ $+$14.2   \\ [-0.2cm] $-$10.3   } \\ \cline{2-8}
&  50-58 &   \scriptsize \specialcell[c]{ $+$22.6   \\ [-0.2cm] $-$14.3  } & 4.0   & 1.9   & 3.5 &2.6  & \scriptsize \specialcell[c]{ $+$23.4   \\ [-0.2cm] $-$15.6   } \\ \cline{2-8} 
& 86-100 &   \scriptsize \specialcell[c]{ $+$23.7   \\ [-0.2cm] $-$13.7  } & 4.0   & 6.0   & 3.5 &2.7  & \scriptsize \specialcell[c]{ $+$25.1   \\ [-0.2cm] $-$16.1   } \\ \cline{2-8}

\hline 
\hline
\end{tabular}
\captionsetup{width = 1.0\textwidth}
\caption{\normalsize Summary of systematic uncertainties for selected bins in selected cross section distributions.}

\normalsize 

\label{sysTableXsec}
\end{table}

\section{Estimation of systematic uncertainties}\label{sec5syserrors}
A summary of all systematic uncertainties for selected bins is given in Table~\ref{sysTableXsec} for the cross section measurements, 
and in Table~\ref{sysTableJSFF} for the $\langle N_{\rm ch} \rangle $,  $\langle R_{\rm 80} \rangle$, 
$\langle \rm{d}{\it p}_{\rm T}^{\rm sum}/\rm{d}{\it r} \rangle$, $F^{p_{\rm
    T}}$, $F^{p_{\rm T}}$ and $F^{z}$ distributions. 
The uncertainties given in each column of the table are described in this section.  
\subsection{Tracking efficiency and resolution}\label{subsec7.1}
Uncertainties associated with the momentum resolution and charged track reconstruction 
efficiency  lead to systematic uncertainties in measurements of the  jet cross section, 
jet shapes, and  jet fragmentation functions.\par
The relative systematic uncertainty on tracking efficiency is estimated to be
5\% based on several variations of cuts used in the track selection
introduced earlier.
The relative systematic uncertainty on the track momentum resolution 
amounts to 20\%~\cite{A52_R_AA_momentumResolution}.\par
In order to evaluate the effect of these uncertainties on the measured jet cross sections,
the corresponding rescaled response matrix is used to unfold the spectra.
For the jet shape and fragmentation observables, the impact of the finite detector efficiency 
and momentum resolution on the bin-by-bin correction factors is estimated by applying parametrized 
detector response to PYTHIA 
events clustered with FastJet, and varying the efficiency and resolution independently. 
Systematic uncertainties for the jet particle multiplicity and jet shape observables are given in Table~\ref{sysTableJSFF} 
for a resolution parameter $R$ = 0.4. For larger (smaller) $R$, a moderate increase (decrease) of the uncertainties 
is observed related to tracking efficiency. For the fragmentation distributions, variations of the momentum resolution induce the most significant changes at high track $p_{\rm T}$. 
The systematic uncertainties due to the efficiency variations are largest at the highest $z^{\rm ch}$ and smallest at intermediate values. 
\subsection{Bin-by-bin correction}
\label{sec:sysEvGen}
The data correction methods used in this work are largely based on tune Perugia-0 of the PYTHIA event generator. The 
particular structure of jets produced by PYTHIA might however conceivably affect the magnitude, and dependencies 
of the correction factors on the jet momentum, particle momentum, or radial dependence $r$. The possible 
impact of such event generator dependencies is examined by comparing the amplitude of the bin-by-bin corrections obtained with PYTHIA
tunes Perugia-0 and Perugia-2011, with those obtained with the HERWIG generator. This is accomplished with a parametrized 
detector response and the anti-$k_{\rm T}$ jet finder. In addition, the impact of modifications of the 
jet fragmentation is studied by artificially duplicating and removing jet particles with a momentum dependent probability. 
The variations are constrained to be at
a similar level as the differences observed between simulations and
data reaching 
up to a factor of 2.5 for values of $z^{\rm ch}$ close to 1 in the fragmentation distributions. The charged 
particle multiplicity is affected by $\sim$30\%. The resulting systematic uncertainties are largest for high 
values of $z^{\rm ch}$ and track $p_{\rm T}$ and small values of $\xi^{\rm ch}$.\par
As an independent check, a closure test with a 2-dimensional folding technique is carried out on the fragmentation distributions 
from an inclusive jet sample (comprising leading and sub-leading jets). A response matrix in bins of 
generated and reconstructed jet $p_{\rm T}$ and particle (scaled)
transverse momentum is used
to fold the corrected results back to the uncorrected level. Since the folding method has negligible dependence on 
the event generator, the comparison of the folded to the original distributions 
reveals possible biases of the bin-by-bin correction. The observed non-closure at the level of few percent is consistent with 
the systematic uncertainty assigned to the bin-by-bin correction from
modifications of the fragmentation pattern. 
\subsection{Response unfolding}
The unfolding techniques used in this work correct the measured jet spectra for the detector response. The limited measurement resolution, 
discussed in Sec.~\ref{secMC}, results in a small, but finite, probability for bin migration of the reconstructed 
jet momentum relative to the true value. 
Consequently, the unfolding introduces a correlation between neighbouring bins of the corrected spectrum, and statistical fluctuations 
in the measured data result in a spectral shape systematic uncertainty. 
To assess this uncertainty, the raw jet spectra are smeared by a Gaussian function 
with a width given by the statistical uncertainty in the given momentum bin. The resulting spectra 
are then unfolded and the systematic uncertainty is evaluated as a spread of the corrected spectra. 
The value of this systematic uncertainty increases roughly linearly
with $p_{\rm T}^{\rm jet,ch}$, reaching
a maximum value of $\sim$7\% at $p_{\rm T}^{\rm jet,ch} \approx$~100~GeV/$c$.
\subsection{Underlying event subtraction} 
In this work, we use perpendicular cones to measure and subtract the UE as described in Sec.~\ref{sec:underlyingEventSub}. 
However, there is no unique prescription on how to determine the UE. In a prior, trigger hadron based, UE analysis by the 
ALICE collaboration~\cite{A51_ALICE_UE}, a geometrically different definition of the transverse region was used. 
The charged particle transverse momentum densities obtained in our analysis are consistent with the saturation values in the 
transverse region measured in~\cite{A51_ALICE_UE}. In~\cite{A54_CDF_Dijet_UE}, the UE was estimated 
from dijet events and imposing an additional veto on a third jet. An alternative simulation to 
estimate and subtract the UE in a similar way is performed using particle level output from a MC event generator. The UE is measured from events with a dijet in the detector
acceptance, to understand if and how the non-leading jet affects the
UE estimate, rejecting events with additional charged jets with a $p_{\rm T}$ exceeding 12~GeV/$c$. The resulting difference on the fragmentation distributions 
is used to assign a 5\% systematic uncertainty to the estimated UE. 
The resulting systematic uncertainty on the fragmentation distributions is highest at 
low transverse momenta. Systematic uncertainties on $\langle \rm{d}{\it
  p}_{\rm T}^{\rm sum}/\rm{d}{\it r} \rangle$
are largest at large distances $r$ in the jet $p_{\rm T}$ interval 20 - 30~GeV/$c$. 
The uncertainty increases for higher values of the resolution parameter $R$. 
Systematic uncertainties on the measured charged jet cross sections
are smaller than 1\% and considered negligible.\par
The anti-$k_{\rm T}$ jet finder typically produces circular jet cones, and the UE contribution to the jet shapes and fragmentation distributions 
is evaluated consistently in circular cones. In individual jets, particles may however be added at a distance $r \geq R$ thereby giving rise to a convex deformation of the cone. 
Concave deformations might also occur. The dependence of the
fragmentation distributions on the cone shape is checked by repeating the analysis using only tracks in an ideal 
cone around the jet axis. In this case no jet area scaling of the UE is applied. The low momentum particle 
yield is most affected: at high jet radii, low $z^{\rm ch}$ fragmentation dominates over high $z^{\rm ch}$ fragmentation. In addition, 
the probability to collect a soft particle from the UE is comparatively higher 
than at small $r$. The observed effect is negligibly small: a maximum depletion 
of 4\% of the particle yield at the highest $\xi^{\rm ch}$ in the smallest jet momentum bin is observed. Considerably smaller 
variations are found for all other jet momenta and $\xi^{\rm ch}$ bins. The effect is reproduced in MC simulations, 
and no systematic uncertainty is associated to the jet cone shape. 
\subsection{Cross section normalization} 
The determination of luminosity and related systematic uncertainties
are discussed in~\cite{A53_RefLumiUnc,A53_RefLumiUnc_1}. A normalization uncertainty of
3.5\% is assigned to the cross section measurement. 
\subsection{Contamination from secondary particles} 
The reconstructed primary particles originate from the main interaction vertex and have a non-zero 
distance of closest approach DCA because of finite resolution effects. The DCA of secondaries 
however spans a much broader range of values. Reducing 
the maximum allowed DCA value reduces contaminations from secondaries but also reduces the detection 
efficiency of primary particles. 
In this analysis, primary particles are selected requiring a small DCA as 
discussed in Sec.~\ref{sec:expmethod}, and a correction for the residual contribution of secondary 
particles is applied, as explained in Sec.~\ref{sec:secondaries}. The
systematic uncertainty associated to the correction is estimated by reducing the maximum allowed DCA 
used in the selection of primary tracks by more than a factor of 9 using a $p_{\rm T}$ dependent 
cut. The resulting fragmentation distributions are corrected 
consistently for contamination and cut efficiency and residual differences in the fully corrected spectra 
are assigned as systematic uncertainty. The highest uncertainty is 
found for large values of $\xi^{\rm ch}$. 

The dependence of the correction on the strange particle yield in the PYTHIA Perugia-0 
simulations is estimated from comparison to data as explained in Sec.~\ref{sec:secondaries}. 
The effect on the jet cross sections is less than 3\% and is assigned as systematic 
uncertainty. For the jet shape observables it is negligible. 

\begin{table}[th!f]
\small 

\centering  

\begin{tabular}{ 
m{3.4cm}             | 
r                       
m{1.25cm}<{\centering}   
m{1.25cm}<{\centering}   
m{1.25cm}<{\centering}   
m{1.25cm}<{\centering}   
m{1.1cm}<{\centering}   
m{1.0cm}<{\centering} } 

\hline \hline 

Distribution &  \rule{1.2cm}{0cm} Bin \rule{0.4cm}{0cm}  & Track
eff. (\%) & Track $p_{\rm T}$ res. (\%) & Bin-by-bin corr. (\%)& UE (\%) & Sec. (\%) & Total (\%) \\ 

\hline 

\multirow {2}{*}{ $\langle N_{\rm ch} \rangle $ }  
& 20-25~GeV/$c$   
& \scriptsize \specialcell[c]{ $+$5.8 \\   [-0.2cm] $-$5.0 } 
& \scriptsize \specialcell[c]{ $+$4.0 \\   [-0.2cm] $-$3.5 } 
& \scriptsize \specialcell[c]{ $+$0.7 \\   [-0.2cm] $-$0.9 } 
& \specialcell[c]{ 0.8 } 
& \scriptsize negligible  
& \scriptsize \specialcell[c]{ $+$7.1 \\   [-0.2cm] $-$6.2 } 
\\\cline{2-8}

& 80-100~GeV/$c$ 
& \scriptsize \specialcell[c]{ $+$5.8 \\   [-0.2cm] $-$5.0 } 
& \scriptsize \specialcell[c]{ $+$4.0 \\   [-0.2cm] $-$3.5 } 
& \scriptsize \specialcell[c]{ $+$0.7 \\   [-0.2cm] $-$0.9 } 
& \specialcell[c]{ 0.5  } 
& \scriptsize negligible 
& \scriptsize \specialcell[c]{ $+$7.1 \\   [-0.2cm] $-$6.2 } 
\\ \cline{2-8}

\hline

\multirow {2}{*}{ $\langle R_{\rm 80} \rangle $ }  
& 20-25~GeV/$c$  
& \scriptsize \specialcell[c]{ $+$6.1 \\ [-0.2cm] $-$5.5 }   
& \scriptsize \specialcell[c]{ $+$3.6 \\ [-0.2cm] $-$4.3 }   
& \scriptsize \specialcell[c]{ $+$1.7 \\ [-0.2cm] $-$1.7 }   
& $-$   
& $-$   
& \scriptsize \specialcell[c]{ $+$7.2 \\ [-0.2cm] $-$7.2 }   
\\ \cline{2-8}

& 80-100~GeV/$c$ 
& \scriptsize \specialcell[c]{ $+$6.1 \\ [-0.2cm] $-$5.5 }   
& \scriptsize \specialcell[c]{ $+$3.6 \\ [-0.2cm] $-$4.3 }   
& \scriptsize \specialcell[c]{ $+$1.7 \\ [-0.2cm] $-$1.7 }   
&  $-$   
&  $-$    
& \scriptsize \specialcell[c]{ $+$7.2 \\ [-0.2cm] $-$7.2 }   
\\ \cline{2-8}
\hline

\multirow {3}{*}{ \specialcell[c]{ $  \displaystyle  \langle \frac{\rm {d}{\it p}_{\rm T}^{\rm sum}}{\rm {d}{\it r}} \rangle $ 
   \\ 20$<p_{T}^{\rm jet, ch}<$30~GeV/$c$ } } 
& 0.00 - 0.04  
& \scriptsize \specialcell[c]{ $+$8.1 \\       [-0.2cm] $-$6.5 }  
& \scriptsize \specialcell[c]{ $+$5.9 \\       [-0.2cm] $-$2.4 }  
& \scriptsize \specialcell[c]{ $+$2.9 \\       [-0.2cm] $-$3.1 }  
& \scriptsize negligible   
& \scriptsize negligible  
& \scriptsize \specialcell[c]{ $+$10.4 \\       [-0.2cm] $-$7.5 }  
\\ \cline{2-8}

& 0.20 - 0.24 
& \scriptsize \specialcell[c]{ $+$8.1 \\       [-0.2cm] $-$6.5 }  
& \scriptsize \specialcell[c]{ $+$5.9 \\       [-0.2cm] $-$2.4 }  
& \scriptsize \specialcell[c]{ $+$2.9 \\       [-0.2cm] $-$3.1 }  
& 0.3   
& \scriptsize negligible  
& \scriptsize \specialcell[c]{ $+$10.5 \\       [-0.2cm] $-$7.6 }  
\\ \cline{2-8}

& 0.36 - 0.40 
& \scriptsize \specialcell[c]{ $+$8.1 \\   [-0.2cm] $-$12.0 }    
& \scriptsize \specialcell[c]{ $+$5.9 \\   [-0.2cm] $-$2.4 }  
& \scriptsize \specialcell[c]{ $+$2.9 \\   [-0.2cm] $-$3.1 }  
& 15.0   
& \scriptsize negligible  
& \scriptsize \specialcell[c]{ $+$18.3 \\ [-0.2cm] $-$19.6 }  
\\ \cline{2-8}

\hline

\multirow {3}{*}{ \specialcell[c]{ $  \displaystyle  \langle \frac{\rm {d}{\it p}_{\rm T}^{\rm sum}}{\rm {d}{\it r}} \rangle $ 
   \\ 60$<p_{T}^{\rm jet, ch}<$80~GeV/$c$ } }
& 0.00 - 0.04 
& \scriptsize \specialcell[c]{ $+$10.6 \\   [-0.2cm] $-$5.1 }    
& \scriptsize \specialcell[c]{ $+$5.6 \\   [-0.2cm] $-$6.5 }    
& \scriptsize \specialcell[c]{ $+$3.7 \\   [-0.2cm] $-$3.4 }    
& \scriptsize negligible   
& \scriptsize negligible     
& \scriptsize \specialcell[c]{ $+$12.6 \\   [-0.2cm] $-$8.9 }    
\\ \cline{2-8}

& 0.20 - 0.24 
& \scriptsize \specialcell[c]{ $+$10.6 \\   [-0.2cm] $-$5.1 }    
& \scriptsize \specialcell[c]{ $+$5.6 \\   [-0.2cm] $-$6.5 }    
& \scriptsize \specialcell[c]{ $+$3.7 \\   [-0.2cm] $-$3.4 }    
& 0.4    
& \scriptsize negligible     
& \scriptsize \specialcell[c]{ $+$12.6 \\   [-0.2cm] $-$9.0 }     
\\ \cline{2-8}

& 0.36 - 0.40 
& \scriptsize \specialcell[c]{ $+$10.6 \\   [-0.2cm] $-$5.1 }    
& \scriptsize \specialcell[c]{ $+$5.6 \\   [-0.2cm] $-$6.5 }    
& \scriptsize \specialcell[c]{ $+$3.7 \\   [-0.2cm] $-$3.4 }  
&  1.6
& \scriptsize negligible       
& \scriptsize \specialcell[c]{ $+$12.7 \\   [-0.2cm] $-$9.1 }      
\\ \cline{2-8}

\hline

\multirow {3}{*}{ \specialcell[c]{$ F^{p_{\rm T}}$  \\ 20$<p_{T}^{\rm jet, ch}<$30~GeV/$c$} } 
&   0 - 1~GeV/$c$   & 5.0 &    0.1    & 0.7  &    3.3    &  3.2 &   6.8  \\ \cline{2-8}
&   6 - 7~GeV/$c$   & 0.8 & \scriptsize negligible & 2.3 & \scriptsize negligible &  0.5 &   2.4  \\ \cline{2-8}
&  18 -20~GeV/$c$ & 9.9&    0.5    & 6.0  & \scriptsize negligible &  0.4 &  11.6  \\ \cline{2-8}

\hline

\multirow {3}{*}{ \specialcell[c]{$ F^{p_{\rm T}}$  \\ 60$<p_{T}^{\rm jet, ch}<$80~GeV/$c$} } 
&   0 -  5~GeV/$c$   &  5.2  &    0.3     & 0.2 &    0.8    &2.1  &   5.7  \\ \cline{2-8}
&  20 - 30~GeV/$c$ &  1.4  & \scriptsize negligible  & 3.7 & \scriptsize negligible & 0.6 &   4.0  \\ \cline{2-8}
&  50 - 60~GeV/$c$ & 10.5 & 3.5        & 9.6 & \scriptsize negligible & 0.6 &  14.6  \\ \cline{2-8}

\hline

\multirow {3}{*}{ \specialcell[c]{$ F^{z}$ \\ 20$<p_{T}^{\rm jet, ch}<$30~GeV/$c$} } 
	& 0   - 0.1  &  4.7 &   1.6  & 0.2 & 1.6   & 1.4 &   5.2  \\ \cline{2-8}
       & 0.3 - 0.4  &  0.4 & \scriptsize negligible & 2.7 & \scriptsize negligible & 0.3 &   2.8  \\ \cline{2-8}
       & 0.9 - 1.0  & 15.5 &    1.1 & 4.8 & \scriptsize negligible & 0.6 &  16.3  \\ \cline{2-8}

\hline

\multirow {3}{*}{ \specialcell[c]{$ F^{z}$  \\ 60$<p_{T}^{\rm jet, ch}<$80~GeV/$c$} } 
	&   0 - 0.1  & 5.0  & 0.3 & 0.3 &    0.7 & 1.3 &   5.3  \\ \cline{2-8}
       & 0.3 - 0.4  & 1.2  & 0.2 & 3.7 & \scriptsize negligible & 0.4 &   3.9  \\ \cline{2-8}
       & 0.8 - 1.0  & 13.8 & 3.1 & 6.1 & \scriptsize negligible & 1.2 &  15.4  \\ \cline{2-8}

\hline

\multirow {3}{*}{ \specialcell[c]{$ F^{\xi}$  \\ 20$<p_{T}^{\rm jet, ch}<$30~GeV/$c$} } 
	&   0 - 0.4  & 9.9 &    0.5 & 4.6 & \scriptsize negligible & 0.7 &  10.9 \\ \cline{2-8}
       & 0.8 - 1.2  & 0.6 & \scriptsize negligible & 3.0 & \scriptsize negligible & 0.5 &   3.1 \\ \cline{2-8}
       & 4.8 - 5.3  & 5.1 &    0.7 & 0.9 &   15.3 & 7.8 &  17.9 \\ \cline{2-8}

\hline

\multirow {3}{*}{ \specialcell[c]{$ F^{\xi}$  \\ 60$<p_{T}^{\rm jet, ch}<$80~GeV/$c$} } 
	&   0 - 1.0  & 5.0 & 0.5 & 3.9 & \scriptsize negligible & 0.7 &   6.4  \\ \cline{2-8}
       & 1.0 - 2.0  & 1.3 & 0.4 & 3.4 & \scriptsize negligible & 0.6 &   3.8  \\ \cline{2-8}
       & 5.0 - 6.2  & 5.7 & 0.2 & 0.7 &    6.5 & 6.2 &  10.6  \\ \cline{2-8}

\hline 
\hline
\end{tabular}
\captionsetup{width = 1.0\textwidth}
\caption{\normalsize Summary of systematic uncertainties for selected bins in
 selected jet shape and fragmentation distributions for $R$ = 0.4.}
\normalsize 
\label{sysTableJSFF}
\end{table}

\begin{figure}[th]
 \centering
 \includegraphics[width=0.5\textwidth]{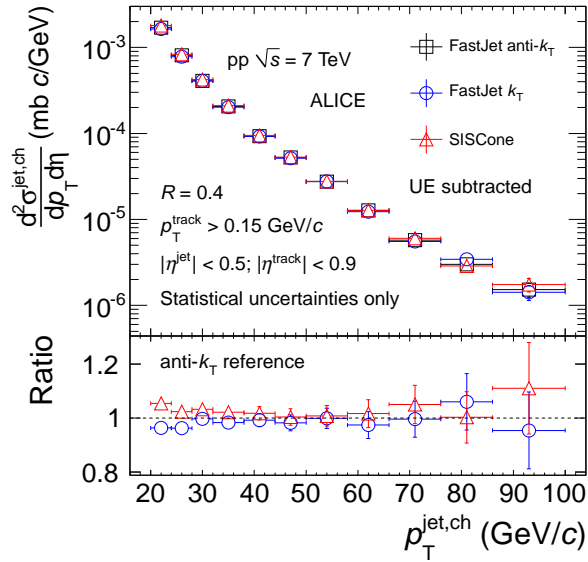}
 \caption{(Color online) Top panel: Charged jet cross sections in pp collisions 
   at $\sqrt{s}$~=~7~TeV.
   Symbols correspond to different algorithms used for
   jet reconstruction. Bottom panel: Ratios between jet cross sections obtained by $k_{\rm T}$, and SISCone 
   to that obtained by anti-$k_{\rm T}$.}
 \label{dndpt4method}
\end{figure} 

\section{Results}\label{sec6results}
\subsection{Comparison of jet finding algorithms}
\label{algocomp} 
Figure~\ref{dndpt4method} (top panel) shows the differential 
cross sections of charged jet production measured in pp collisions at $\sqrt{s}$~=~7~TeV
using the $k_{\rm T}$, anti-$k_{\rm T}$, and SISCone jet finding
algorithms.
The distributions are obtained with a resolution parameter, $R$~=~0.4, for 
jets in the pseudorapidity range $|\eta^{\rm jet}| <$~0.5, and transverse 
momenta from 20 to 100~GeV/$c$. The bottom panel of the figure displays the ratios between
the cross sections obtained with the $k_{\rm T}$, and SISCone algorithms to those obtained 
with the anti-$k_{\rm T}$ as a function of the jet transverse momentum. 
For a correct treatment of statistical correlations between the numerator and denominator, 
the data were divided into fully correlated and uncorrelated subsets. 
The distributions are corrected using the 
bin-by-bin correction
procedure described in Sec.~\ref{binBybin}. 
The ratios of the jet cross sections are 
consistent with unity over nearly the entire range of jet transverse momenta spanned by this 
analysis. A significant 
deviation of 5\% is observed only in the lowest $p_{\rm T}$ 
bin ($p_{\rm T}^{\rm jet,ch}$~=~20-24~GeV/$c$) between the SISCone and anti-$k_{\rm T}$ algorithms.
For larger $p_{\rm T}^{\rm jet,ch}$ SISCone and $k_{\rm T}$ algorithms agree within errors with the anti-$k_{\rm T}$ 
algorithm. These observations are in good agreement with that obtained using PYTHIA Perugia-0 simulation (not shown).\par
The anti-$k_{\rm T}$ algorithm 
initiates particle clustering  around the highest $p_{\rm T}$ particles of an event. 
In contrast, the $k_{\rm T}$ algorithm initiates jet finding by clustering particles with 
the lowest momenta. It is thus rather sensitive to events with a large, fluctuating 
density of low momentum particles 
as produced in A--A collisions. The anti-$k_{\rm T}$ algorithm does not exhibit such 
sensitivity and is thus favored for studies of jet production in A--A
collisions. 
Since there are no large differences observed between the spectra obtained
with the three jet finders discussed above, and considering the fact that the
results of this work will be used as a reference for similar
measurements in A--A and p--A collisions, 
the remainder of the analyses presented in this work are performed with the  anti-$k_{\rm T}$ 
algorithm exclusively. 

\begin{figure}[th]
 \begin{center}
   \includegraphics[width=0.5\textwidth]{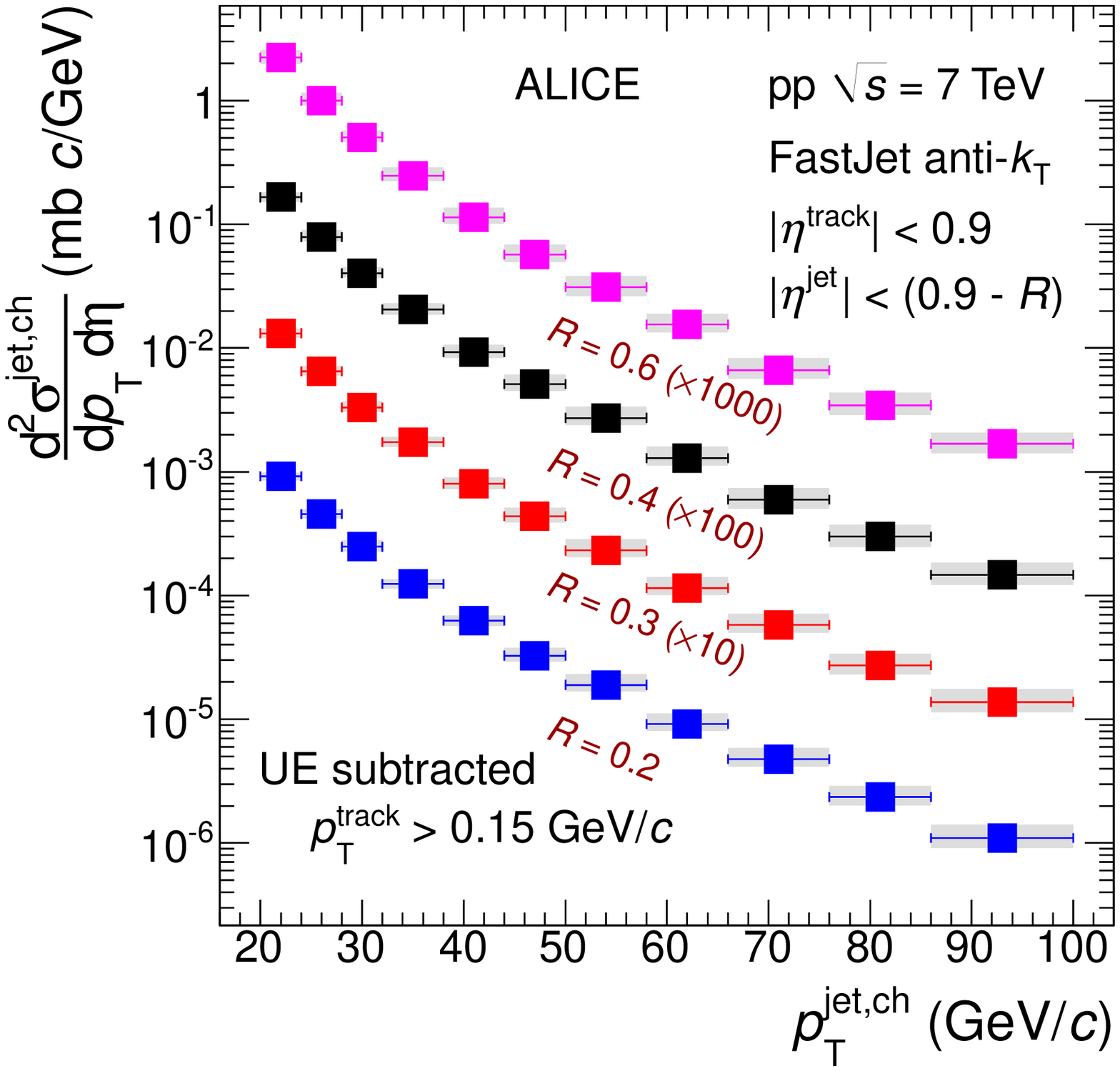} 
   \caption{(Color online) Inclusive charged jet cross sections in pp collisions at $\rm{\sqrt{s}=7}$~TeV using 
     the anti-$k_{\rm T}$ algorithm with $R$~=~0.2 (0.3, 0.4, and
     0.6) within $\left| \eta^{\rm jet} \right| \leq {\rm 0.7}$ ($\left| \eta^{\rm jet} \right| \leq {\rm 0.6}$, $\left|
       \eta^{\rm jet} \right| \leq {\rm 0.5}$, and $\left| \eta^{\rm jet} \right| \leq {\rm 0.3}$).} 
   \label{result-spectra-different-R}
 \end{center}
\end{figure}

\begin{figure}[th!f]
   \begin{center}
     \includegraphics[width=0.9\textwidth]{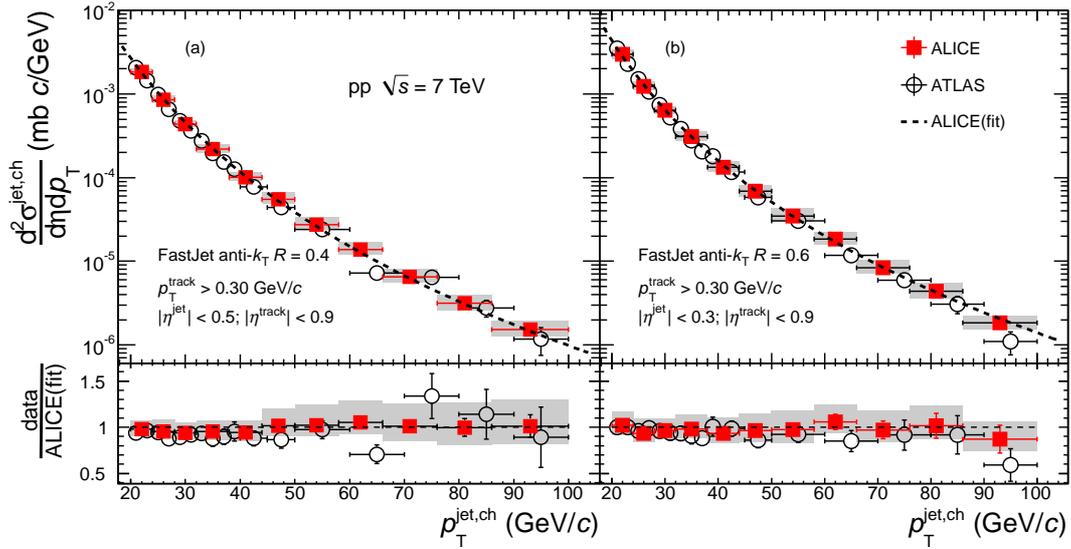} 
     \caption{(Color online) Top panels: Comparison of the charged jet cross section in the ALICE and the ATLAS~\cite{A3_ATLASchJets} 
       experiments in pp collisions at $\rm{\sqrt{s}=7}$~TeV. Statistical and systematic uncertainties are shown separately
       for ALICE data points, the gray bands indicating the systematic
       uncertainties, while for the ATLAS data points, the error bars
       show the statistical
       and systematic uncertainties summed in quadrature.
       The dotted line represents a Tsallis fit used to parametrize the ALICE data. 
       Bottom panels: The ratio of the ALICE and ATLAS charged jet spectrum to the parametrized ALICE data. 
       Note that the labels in the figures correspond to the ALICE
       measurements (see text for details).}
     \label{result-spectra-ATLAScomparison}
   \end{center}
 \end{figure}

\subsection{Charged jet cross section}\label{jetspec} 
Figure~\ref{result-spectra-different-R} presents the fully corrected inclusive charged jet cross section measured in 
pp collisions at $\rm{\sqrt{s}~=}$~7~TeV using the anti-$k_{\rm T}$
jet finder. 
Corrections for the detector response and instrumental effects are carried out using the Bayesian unfolding 
method presented in Sec.~\ref{sec:unfolding}. The distributions are also corrected for UE contamination on 
an event-by-event basis according to the method described in Sec.~\ref{sec:underlyingEventSub}. 
Inclusive charged jet cross sections are reported for resolution parameter values $R$~=~0.2, 0.3, 0.4 and 0.6,
and limited to pseudorapidity ranges $|\eta|~\textless$~(0.9 - $R$) in order to avoid losses due to partially 
reconstructed jets at the edge of the pseudorapidity acceptance. Statistical uncertainties are displayed as vertical 
error bars. Individual sources of systematic uncertainties are $p_{\rm
  T}$ dependent. 
 \begin{figure}[th!f]
  \begin{center}
    \includegraphics[width=1.0\textwidth]{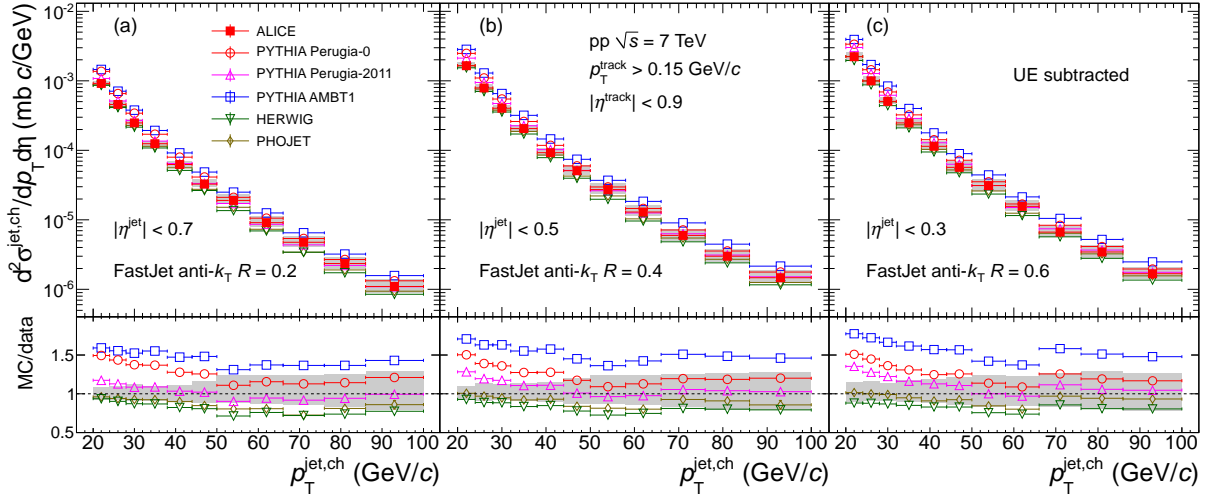}
    \caption{(Color online) Top panels: Charged jet cross sections
      measured in the ALICE experiment in pp collisions at
      $\sqrt{s}$ = 7 TeV compared to several MC generators: PYTHIA AMBT1,
      PYTHIA Perugia-0 tune, PYTHIA
      Perugia-2011 tune, HERWIG, and PHOJET. Bottom panels: Ratios
      MC/Data. Shaded bands show quadratic sum of statistical and
       systematic uncertainties on the data drawn at unity.}
     \label{result-spectra-generatorcomparison}
   \end{center}
\end{figure}
\begin{figure}[ht]	
 \begin{center}
   \includegraphics[width=0.49\textwidth]{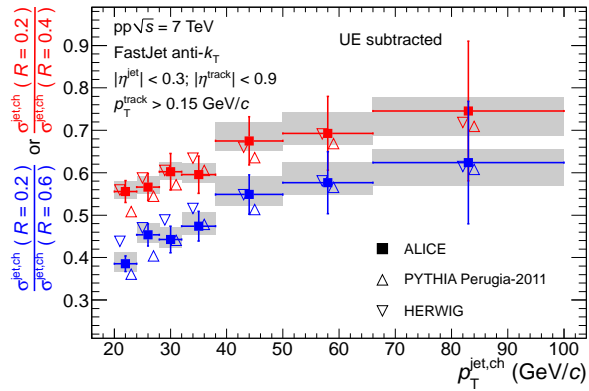}
   \caption{(Color online) Ratios of jet cross sections for charged jets reconstructed using anti-$k_{\rm T}$ algorithm with resolution  
     parameters 0.2 and 0.4 and  0.2 and 0.6. The jet acceptance is restricted 
     to $\left|\eta^{\rm jet} \right| \leq {\rm 0.3}$. The ratios in data are compared to PYTHIA Perugia-2011 
     and HERWIG simulations.}
   \label{R-ratios}	
 \end{center}
\end{figure}
In Fig.~\ref{result-spectra-different-R} as well as
in all other figures the data points are placed at the bin centre
along the abscissa and the horizontal error bars indicate the bin
width while the vertical error bars indicate the statistical
uncertainties. The total systematic uncertainties are obtained as a quadratic sum of
individual systematic uncertainties, as described in
Sec.~\ref{sec5syserrors}, and are shown as shaded bands around the
data points in Fig.~\ref{result-spectra-different-R} as well as in all
other figures. \par
The measured charged jet cross sections are compared to those 
reported by the ATLAS experiment~\cite{A3_ATLASchJets} at $R$~=~0.4
and 0.6 in Fig.~\ref{result-spectra-ATLAScomparison}. 
The ATLAS charged jets are measured in the rapidity
$\rm{\left| \emph{y}  \right| \leq 0.5}$ at both $R$ = 0.4 and
0.6, using charged tracks with 
$p_{\rm T} \geq$~0.3~GeV/$c$ without underlying event
subtraction. The ALICE therefore also uses the same track $p_{\rm
  T}$ selection without underlying event subtraction unlike
Fig.~\ref{result-spectra-different-R}.
To quantify the level of agreement between the ALICE and ATLAS jet
cross section measurements, 
the ALICE data are fitted with a modified 
Tsallis~\cite{A55_Tsallis,A55_Tsallis_1} distribution ($f\left(p_{\rm T}\right) = a \cdot
\left(1 + \frac{p_{\rm T}}{b} \right)^{-c}$).
The Tsallis fits are shown as dotted black curves in the 
top panels of Fig.~\ref{result-spectra-ATLAScomparison}. The
$\chi^2/dof$ of the fits are $2.97/8$ and $4.27/8$ for $R$~=~0.4 and 0.6 
respectively.
The bottom panels of Fig.~\ref{result-spectra-ATLAScomparison} show 
the ratios of the ALICE and ATLAS data points to the fit function. 
\begin{figure}[th]
 \includegraphics[scale=0.402]{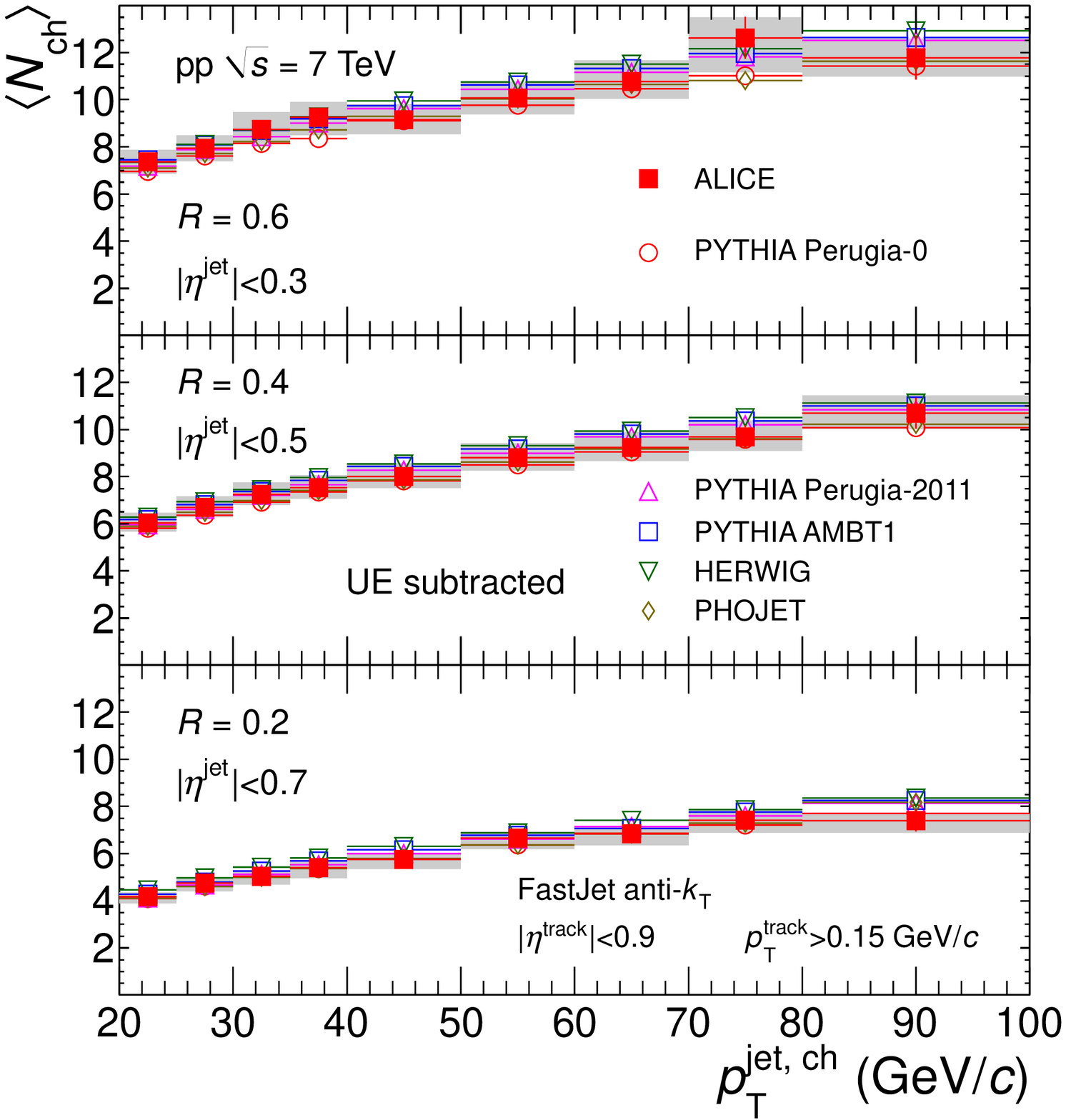}
 \includegraphics[scale=0.402]{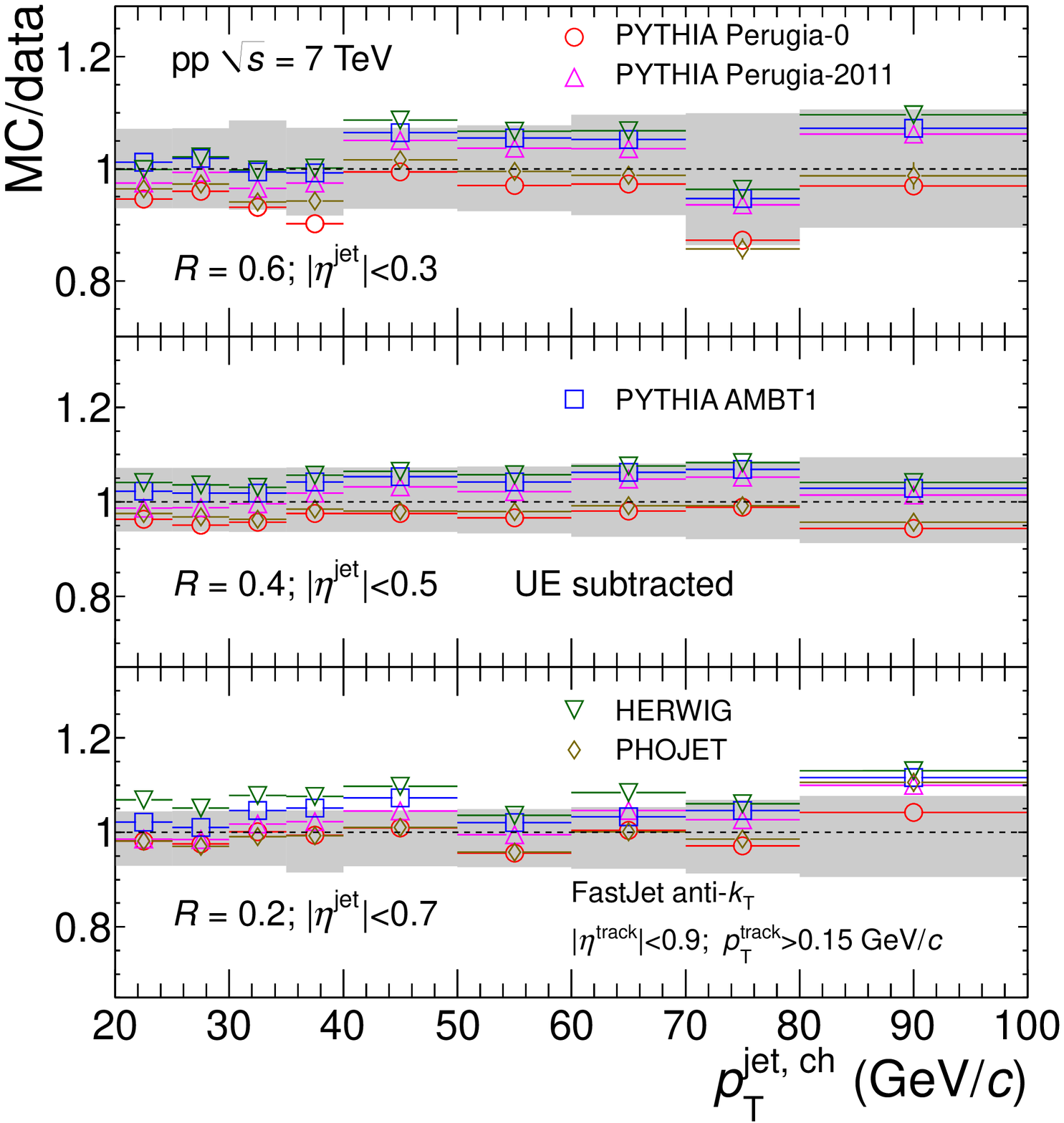}
 \caption{(Color online) Left panel: Mean charged particle multiplicity in the leading charged jet as a
   function of jet $p_{\rm T}$ compared to MC models for pp
   collisions at $\sqrt{s}$ = 7~TeV for various jet resolution
   parameters ($R$ = 0.6 (left top), $R$ = 0.4 (left middle) and $R$ = 0.2
   (left bottom)). UE contributions are subtracted from
   both data and MC. Right panel: Ratios MC/data. Shaded bands show the quadratic sum
   of statistical and systematic
   uncertainties on the data drawn at unity.}
 \label{Fig.nCh}
\end{figure}
The gray bands represent the systematic uncertainties on ALICE data points.
Despite fluctuations in the high $p_{\rm T}$ range of the ATLAS data, both datasets are in excellent agreement.\par
In the top panels of Fig.~\ref{result-spectra-generatorcomparison},
the measured
charged jet cross sections are compared to predictions from
PYTHIA (tunes Perugia-0, Perugia-2011, and AMBT1), PHOJET,
and HERWIG for $R$~=~0.2, 0.4 and 0.6.
The ratios of the MC simulations to measured data are shown in the bottom
panels of Fig.~\ref{result-spectra-generatorcomparison}.
In the high $p_{\rm T}$ range, PYTHIA Perugia-2011 describes
the data best, while in the low $p_{\rm T}$ range data is
best described by HERWIG and PHOJET. All PYTHIA tunes systematically
overestimate the measured data in the low transverse momentum
range and the discrepancy increases with increasing cone size.
The worst discrepancy 
with the data is observed for the PYTHIA tune AMBT1, which
overestimates the data by factors ranging from 25\% to 75\% 
over the studied $p_{\rm T}$ range 
for $R$~=~0.2. The disagreement grows with
increasing resolution parameter, and is worst for $R$~=~0.6. \par
Figure~\ref{R-ratios} shows the ratios of cross sections for jets with 
resolution parameters $R$~=~0.2, $R$~=~0.4 and $R$~=~0.2, $R$~=~0.6.
The ratio of jet spectra~\cite{A15_FullJetPaper} is sensitive to the collimation of particles
around the jet axis and serves as an indirect measure of the jet structure
used particularly in A--A 
collisions~\cite{A56_refAlicePbPbChargedJetPaper}, where large background
fluctuations greatly complicate jet shape studies. 
In order to compare the ratios within the same jet pseudorapidity
range, the ratios are studied 
within $|\eta|<$~0.3, which coincides with the fiducial jet acceptance for the largest 
resolution parameter studied ($R$~=~0.6). To avoid statistical 
correlations between the numerator and denominator, disjoint subsets of the  
data are used. The measured ratios are also compared to those from 
PYTHIA Perugia-2011 and HERWIG simulations. The measured ratios confirm the expected trend of increased collimation with increasing transverse 
momentum of jets, corroborated also by the simulation results. 
At high $p_{\rm T}$ ($>$~30~GeV/$c$), both PYTHIA and HERWIG are in
good agreement with the data within uncertainties. However at low
$p_{\rm T}$ ($<$ 30~GeV/$c$) PYTHIA tends to underpredict the data for
both the ratios 
whereas HERWIG tends to
overpredict the data for the ratio $\sigma^{\rm jet, ch}$($R$ = 0.2) /
$\sigma^{\rm jet, ch}$ ($R$ = 0.6).

\begin{figure}[th!f]
 \includegraphics[scale=0.52]{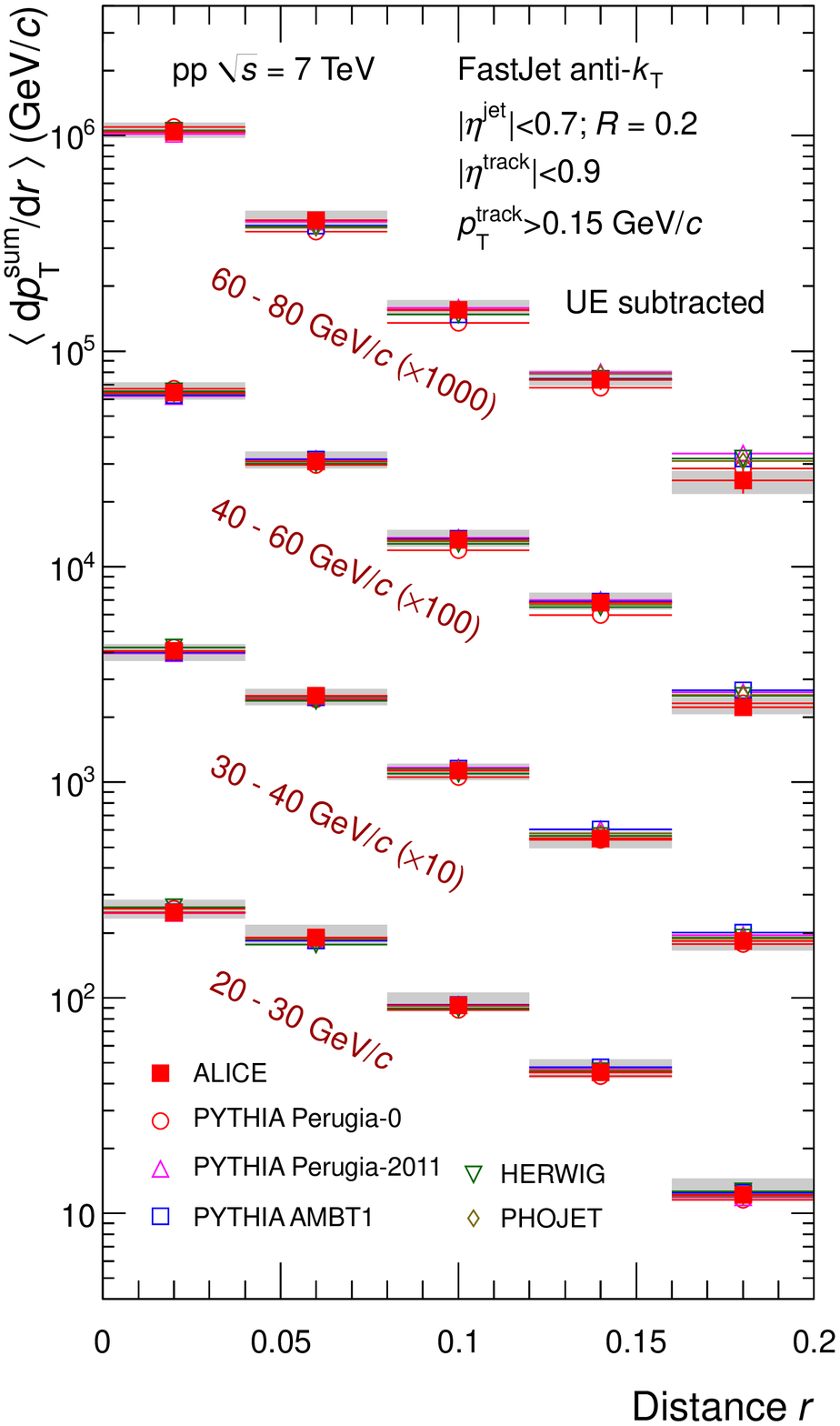}
 \includegraphics[scale=0.52]{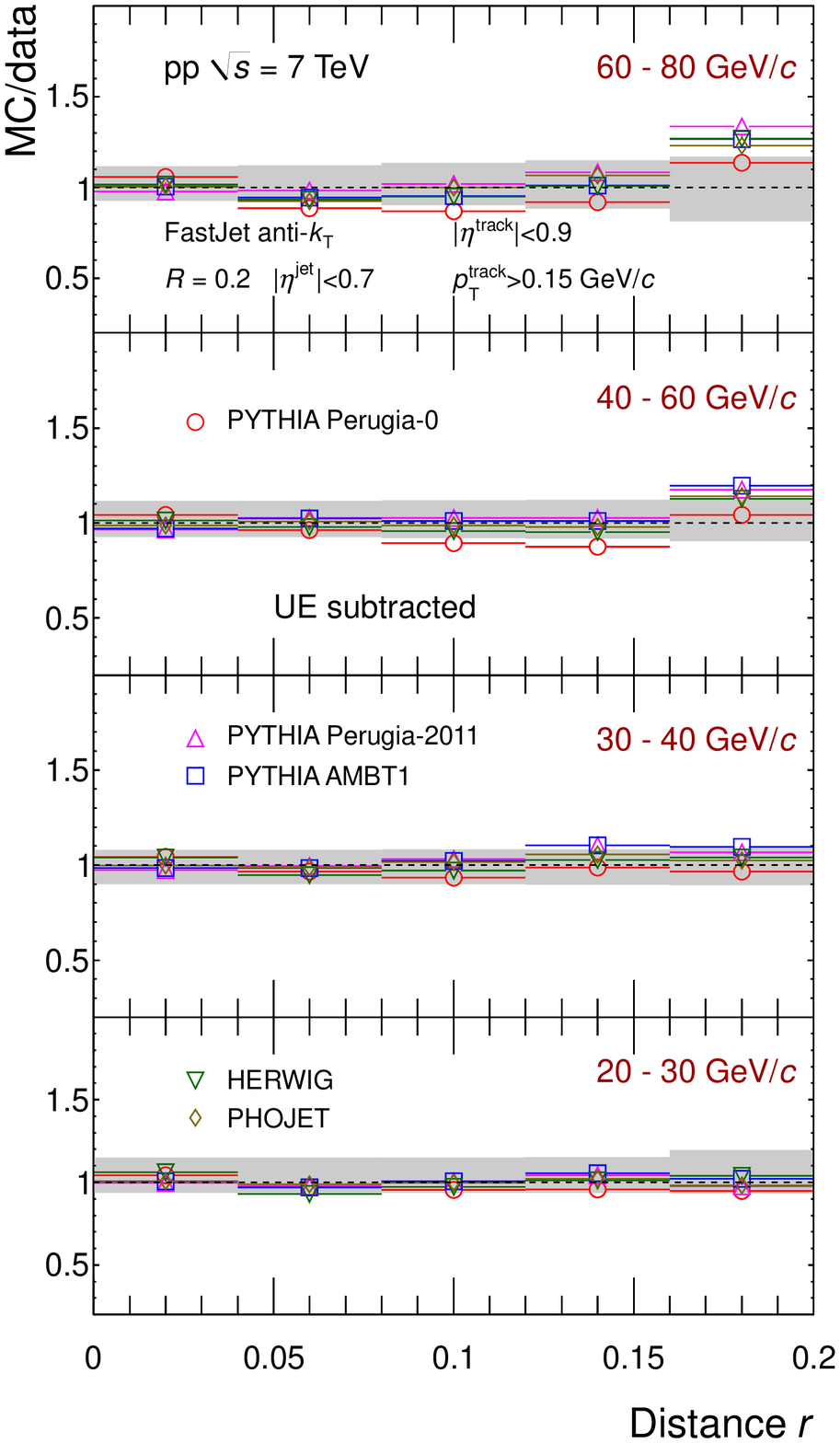}
\caption{(Color online) Left panel: Radial distributions of $p_{\rm T}$ density 
  as a function radial distance 'r' from the
  jet direction for leading charged jets reconstructed with resolution parameter $R$ =
  0.2 for selected  jet $p_{\rm T}$ 
  ranges in pp collisions at $\sqrt{s}$ = 7~TeV.
  Measured distributions are compared to MC model calculations.
  UE contributions
  are subtracted from both data and MC. 
  Right panel: Ratios MC/data. Shaded bands show the quadratic sum
  of statistical and systematic
  uncertainties of the data drawn at unity.}
\label{Fig.raddist_r2}
\end{figure} 

\subsection{Charged particle multiplicity in the leading jet}\label{sec:JetShape}
The corrected mean charged particle multiplicity distributions $\langle N_{\rm ch} \rangle$
in  the leading jet are shown in Fig.~\ref{Fig.nCh} (left panel) as a function of jet $p_{\rm T}$ for
$R$ = 0.2, 0.4, and 0.6. 
The $\langle N_{\rm ch} \rangle$ rises 
monotonically with increasing jet $p_{\rm T}$  as well as with increasing $R$. These results 
are in qualitative agreement with those reported  by the CDF~\cite{cdfprd65} collaboration and
more recently by the CMS~\cite{A11_cmsanalysis} collaboration based on slightly different kinematic 
track cuts.\par 
In the left panel of Fig.~\ref{Fig.nCh}, the measurements are compared to 
predictions by the MC models PYTHIA (tunes Perugia-0, Perugia-2011, AMBT1), PHOJET, and HERWIG.
Ratios of the predictions to the data are displayed in the right
panel. 
The model predictions are well within 10\% of the measured data with
largest deviations of $\sim$15\% at $R$~=~0.6 and 0.2 towards large jet $p_{\rm T}$. 
The PYTHIA tune Perugia-0 
tends to systematically underestimate the 
measured particle multiplicities particularly at the largest $R$ for smaller jet momentum,
whereas HERWIG tends to overpredict the data 
at smaller $R$. An 
overall agreement between the data and MC predictions is found to be
best with the Perugia-2011 tune and PHOJET.

\begin{figure}[th!f]
 \includegraphics[scale=0.52]{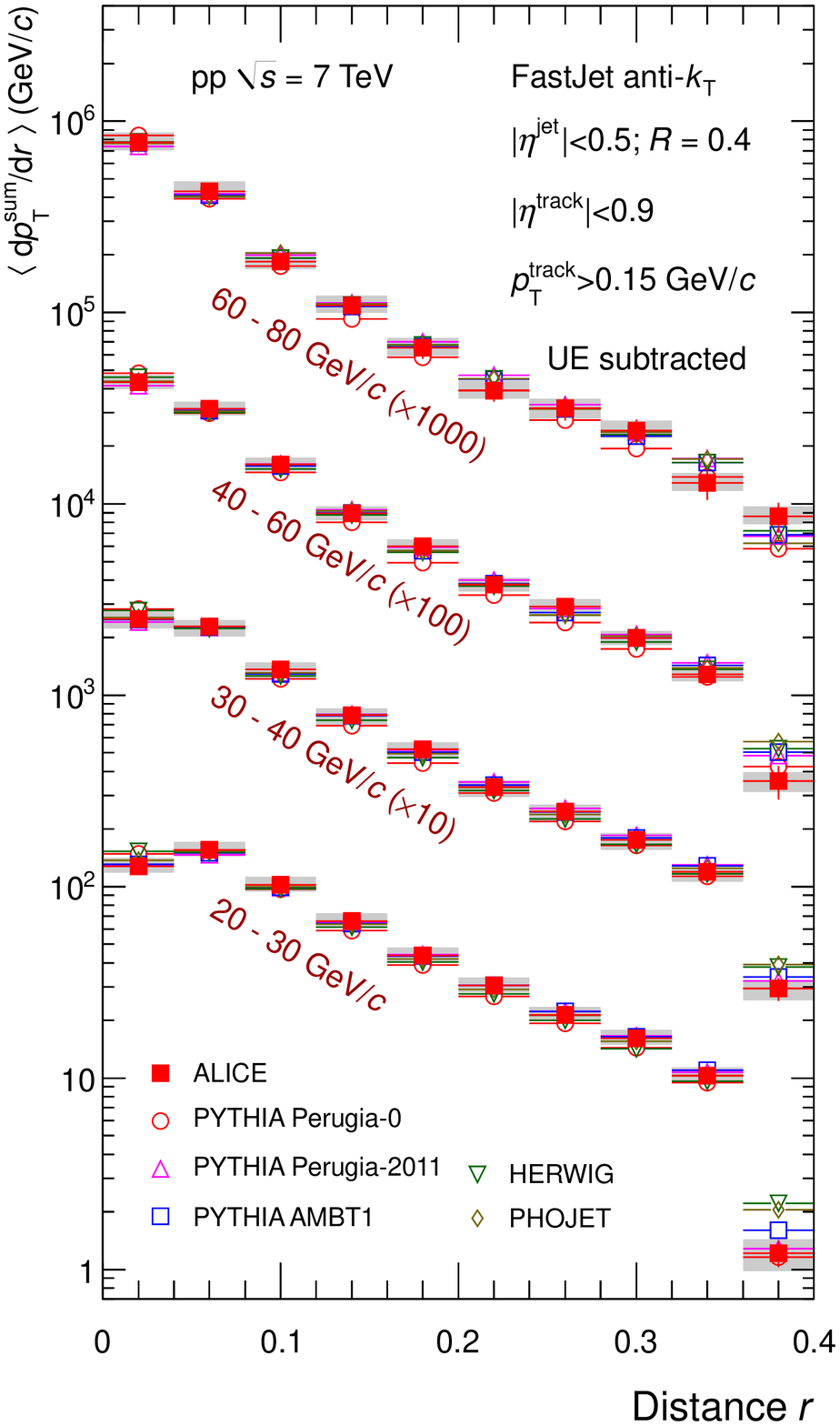}
 \includegraphics[scale=0.52]{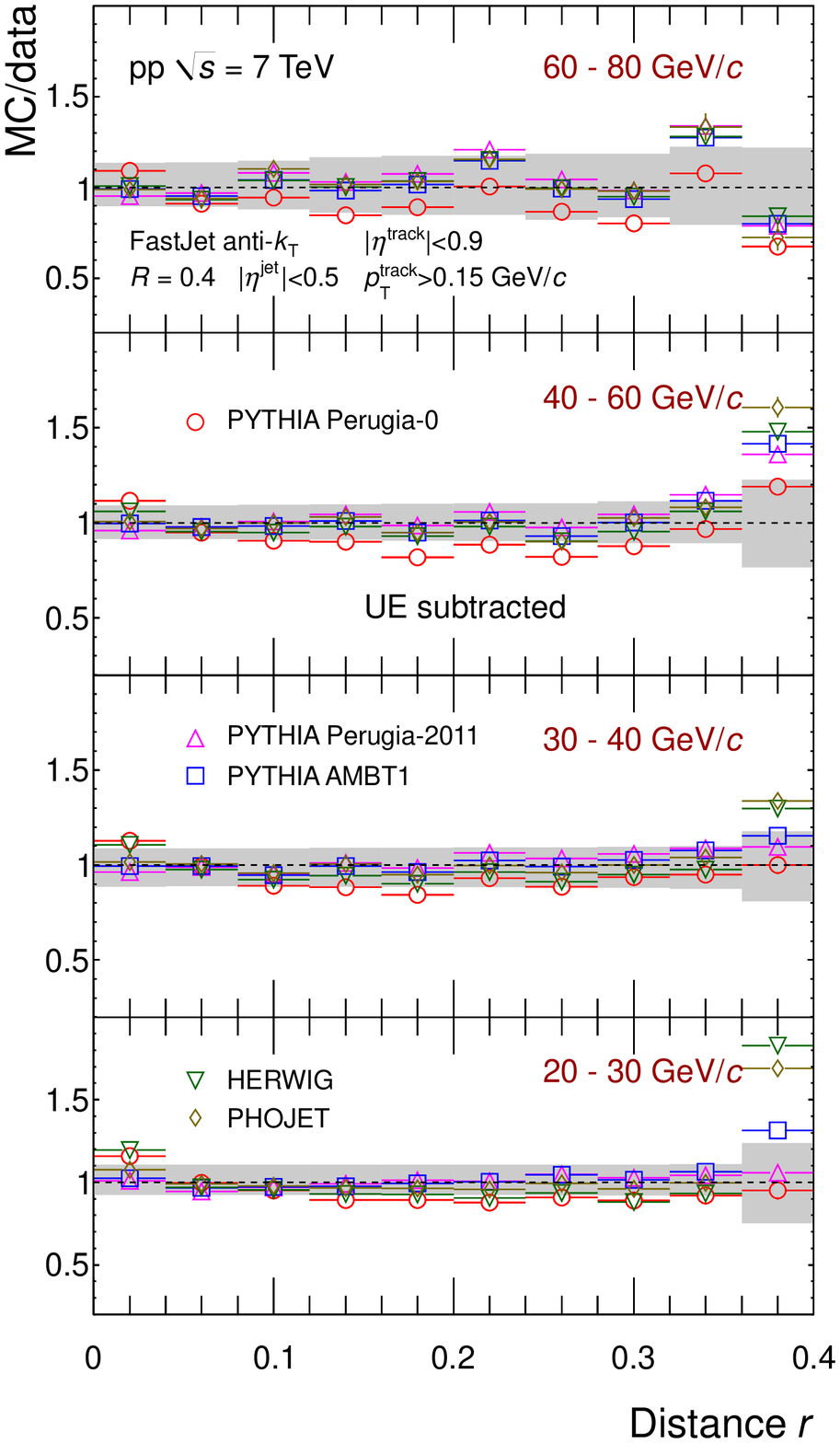}
\caption{(Color online) Same as Fig.~\ref{Fig.raddist_r2} for a resolution parameter $R$
 = 0.4.}
\label{Fig.raddist_r4}
\end{figure} 
\begin{figure}[th!f]
 \includegraphics[scale=0.52]{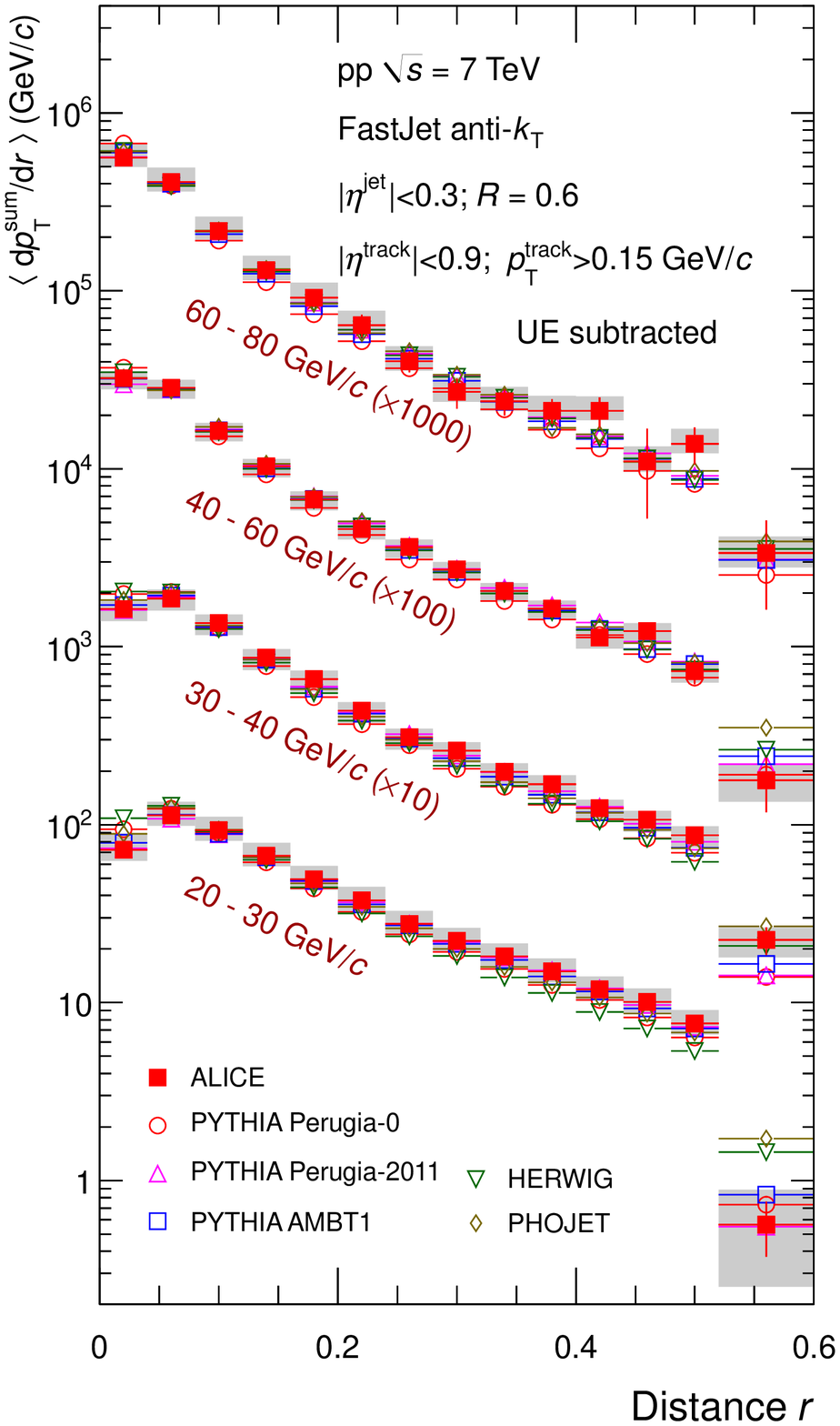}
 \includegraphics[scale=0.52]{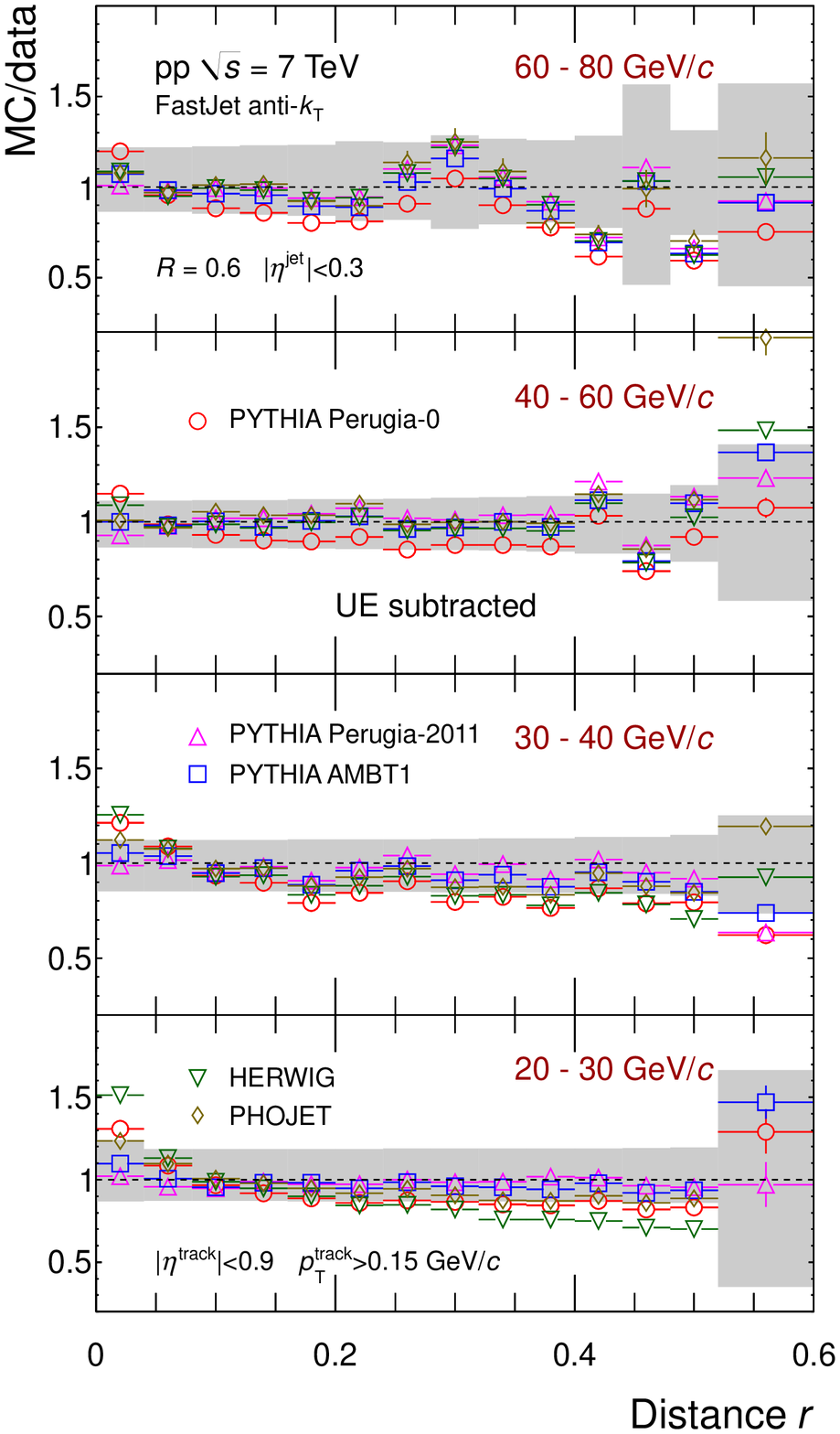}
\caption{(Color online) Same as Fig.~\ref{Fig.raddist_r2} for a resolution parameter
  $R$ = 0.6.}
\label{Fig.raddist_r6}
\end{figure}  

\subsection{Transverse momentum density  distributions within the leading jet}
The left panels of Figs.~\ref{Fig.raddist_r2},~\ref{Fig.raddist_r4}, and~\ref{Fig.raddist_r6}
show  leading jets average $p_{\rm T}$ density radial distributions $\langle \rm{d}{\it p}_{\rm T}^{\rm sum}/\rm{d}{\it r} \rangle$ 
measured with resolution parameters $R$ = 0.2, 0.4, and 0.6,
respectively. 
The
distributions are plotted separately for jets in the $p_{\rm T}$  intervals 
20 - 30, 30 - 40, 40 - 60,  and 60 to 80~GeV/$c$. The latter three distributions are scaled by factors 
of 10, 100, and 1000 respectively for clarity. The transverse momentum
density is largest near the jet axis and decreases approximately exponentially with increasing $r$.
Densities are largest at the highest jet $p_{\rm T}$ where they are also found to have the steepest 
dependence on $r$. This indicates that high $p_{\rm T}$ jets are on average more
collimated than low $p_{\rm T}$ jets as already hinted in Fig.~\ref{R-ratios}.\par
The measured distributions are compared to predictions with 
MC models.~The right panels of Figs.~\ref{Fig.raddist_r2},
\ref{Fig.raddist_r4}, and \ref{Fig.raddist_r6}
display ratios of the model calculations to measured data. 
The MC models qualitatively reproduce the magnitude of the 
measured densities as well as their radial dependence. 
The agreement between the MC model calculations and data is
better at smaller $R$ (= 0.2). At $R$ = 0.4 and 0.6 HERWIG and Perugia-0
tune of PYTHIA tend to underpredict the measured transverse momentum
density except at small $r$ for the two lowest jet $p_{\rm T}$ bins. 
The excess over the data for the smallest r and the slope of the ratio of simulations to data 
observed for $R$ = 0.6 indicates stronger jet collimation for low
$p_{\rm T}$ jets than observed in the 
data. This observation is consistent with the discrepancy of the Herwig model 
to the measured cross section ratio discussed in 
Sec.~\ref{jetspec} (see also Fig.~\ref{R-ratios}). In the last bin of Figs.~\ref{Fig.raddist_r4}, and~\ref{Fig.raddist_r6} (right panel), large deviations of MC models (PHOJET and HERWIG) from the data are found, whereas good agreement 
is observed when data and simulations are not corrected for the UE
contribution (see Appendix~\ref{results:NoUEsubtraction}). This indicates that the 
UE is underestimated by these models, as reported in~\cite{A51_ALICE_UE} for PHOJET and in~\cite{A49_ATLAS_UE_charged+neutral} 
for HERWIG simulations of the UE density of charged and neutral
particles with $p_{\rm T}>$ 0.5~GeV/$c$. 
\subsection{Leading charged jet size}
\begin{figure}[ht]
\includegraphics[scale=0.402]{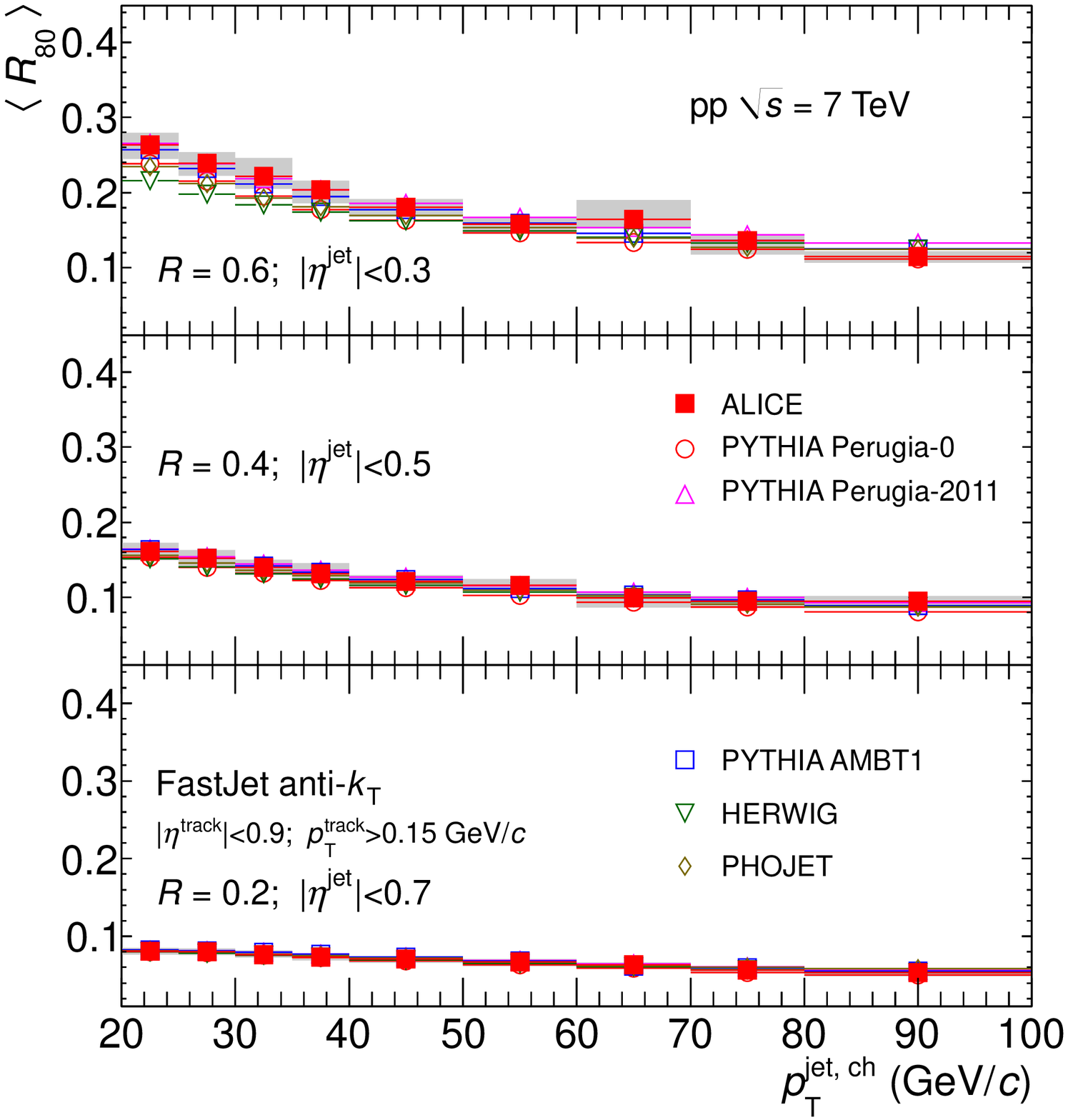}
 \includegraphics[scale=0.402]{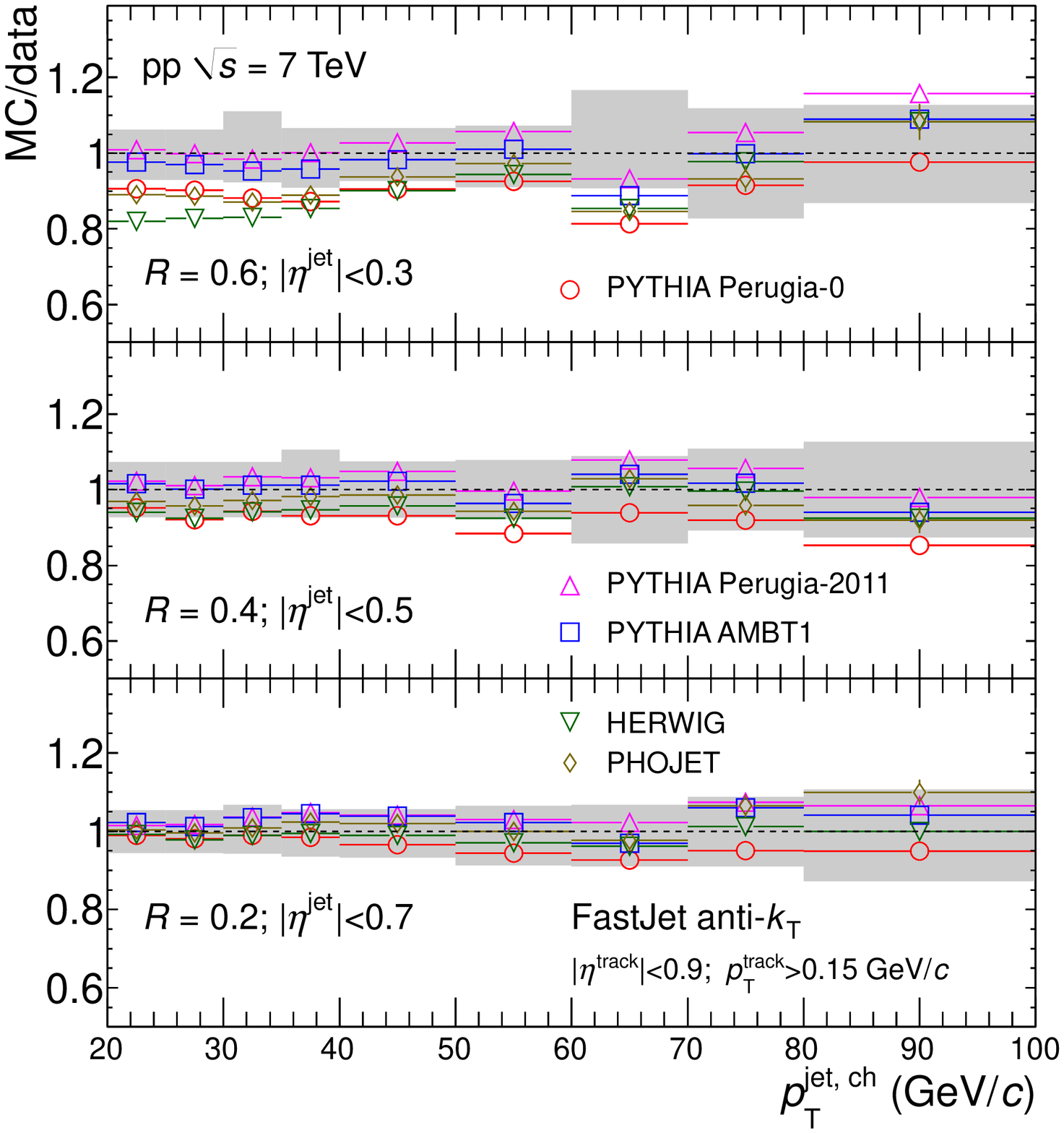}
 \caption{(Color online) Left panel: Distributions of average radius '$\langle R_{\rm 80} \rangle$' containing
   80\% of the $p_{\rm T}$ with respect to the total
   reconstructed jet $p_{\rm T}$ as a function of jet
   $p_{\rm T}$ compared to MC models for pp
   collisions at $\sqrt{s}$ = 7~TeV for various jet resolution
   parameters ($R$ = 0.6 (left top), $R$ = 0.4 (left middle) and $R$ = 0.2
   (left bottom)). Right panel: Ratios MC/data. Shaded bands show quadratic sum of
   the statistical and systematic
   uncertainties of the data drawn at unity.}
 \label{Fig.r80}
\end{figure}

The left panel of Fig.~\ref{Fig.r80} displays measured
distributions of the average radius, $\langle R_{\rm 80}\rangle$, containing 80\% of 
the total jet
$p_{\rm T}$ observed in  jet cones with $R$ = 0.2, 0.4, and 0.6.
The distributions are corrected using the bin-by-bin method described in Sec.~\ref{binBybin}
to account for instrumental effects. No corrections are 
applied for UE contributions, 
which are estimated to have a negligible effects on 
measured  $\langle R_{\rm 80}\rangle$ values. 
Jet widths are largest at the lowest measured $p_{\rm T}$  and
decrease monotonically with increasing $p_{\rm T}$, indicating that
high $p_{\rm T}$ jets are more collimated than low $p_{\rm T}$ jets
(as observed in Figs.~\ref{R-ratios},~\ref{Fig.raddist_r2},~\ref{Fig.raddist_r4},~and~\ref{Fig.raddist_r6}) in
a similar way as
predicted by various MC models and in 
qualitative agreement with prior measurement by the 
CDF~\cite{cdfprd65} collaboration.

Figure~\ref{Fig.r80} also displays $\langle R_{\rm 80}\rangle$ distributions
predicted by PYTHIA (tunes Perugia-0, Perugia-2011, AMBT1),
PHOJET, and HERWIG. 
All five models qualitatively reproduce the observed 
magnitude and $p_{\rm T}$ dependence of $\langle R_{\rm 80}\rangle$ at
$R$ = 0.2 and 0.4. However, at $R$ = 0.6, HERWIG, PHOJET, and PYTHIA Perugia-0 tune 
systematically underpredict the data at low $p_{\rm T}$. The PYTHIA tunes Perugia-2011 and
AMBT1 are in best agreement
with the measured values.
\subsection{Jet fragmentation}

\begin{figure}[ht]
\includegraphics[scale=0.45]{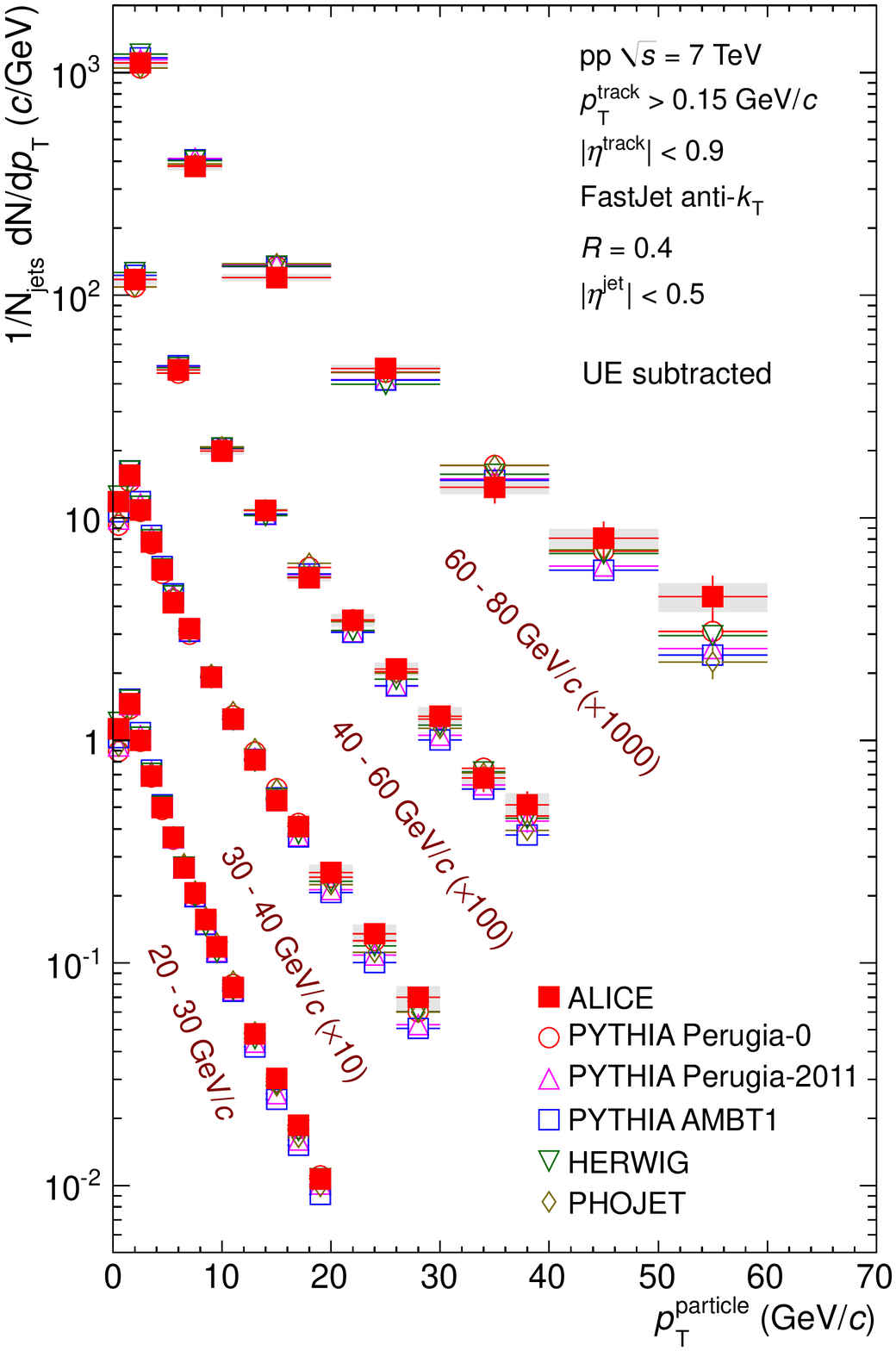}
\includegraphics[scale=0.45]{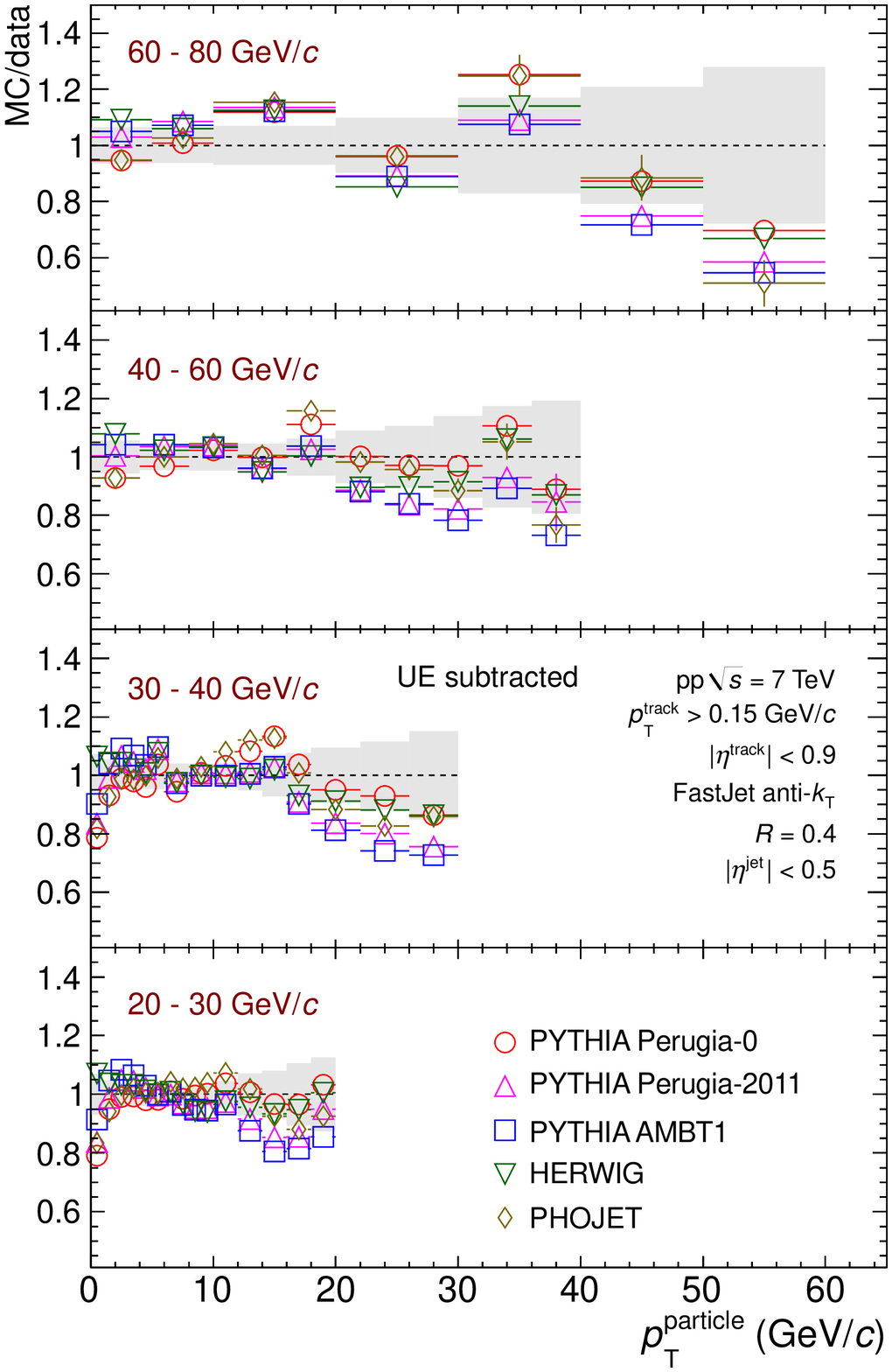}
  \caption{(Color online) Left panel: Charged particle $p_{\rm T}$ spectra ${\rm d}N/{\rm d}p_{\rm T}$ in leading jets 
   for different bins in jet transverse momentum, compared to
   simulations. For simulations and data, the UE contribution is subtracted. 
   Right panel: Ratio of simulations to data. The shaded band indicates the quadratic sum of statistical and systematic  
   uncertainties on the data.}
   \label{fig:FFpt}
\end{figure}

\begin{figure}[ht]
  \includegraphics[scale=0.45]{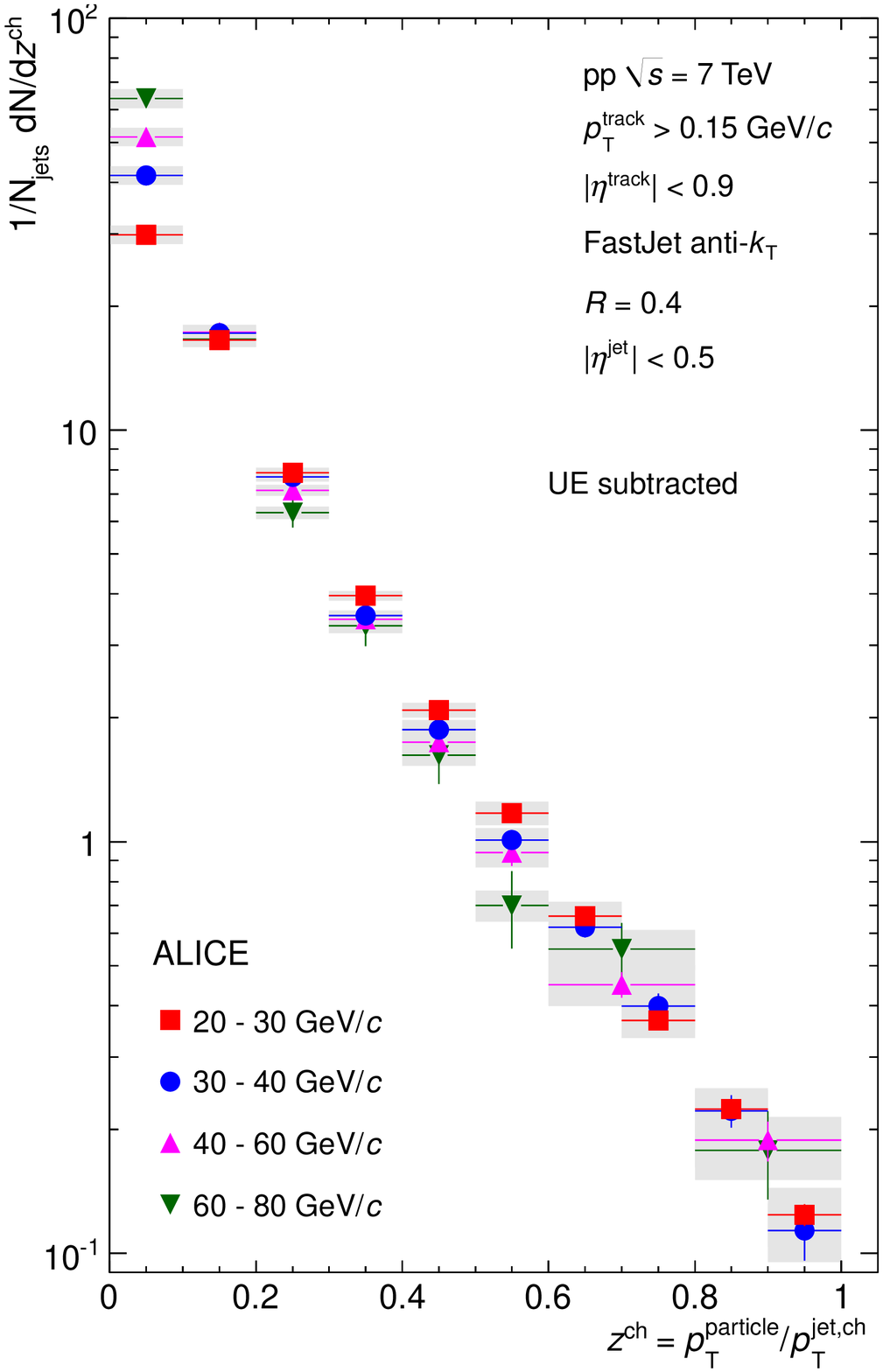}
  \includegraphics[scale=0.45]{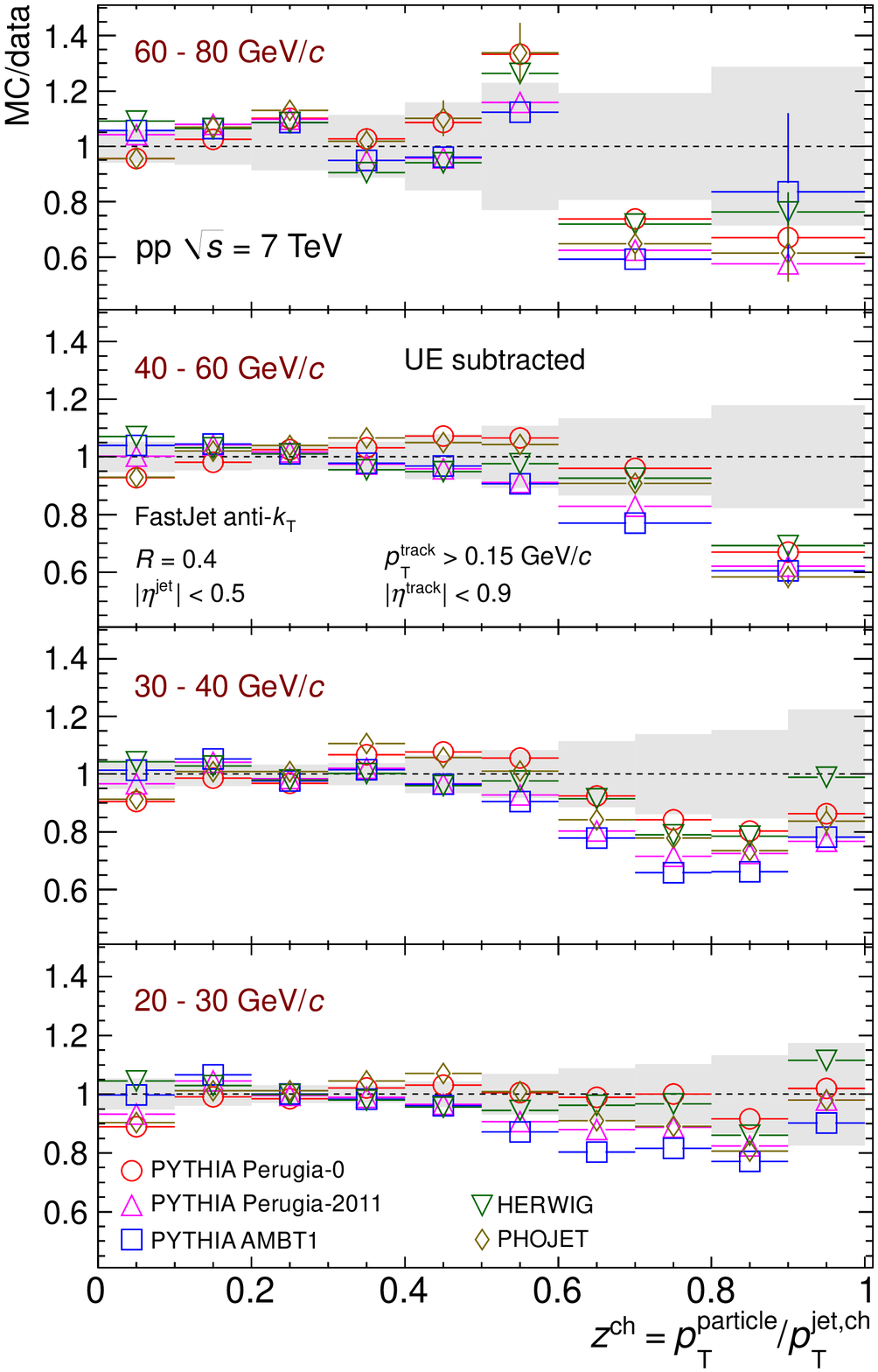}
  \caption{(Color online) Left panel: Charged particle scaled $p_{\rm
      T}$ spectra ${\rm d}N/{\rm d}z^{\rm ch}$ in leading jets 
    for different bins in jet transverse momentum. Right panel: Ratio of simulations to data. The shaded band indicates the quadratic sum of statistical and systematic uncertainties on the data. UE contributions are subtracted from both data and simulations.}
   \label{fig:FFz}
\end{figure}

\begin{figure}[ht]
\includegraphics[scale=0.45]{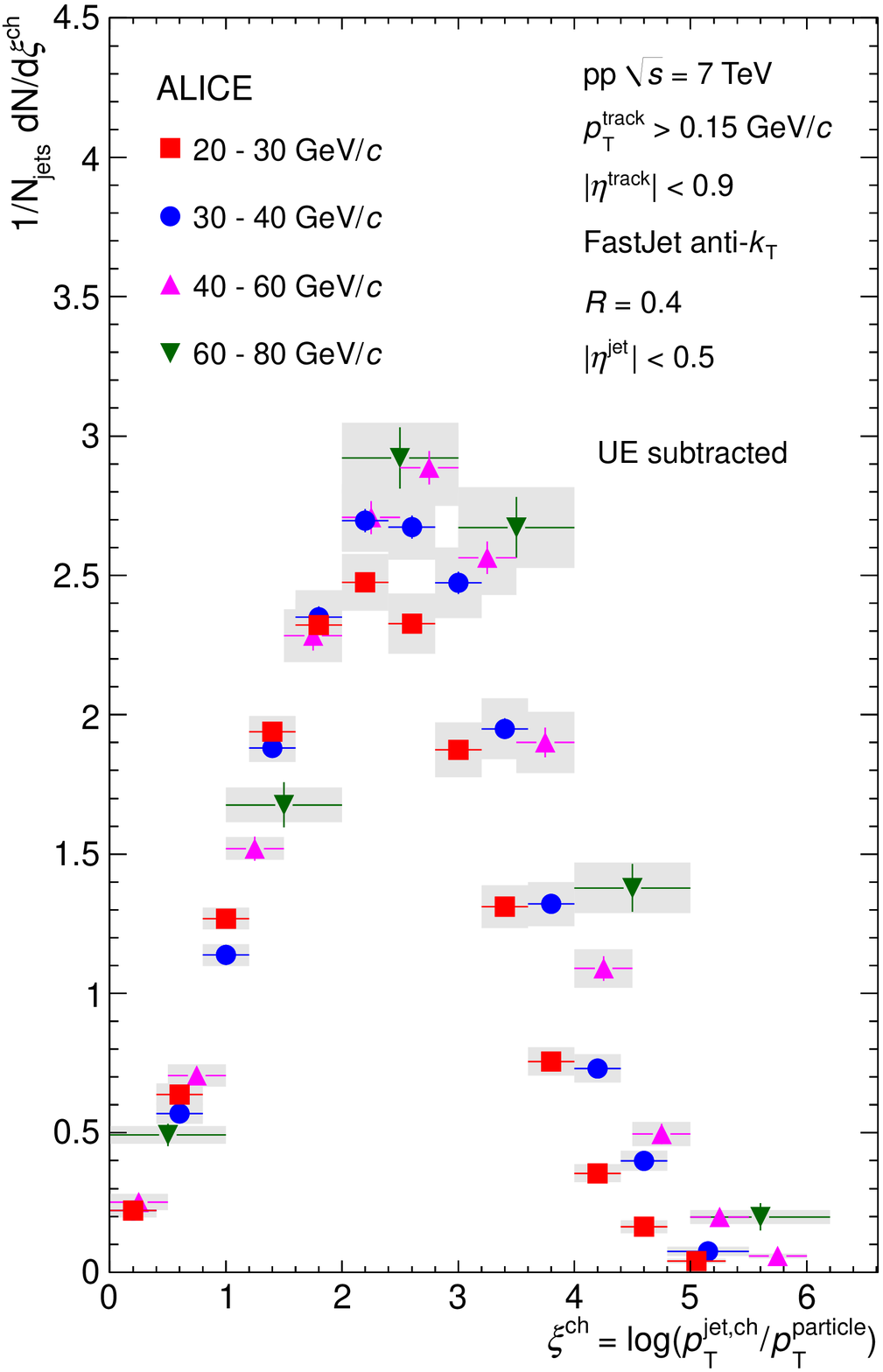}
 \includegraphics[scale=0.45]{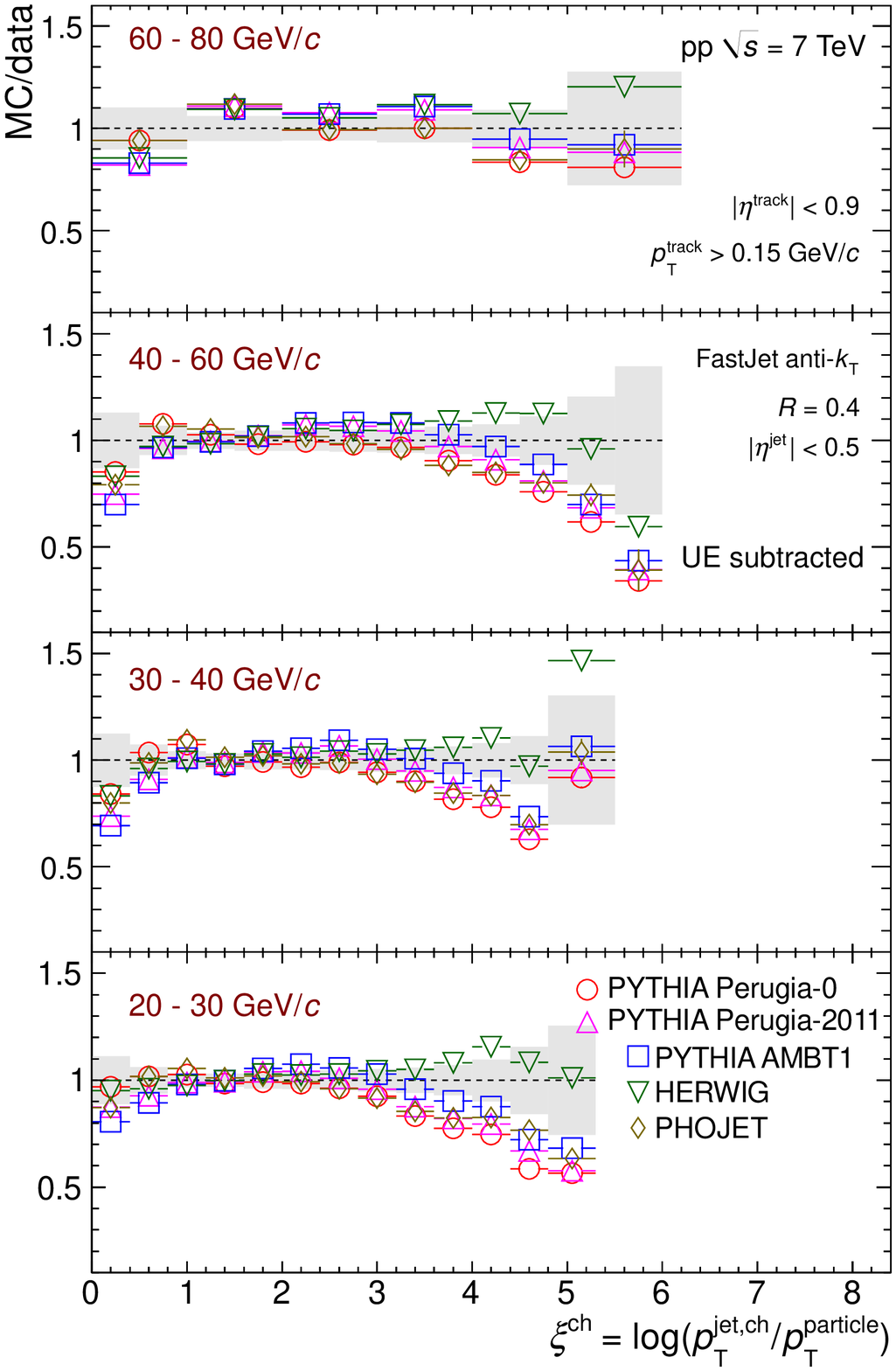}

 \caption{(Color online) Left panel: Charged particle scaled $p_{\rm T}$ spectra ${\rm d}N/{\rm d}{\xi^{\rm ch}}$ in leading jets 
   for different bins in jet transverse momentum.
   Right panel: Ratio of simulations to data. The shaded band indicates the quadratic sum of statistical and systematic  
   uncertainties on the data. UE contributions are subtracted from both data and simulations.}

 \label{fig:FFxi}
\end{figure}

The left panels of Figs.~\ref{fig:FFpt},~\ref{fig:FFz}, and~\ref{fig:FFxi} present the measured $p_{\rm T}$ spectra $F^{p_{\rm T}}$
and scaled $p_{\rm T}$ spectra $F^{z}$ and $F^{\xi}$ of charged particles in leading charged jets reconstructed 
with a resolution parameter $R$ = 0.4. 
The data are corrected for instrumental effects, UE background, and contamination from secondary particles. Systematic uncertainties, 
indicated by the shaded bands, include the detector response, UE subtraction, correction for secondaries and 
event generator dependence. \par 

The particle momentum distributions $F^{p_{\rm T}}$ are shown for four bins in jet transverse momentum:
20 - 30, 30 - 40, 40 - 60, and 60 - 80~GeV/$c$. The latter three are scaled by factors of 
10, 100, and 1000 respectively for clarity. The 
$p_{\rm T}$ spectra of the jet constituents 
span 2 - 3 orders of magnitude. 
The slopes are steepest for the lowest $p_{T}$ jets and progressively
flatter 
with increasing jet $p_{\rm T}$. 
This dependence is essentially driven by the jet energy scale, as illustrated in Fig.~\ref{fig:FFz}, which displays fragmentation 
distributions $F^{z}$ for jets in the same four jet momentum
ranges. 
For $z^{\rm ch} > 0.1$ all measured distributions are consistent 
within uncertainties, indicating a scaling of charged jet fragmentation with charged jet transverse momentum.  

The fragmentation distributions 
$F^{\xi}$, shown in  Fig.~\ref{fig:FFxi}, resolve in more detail the differences observed for small values of $z^{\rm ch}$.  
For small values of $\xi^{\rm ch} \lesssim 2$, the distributions exhibit the approximate scaling already seen for $F^{z}$, whereas at 
higher $\xi^{\rm ch}$, corresponding to small $z^{\rm ch}$, a pronounced maximum ('hump-backed plateau') is observed, indicating the 
suppression of low momentum particle production by QCD coherence~\cite{A29_QCD_coherence,A29_QCD_coherence_1}. With increasing 
jet transverse momentum, the area of the distributions increases, showing the rise of particle multiplicity in 
jets (as observed in Fig.~\ref{Fig.nCh}), and the maximum shifts to higher values of $\xi^{\rm ch}$. 
This observation is in qualitative agreement with full di-jet fragmentation functions measured in p$\bar{\rm p}$ collisions 
at $\sqrt{s}$ = 1.8~TeV~\cite{A12_CDF_FF} and with expectations from QCD calculations based on the Modified Leading Logarithmic Approximation (MLLA)~\cite{A58_MLLA}. \par 

The measured fragmentation distributions are compared to calculations
obtained from the HERWIG~\cite{A36_Herwig,A36_Herwig_1}, PHOJET~\cite{A35_Phojet} 
and PYTHIA~\cite{A30_Pythia} event generators 
and the ratios of the calculated MC distributions to measured
distributions are shown in the right panels of Figs.~\ref{fig:FFpt},
~\ref{fig:FFz}, and~\ref{fig:FFxi}.
The UE contributions to MC events are estimated and subtracted using perpendicular cones pointing into the event transverse region as 
described in Sec.~\ref{sec:underlyingEventSub}. 
At high particle transverse 
momenta and high $z^{\rm ch}$, the data and simulations agree within uncertainties, except for the two 
lowest jet $p_{\rm T}$ bins, where the measured yield seems to be systematically higher than the simulations with PYTHIA tunes Perugia-2011 and AMBT1 for $z^{\rm ch} >$ 0.6.  
In the low momentum / high $\xi^{\rm ch}$ region, the measured yield is systematically larger than produced by the 
PYTHIA and PHOJET simulations. To investigate the discrepancy at low particle momentum, data and 
simulations are also compared without subtraction of the UE (see Appendix~\ref{results:NoUEsubtraction}). In this case, the excess of 
low momentum constituents in data over PYTHIA simulations 
is still significant, however reduced in magnitude and comparable to other measurements at higher constituent 
momenta~\cite{A3_ATLASchJets}. It is thus concluded that in the PYTHIA tunes investigated in this work 
the UE contribution to the low momentum particle yield is overestimated relative to the contribution from hard parton fragmentation. 
The data at low $p_{\rm T}$ are best described by the HERWIG 
event generator, which hints to a sensitivity of the low momentum fragmentation to the details of the parton shower 
model in the simulations. 

\section{Summary and conclusion}\label{sec7summary}
In summary, we reported measurements of the inclusive charged particle
jet cross section, jet fragmentation and jet shapes at midrapidity in
pp collisions at $\sqrt{s}$~=~7~TeV using the ALICE detector at the LHC.

Jets were reconstructed with infrared and collinear safe jet finding algorithms, $k_{\rm T}$, anti-$k_{\rm T}$ and a seedless infrared safe iterative cone
based algorithm, SISCone. 
As the measured inclusive jet spectra did not show any significant
dependence on the jet algorithm used, all observables 
discussed throughout the paper were based on jets reconstructed with the anti-$k_{\rm T}$ sequential recombination algorithm, 
commonly utilized in the LHC community. In order to gain as much
information as possible , the anti-$k_{\rm T}$ algorithm 
was run with several resolution parameters $R$ ranging from 0.2 to 0.6.

The inclusive charged jet cross section was measured in the $p_{\rm
  T}^{\rm jet,ch}$ interval from 20 to 100~GeV/$c$ and 
found to be consistent
with the ATLAS measurement at the same collision energy. The ratios of jet cross sections 
for resolution parameter $R$~=~0.2 over $R$~=~0.4 and 0.6, respectively, are found to increase with 
increasing $p_{\rm T}$ of jets, pointing toward an increasing collimation of particles in jets around the jet axis. 
This finding, expected by pQCD calculations, is corroborated by a detailed study of $\langle R_{\rm 80} \rangle$ variable defined 
as the average radius containing 80\% of total charged jet $p_{\rm T}$. The $p_{\rm T}$ density is found to be the largest near the jet axis and decreases 
radially away from the jet axis. This radial decrease is found to be larger for high $p_{\rm T}$ jets 
which are more collimated. The averaged charged particle multiplicity in jets ($\langle N_{\rm ch} \rangle$) increases with jet momentum and 
resolution parameter $R$. We studied charged particle fragmentation in leading charged jets.
The scaled $p_{\rm T}$ spectra of charged particles associated with jets 
exhibit a pronounced maximum commonly referred to as `hump-backed plateau' consistent with the suppression of low momentum particle production by QCD coherence. 
The area of the distribution increases with 
jet $p_{\rm T}$ and reflects the observed increase of $\langle N_{ch} \rangle$ discussed above. 
The observed behaviour is in qualitative agreement with MLLA~\cite{A58_MLLA} calculations and earlier measurements~\cite{A12_CDF_FF} in
p$\bar{\rm p}$ collisions at the Tevatron ($\sqrt{s}$~=~1.8~TeV). The
jet fragmentation distributions for the measured jet $p_{\rm T}$
ranges indicate a scaling of charged jet fragmentation with
jet $p_{\rm T}$ for $z^{\rm ch} > 0.1$. \par

All measured observables were also compared to several MC generators (PYTHIA, PHOJET, HERWIG). 
None of the generators gives a perfect description of the measured charged jet cross section. 
PHOJET and most of the PYTHIA tunes used in this work overestimate the cross section.  
PYTHIA Perugia-2011 agrees reasonably well with the data for
intermediate and high charged jet $p_{\rm T}$, whereas HERWIG
reproduces best the cross section at low jet $p_{\rm T}$. 
The jet properties are reproduced rather well by the MC generators. The agreement of the calculations with the 
data for observables $\langle N_{\rm ch} \rangle$, $\langle R_{\rm 80} \rangle$, and radial $p_{\rm T}$ density is typically 
at the level of 5-10\%. 
In case of the fragmentation functions, the data are better described
by the HERWIG event generator. 
The high 
momentum (low $\xi^{\rm ch}$) region is relatively well described by the generators, while for the 
low momenta (high $\xi^{\rm ch}$), the measured yield significantly
exceeds PHOJET and PYTHIA predictions. 
We emphasize the relevance 
of this observation for the choice of a generator based pp reference for future measurements 
of jet fragmentation in nuclear collisions, where similar effects are predicted 
as a signature of parton energy loss in the hot and dense strongly-interacting medium.

%
%

\newenvironment{acknowledgement}{\relax}{\relax}
\begin{acknowledgement}
\section*{Acknowledgements}
The ALICE Collaboration would like to thank all its engineers and technicians for their invaluable contributions to the construction of the experiment and the CERN accelerator teams for the outstanding performance of the LHC complex.
The ALICE Collaboration gratefully acknowledges the resources and support provided by all Grid centres and the Worldwide LHC Computing Grid (WLCG) collaboration.
The ALICE Collaboration acknowledges the following funding agencies for their support in building and
running the ALICE detector:
State Committee of Science,  World Federation of Scientists (WFS)
and Swiss Fonds Kidagan, Armenia,
Conselho Nacional de Desenvolvimento Cient\'{\i}fico e Tecnol\'{o}gico (CNPq), Financiadora de Estudos e Projetos (FINEP),
Funda\c{c}\~{a}o de Amparo \`{a} Pesquisa do Estado de S\~{a}o Paulo (FAPESP);
National Natural Science Foundation of China (NSFC), the Chinese Ministry of Education (CMOE)
and the Ministry of Science and Technology of China (MSTC);
Ministry of Education and Youth of the Czech Republic;
Danish Natural Science Research Council, the Carlsberg Foundation and the Danish National Research Foundation;
The European Research Council under the European Community's Seventh Framework Programme;
Helsinki Institute of Physics and the Academy of Finland;
French CNRS-IN2P3, the `Region Pays de Loire', `Region Alsace', `Region Auvergne' and CEA, France;
German BMBF and the Helmholtz Association;
General Secretariat for Research and Technology, Ministry of
Development, Greece;
Hungarian OTKA and National Office for Research and Technology (NKTH);
Department of Atomic Energy and Department of Science and Technology of the Government of India;
Istituto Nazionale di Fisica Nucleare (INFN) and Centro Fermi -
Museo Storico della Fisica e Centro Studi e Ricerche "Enrico
Fermi", Italy;
MEXT Grant-in-Aid for Specially Promoted Research, Ja\-pan;
Joint Institute for Nuclear Research, Dubna;
National Research Foundation of Korea (NRF);
CONACYT, DGAPA, M\'{e}xico, ALFA-EC and the EPLANET Program
(European Particle Physics Latin American Network)
Stichting voor Fundamenteel Onderzoek der Materie (FOM) and the Nederlandse Organisatie voor Wetenschappelijk Onderzoek (NWO), Netherlands;
Research Council of Norway (NFR);
Polish Ministry of Science and Higher Education;
National Science Centre, Poland;
Ministry of National Education/Institute for Atomic Physics and CNCS-UEFISCDI - Romania;
Ministry of Education and Science of Russian Federation, Russian
Academy of Sciences, Russian Federal Agency of Atomic Energy,
Russian Federal Agency for Science and Innovations and The Russian
Foundation for Basic Research;
Ministry of Education of Slovakia;
Department of Science and Technology, South Africa;
CIEMAT, EELA, Ministerio de Econom\'{i}a y Competitividad (MINECO) of Spain, Xunta de Galicia (Conseller\'{\i}a de Educaci\'{o}n),
CEA\-DEN, Cubaenerg\'{\i}a, Cuba, and IAEA (International Atomic Energy Agency);
Swedish Research Council (VR) and Knut $\&$ Alice Wallenberg
Foundation (KAW);
Ukraine Ministry of Education and Science;
United Kingdom Science and Technology Facilities Council (STFC);
The United States Department of Energy, the United States National
Science Foundation, the State of Texas, and the State of Ohio;
Ministry of Science, Education and Sports of Croatia and  Unity through Knowledge Fund, Croatia.
\end{acknowledgement}

\bibliographystyle{utphys}   
\bibliography{master_ChargedJetPP}

\newpage
\appendix
\section{Results without UE subtraction}\label{results:NoUEsubtraction}
The results are presented for charged jet properties including inclusive 
differential jet cross section, $\langle N_{\rm ch} \rangle$, $\langle
\rm{d}{\it p}_{\rm T}^{\rm sum}/\rm{d}{\it r} \rangle$, $F^{p_{\rm
    T}}$, $F^{z}$ and $F^{\xi}$ without subtraction of UE in
comparison to MC generators.\par
\begin{figure*}[ht]
  \begin{center}
    \includegraphics[width=1.0\textwidth]{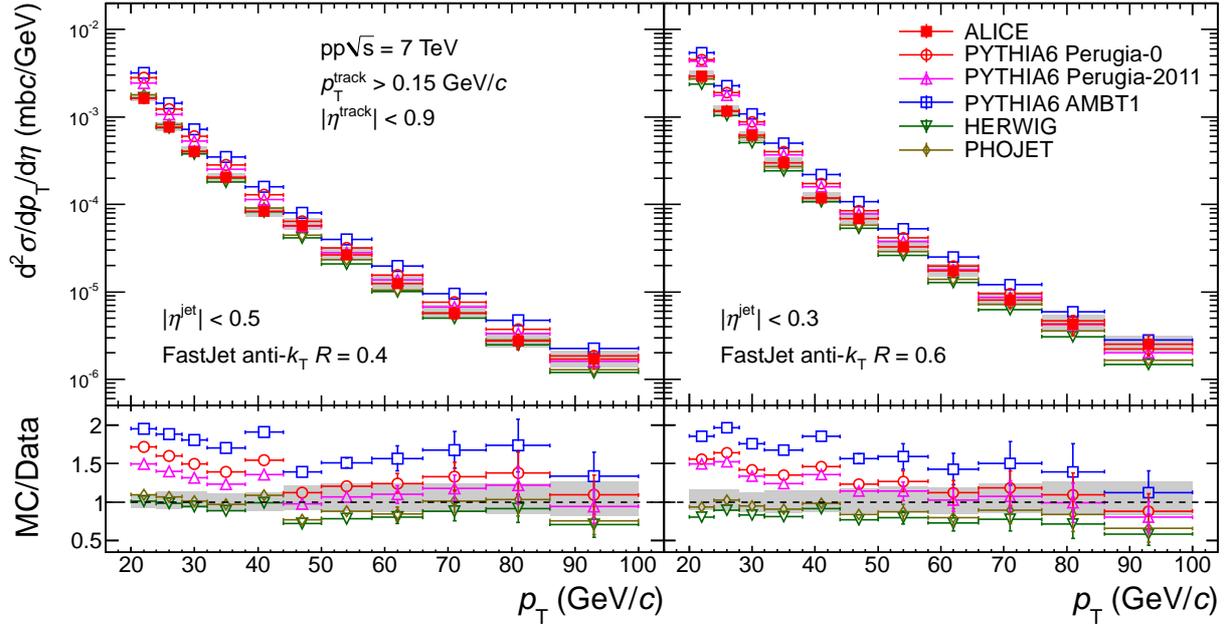}
    \caption{(Color online) Top panels: Charged jet cross sections
      measured in the ALICE experiment in pp collisions at
      $\sqrt{s}$ = 7 TeV without UE subtraction compared to several MC generators: PYTHIA AMBT1,
      PYTHIA Perugia-0 tune, PYTHIA
      Perugia-2011 tune, HERWIG, and PHOJET. Bottom panels: Ratios
      MC/Data. Shaded bands show quadratic sum of statistical and
      systematic uncertainties on the data drawn at unity.}
    \label{result-spectra-generatorcomparison-woUEsub}
  \end{center}
\end{figure*}

In the top panels of
Fig.~\ref{result-spectra-generatorcomparison-woUEsub}, the measured
charged jet cross sections are compared to predictions from
PYTHIA (tunes Perugia-0, Perugia-2011, and AMBT1), PHOJET, and HERWIG
for $R$~=~0.4 and 0.6. The UE is not subtracted for both data and
MC. The ratios of the MC simulations to measured data are shown in the
bottom panels of Fig.~\ref{result-spectra-generatorcomparison-woUEsub}.

\begin{figure*}[ht]
  \includegraphics[scale=0.402]{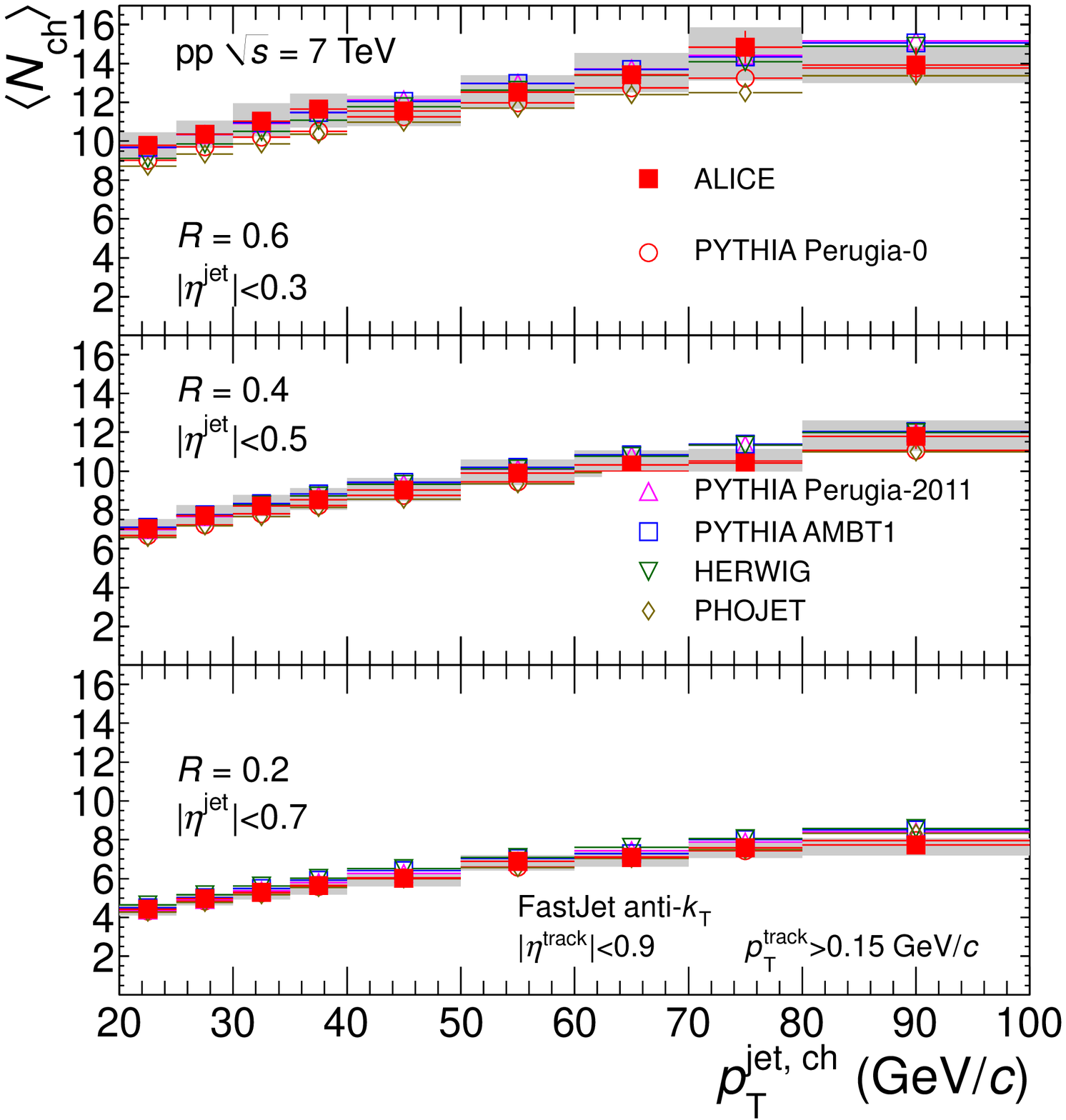}
  \includegraphics[scale=0.402]{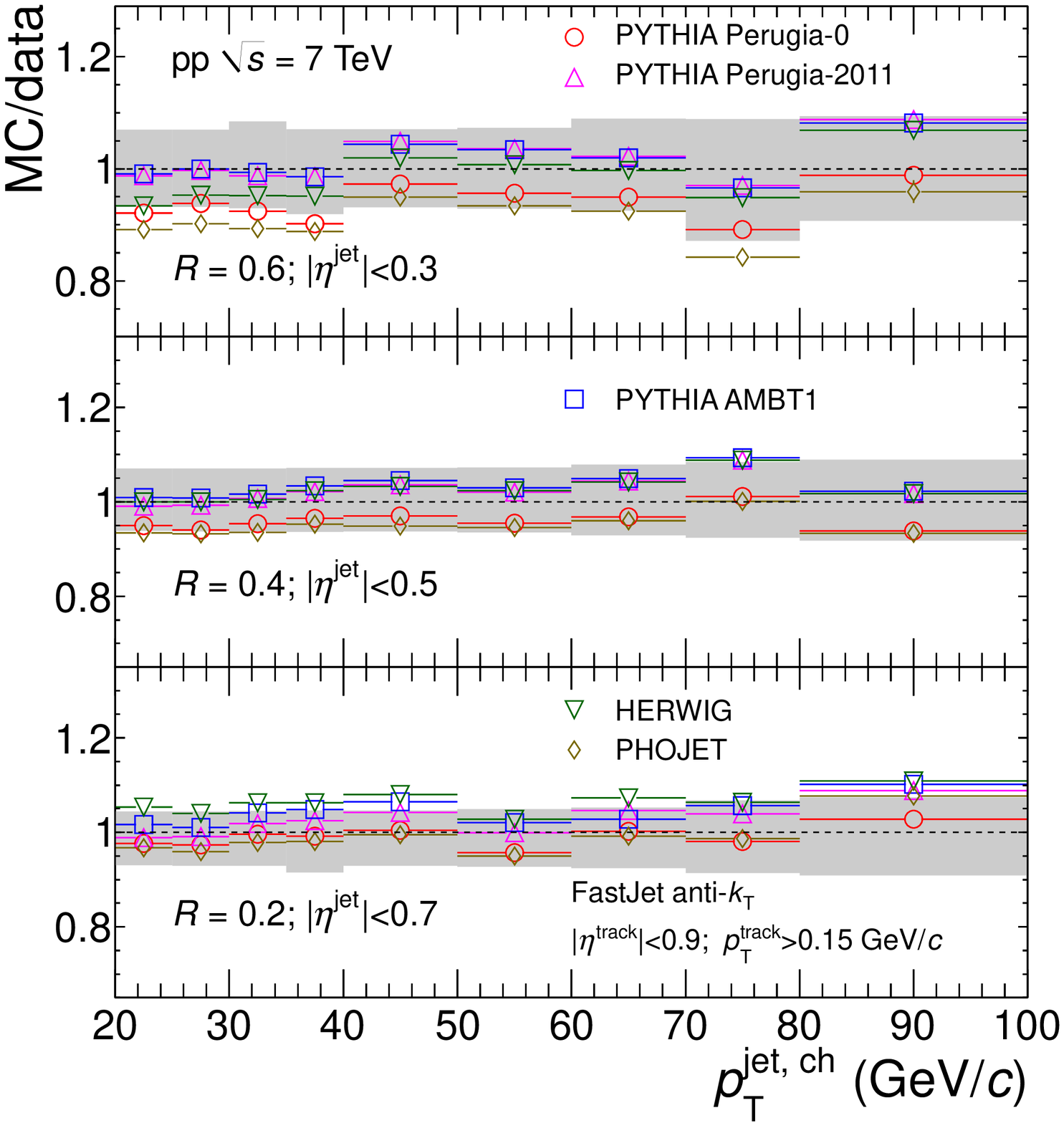}
  \caption{(Color online) Same as Fig.~\ref{Fig.nCh} without UE subtraction.}
  \label{Fig.nCh-woUEsub}
\end{figure*}

The corrected mean charged particle multiplicity distributions
$\langle N_{\rm ch} \rangle$ in the leading jet are shown in
Fig.~\ref{Fig.nCh-woUEsub} (left panel) as a function of jet $p_{\rm T}$ for
$R$ = 0.2, 0.4, and 0.6. The UE is not subtracted for both data and
MC. Ratios of the predictions to the data are displayed in the right panel.

\begin{figure*}[ht]
  \includegraphics[scale=0.52]{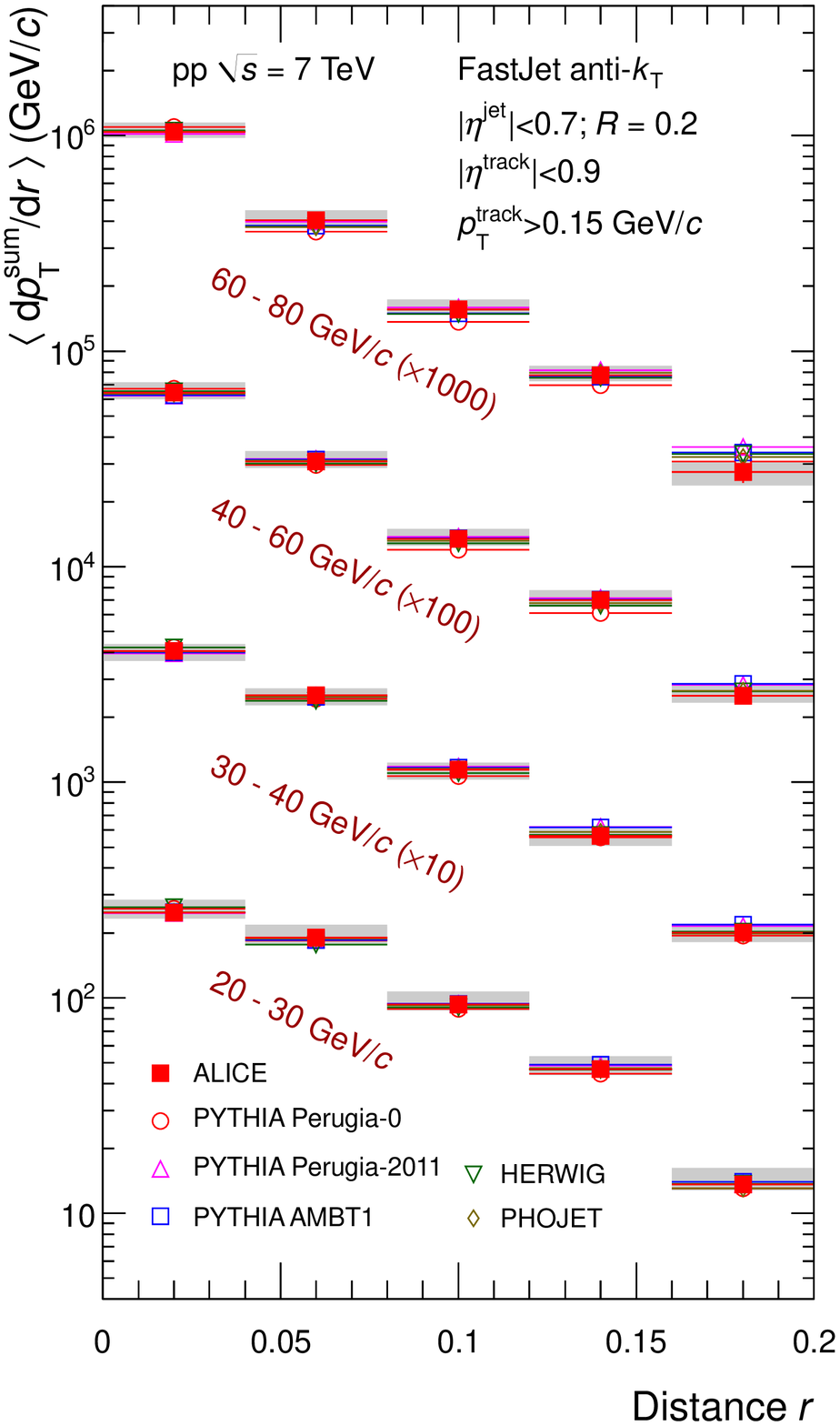}
  \includegraphics[scale=0.52]{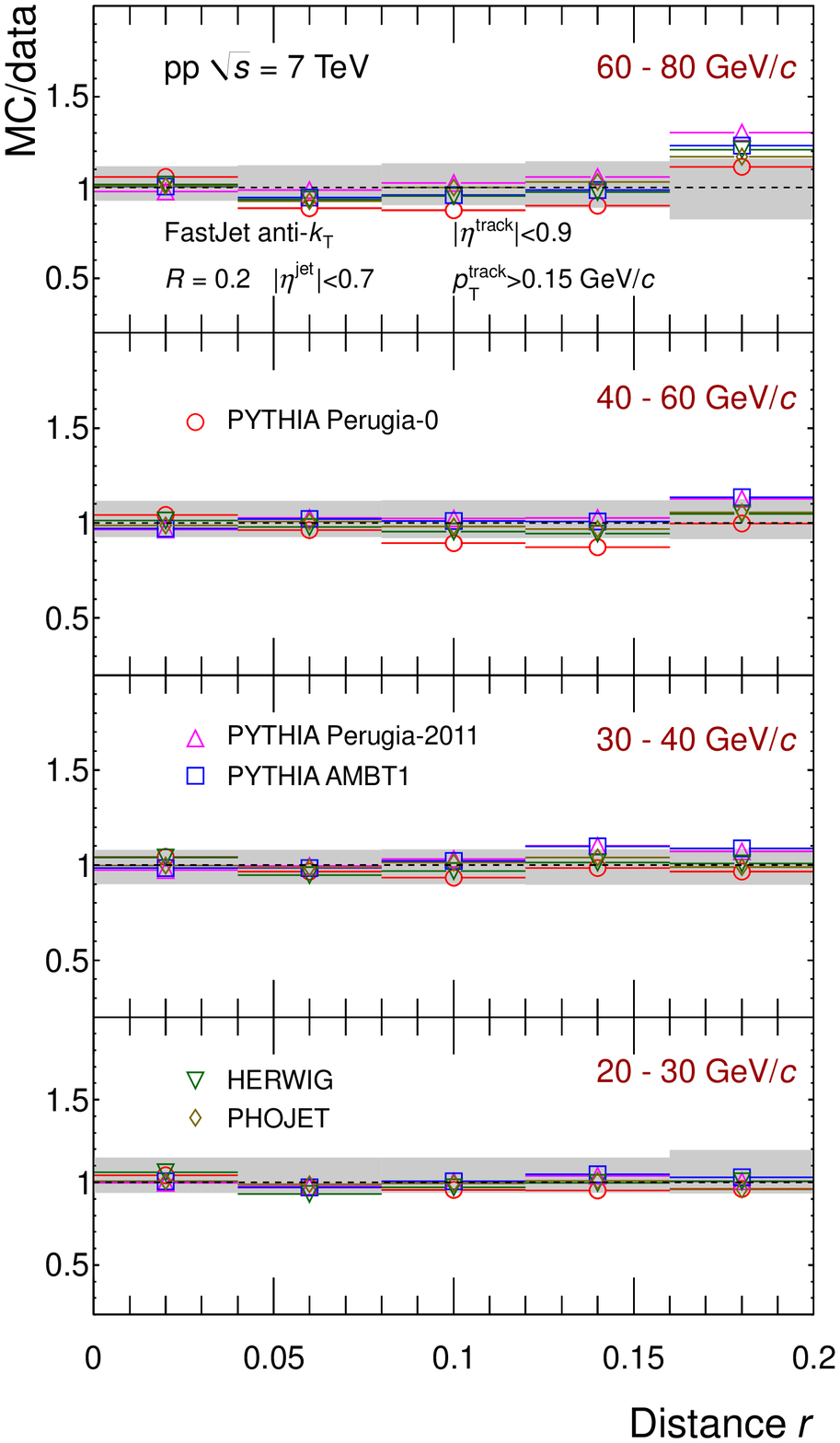}
  \caption{(Color online) Same as Fig.~\ref{Fig.raddist_r2} without UE subtraction.}
  \label{Fig.raddist_r2-woUEsub}
\end{figure*}

\begin{figure*}[ht]
  \includegraphics[scale=0.52]{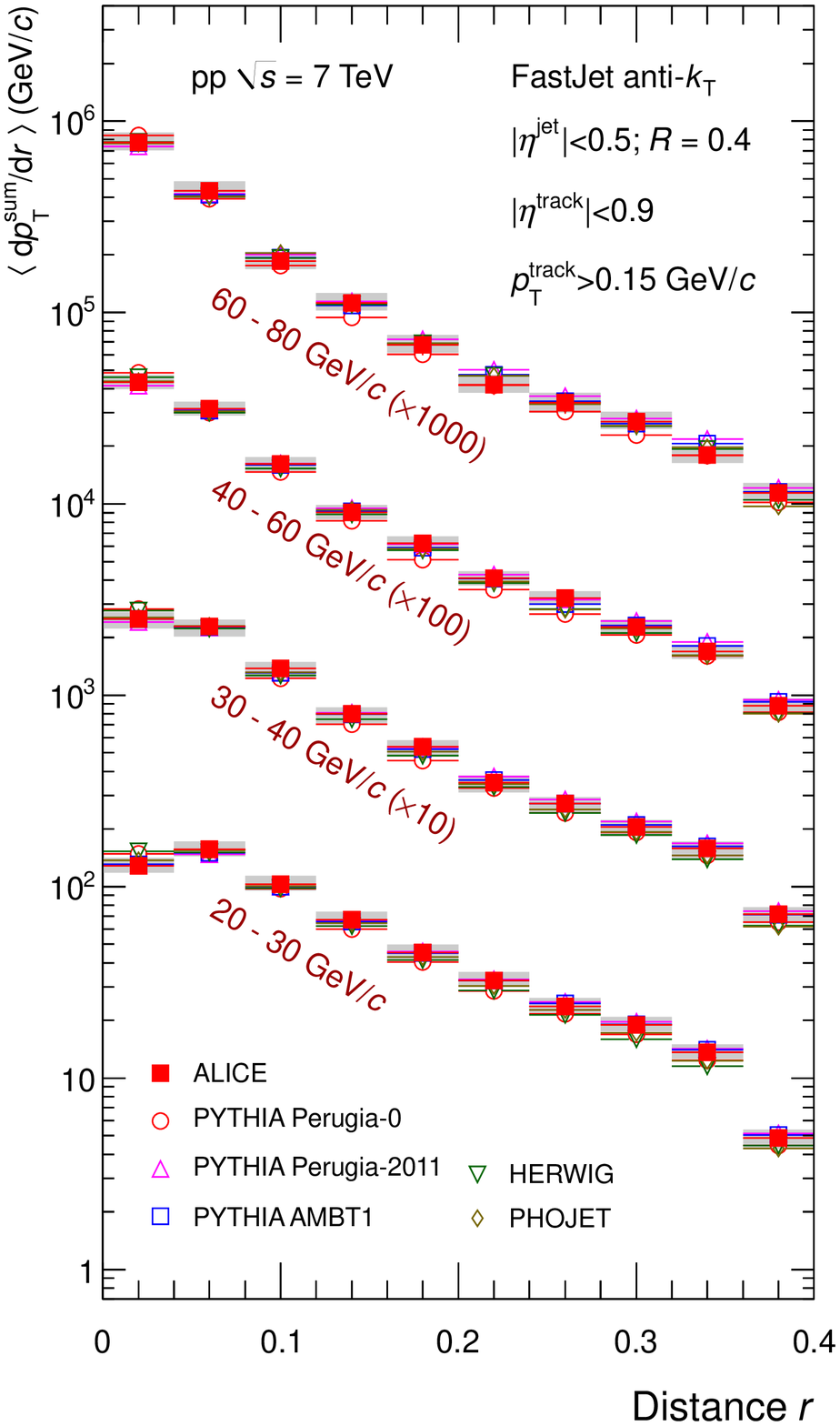}
  \includegraphics[scale=0.52]{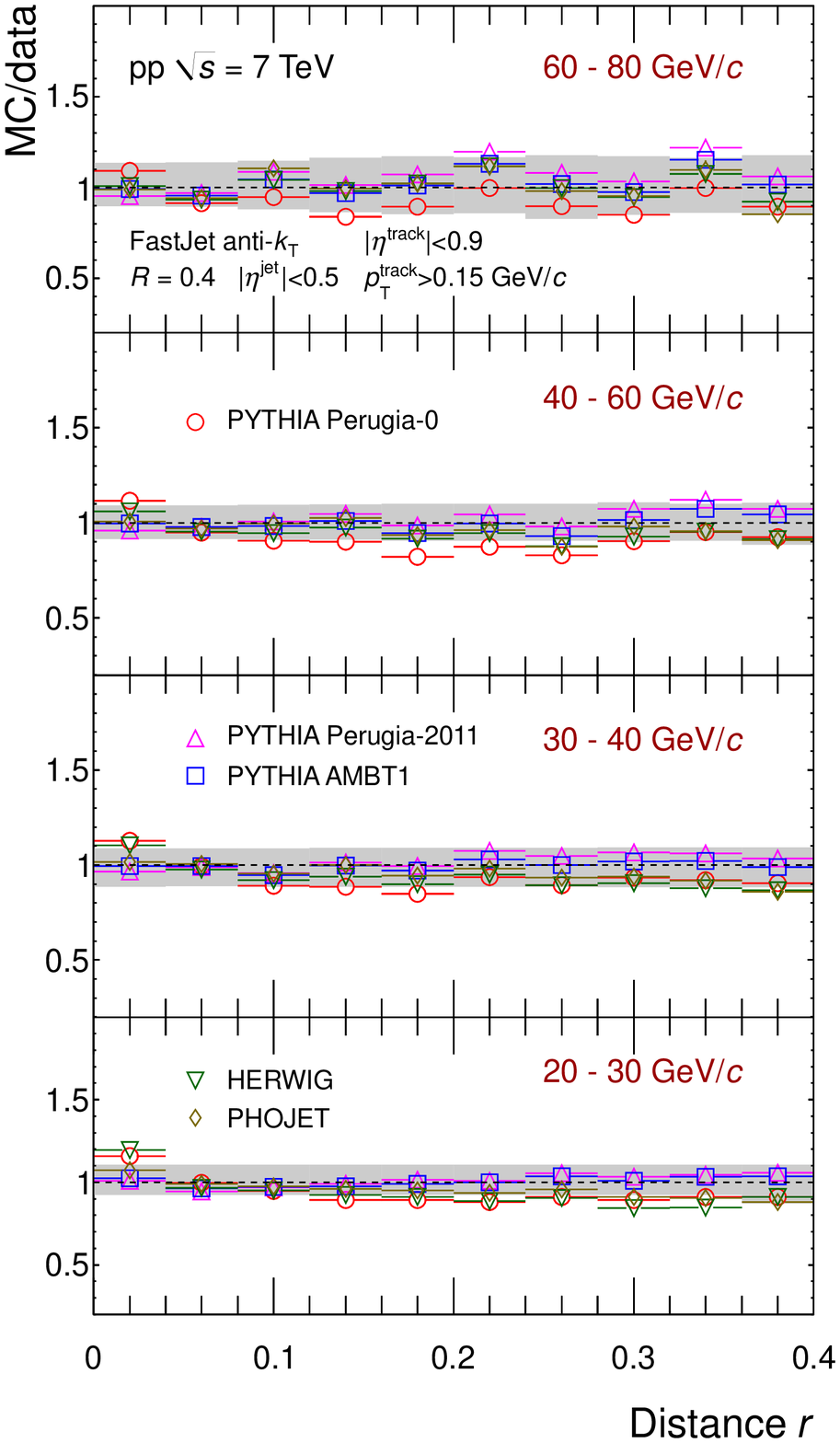}
  \caption{(Color online) Same as Fig.~\ref{Fig.raddist_r4} without UE subtraction.}
  \label{Fig.raddist_r4-woUEsub}
\end{figure*}

\begin{figure*}[ht]
  \includegraphics[scale=0.52]{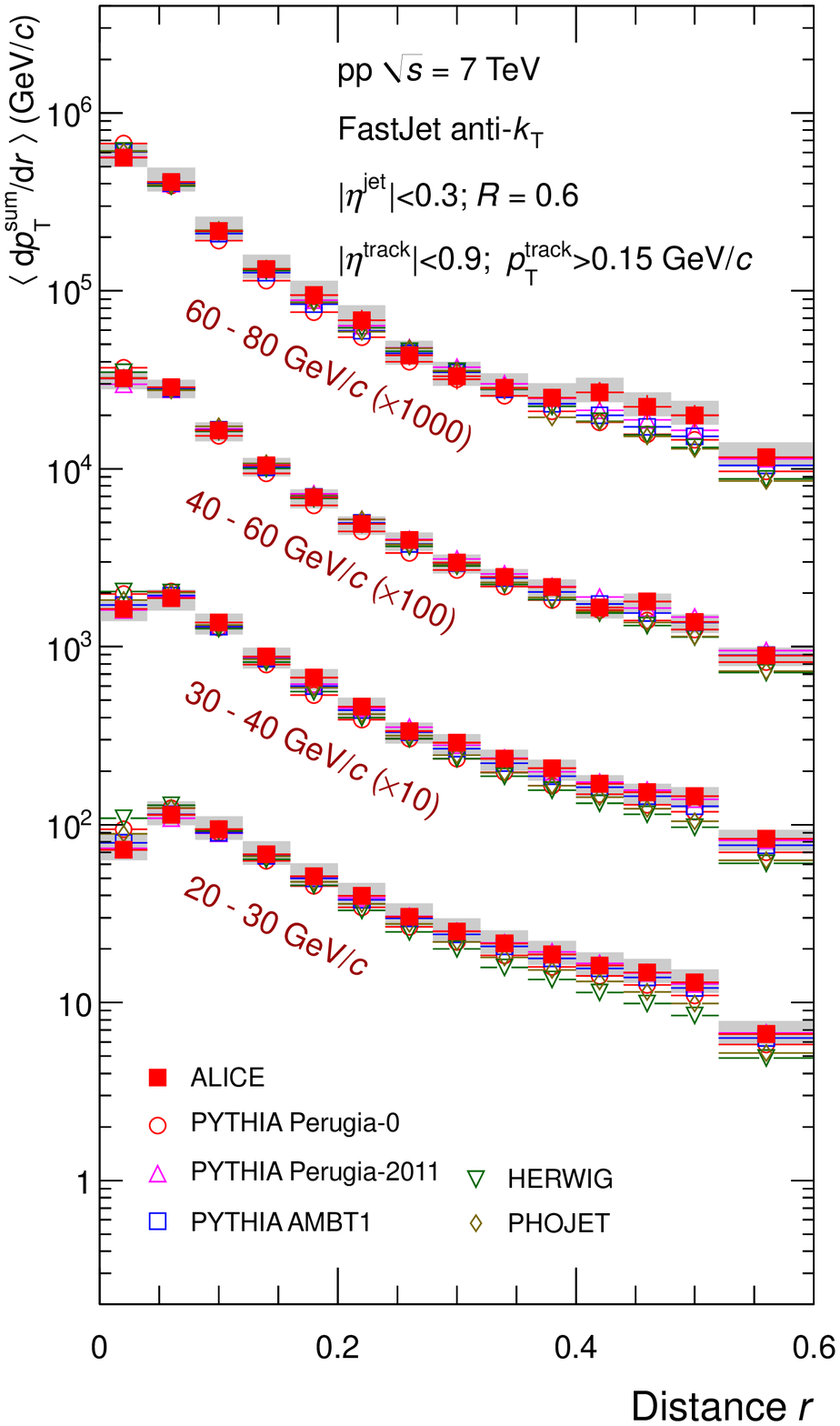}
  \includegraphics[scale=0.52]{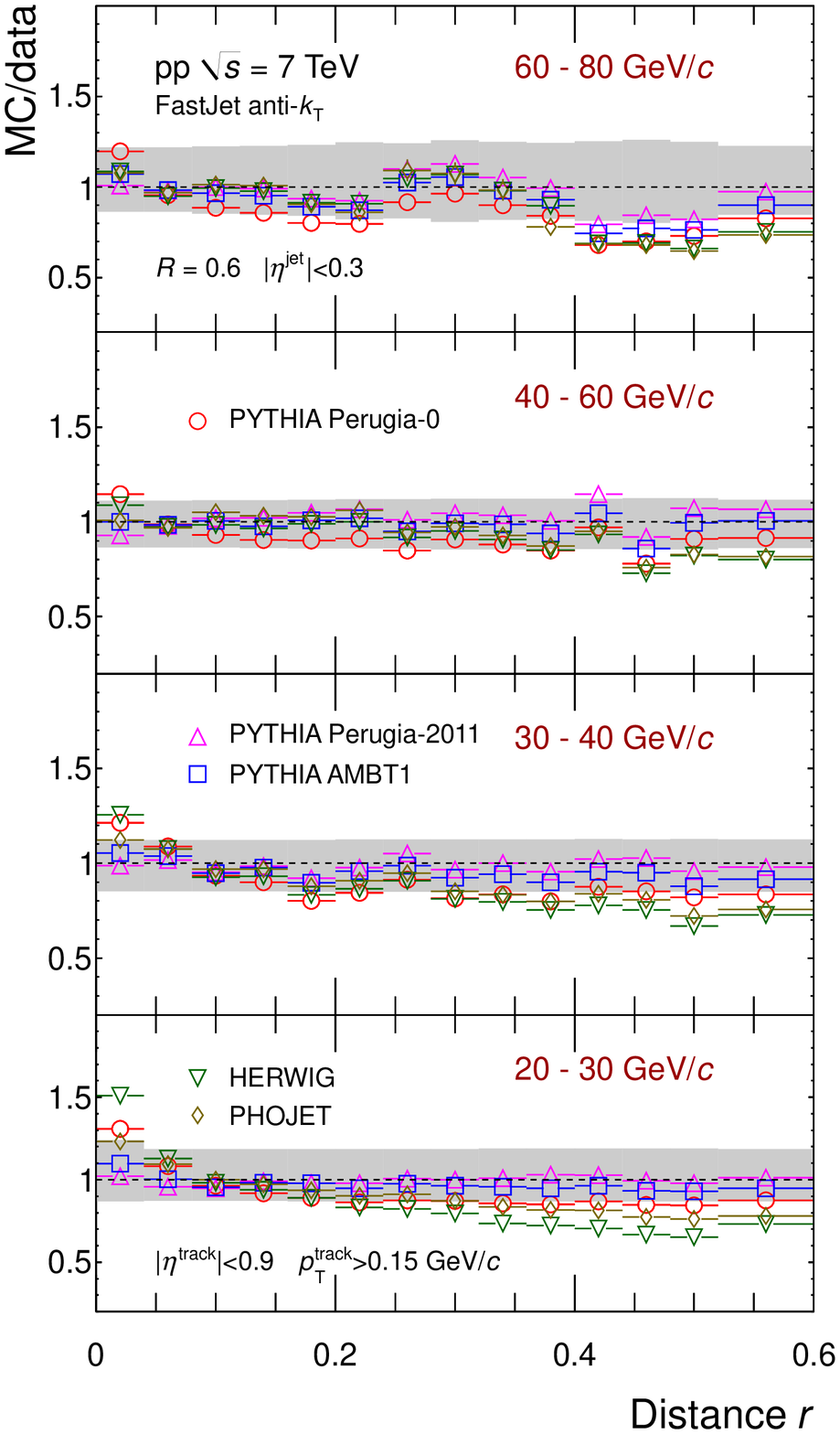}
  \caption{(Color online) Same as Fig.~\ref{Fig.raddist_r6} without UE subtraction.}
  \label{Fig.raddist_r6-woUEsub}
\end{figure*}

The left panels of
Figs.~\ref{Fig.raddist_r2-woUEsub},~\ref{Fig.raddist_r4-woUEsub},
and~\ref{Fig.raddist_r6-woUEsub} show leading jets average $p_{\rm T}$
density radial distributions $\langle \rm{d}{\it p}_{\rm T}^{\rm
  sum}/\rm{d}{\it r} \rangle$ measured with resolution parameters $R$
= 0.2, 0.4, and 0.6, respectively without subtraction of UE (both for
data and MC). The right panels of Figs.~\ref{Fig.raddist_r2-woUEsub},
\ref{Fig.raddist_r4-woUEsub}, and \ref{Fig.raddist_r6-woUEsub} display
ratios of the model calculations to measured data.

\begin{figure*}[ht]
  \includegraphics[scale=0.45]{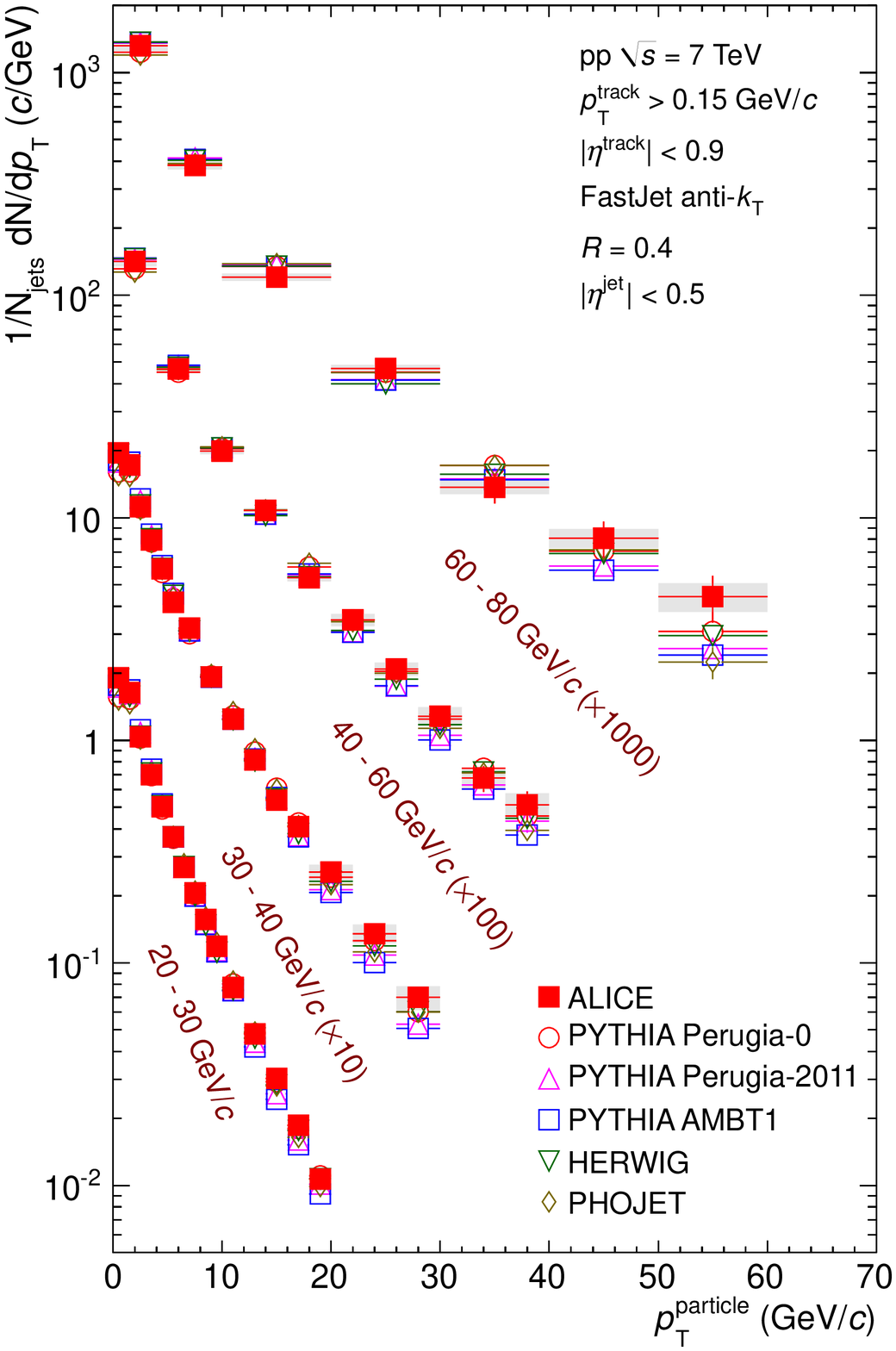}
  \includegraphics[scale=0.45]{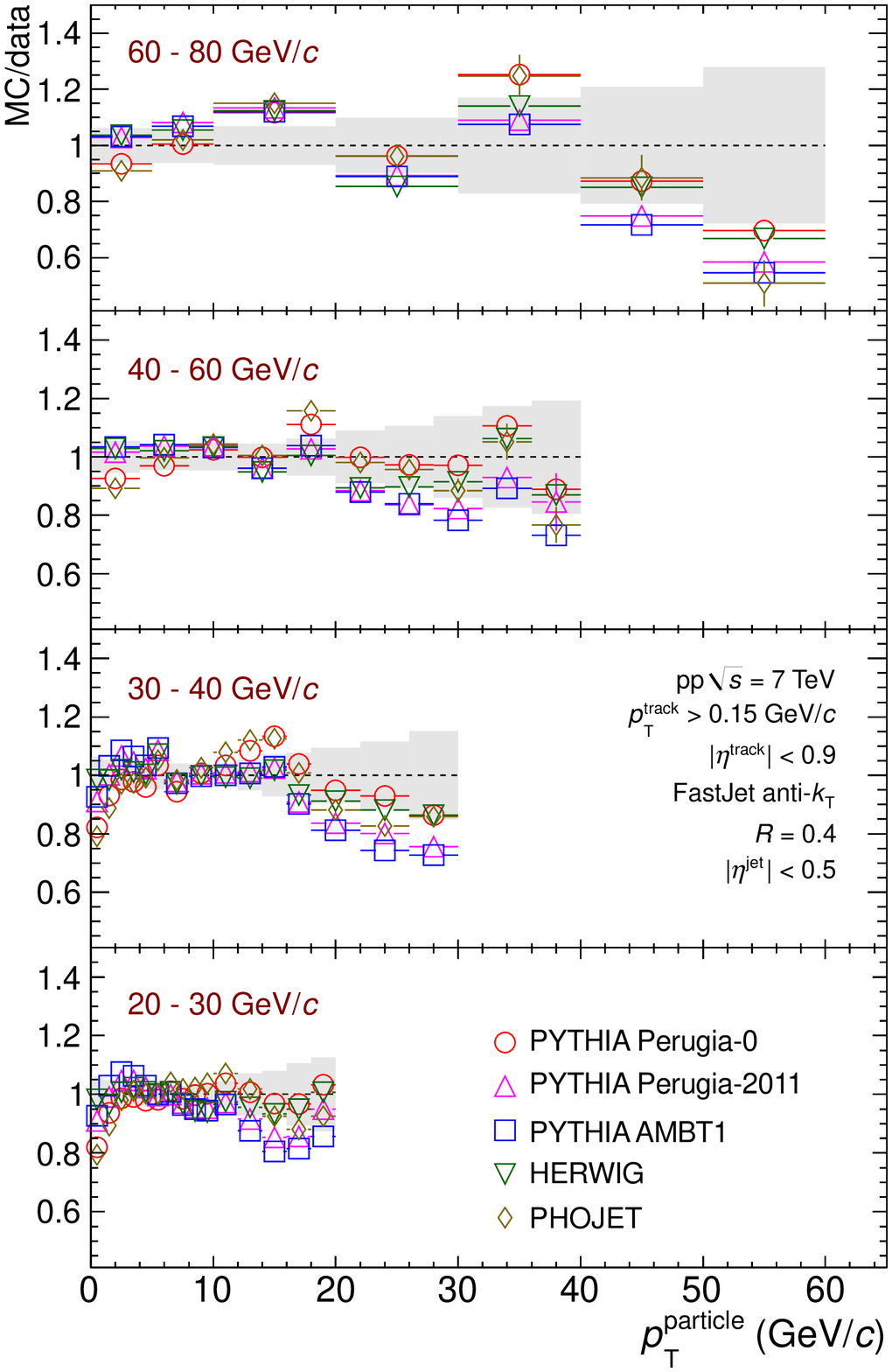}
  \caption{(Color online) Same as Fig.~\ref{fig:FFpt} without UE subtraction.}
  \label{fig:FFpt-woUEsub}
\end{figure*}

\begin{figure*}[ht]
  \includegraphics[scale=0.45]{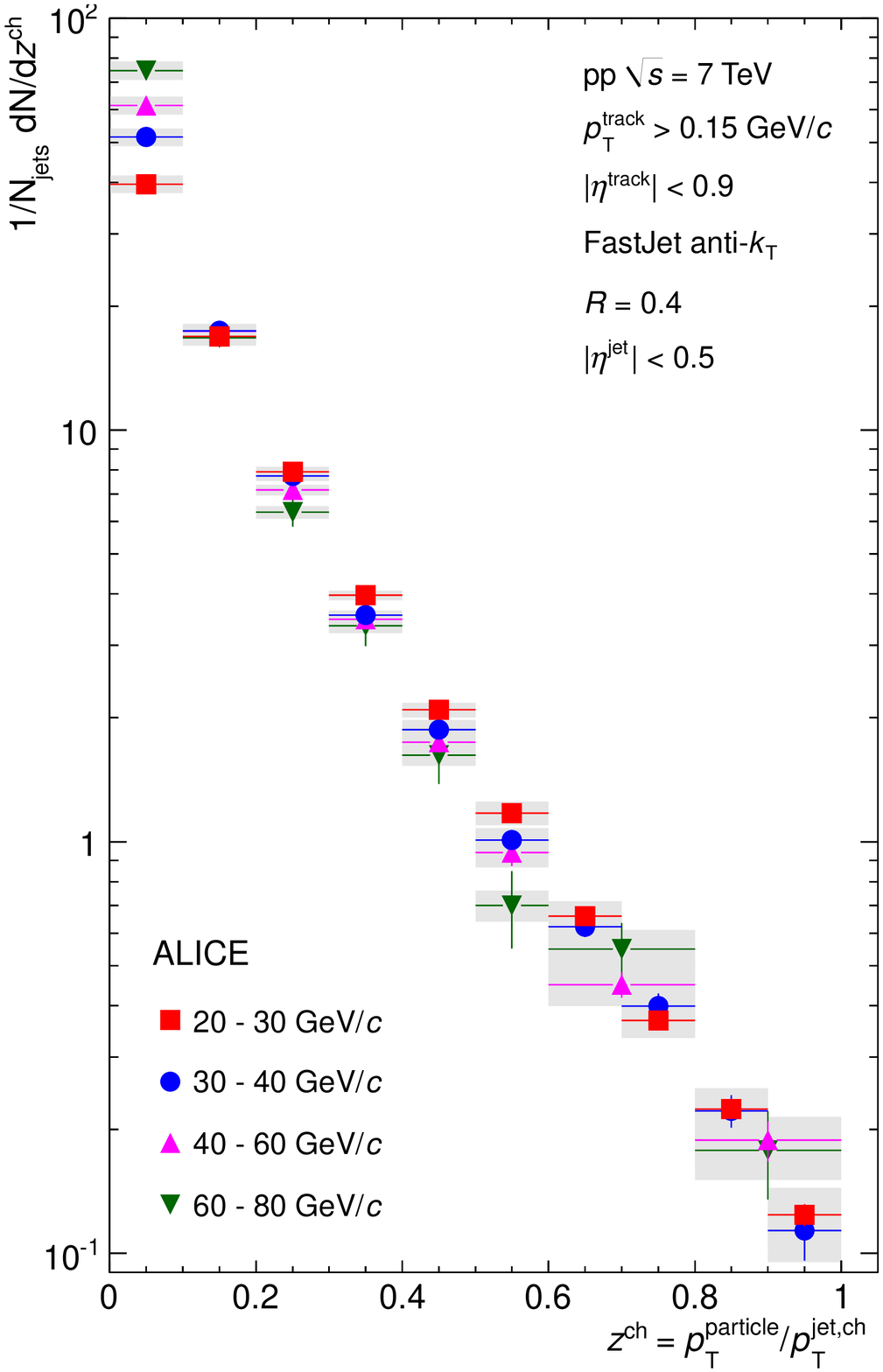}
  \includegraphics[scale=0.45]{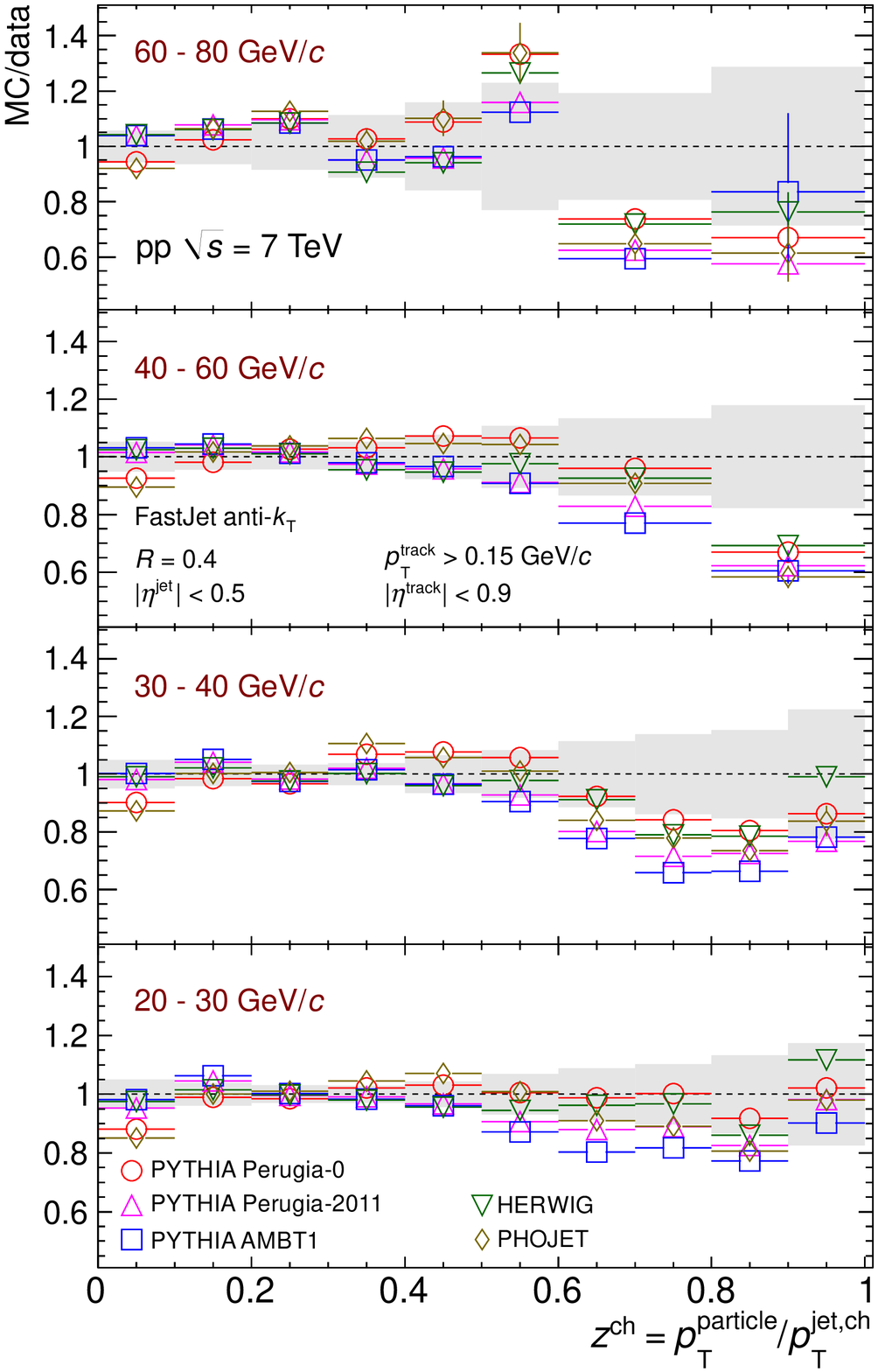}
  \caption{(Color online) Same as Fig.~\ref{fig:FFz} without UE subtraction.}
  \label{fig:FFz-woUEsub}
\end{figure*}

\begin{figure*}[ht]
  \includegraphics[scale=0.45]{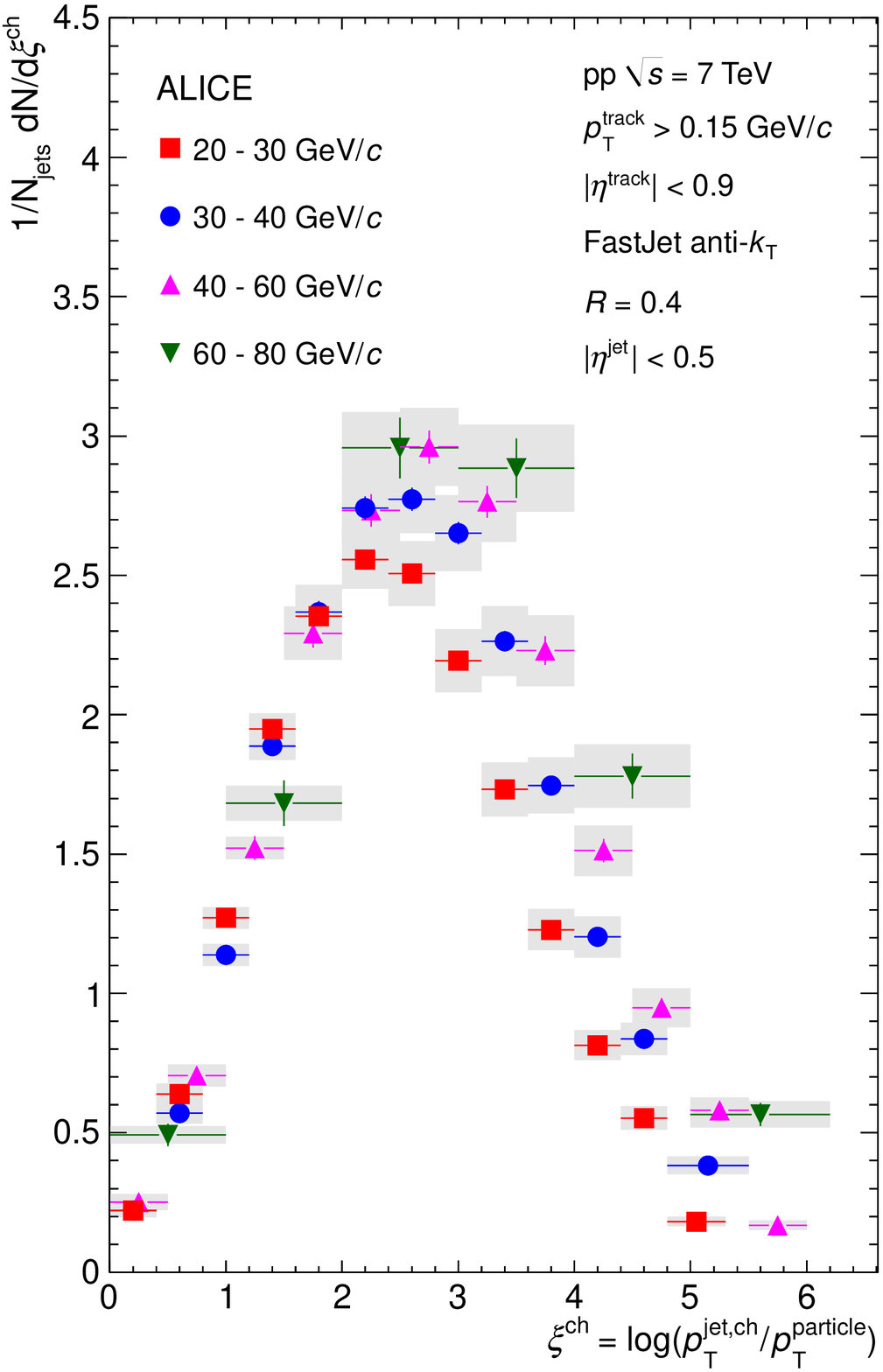}
  \includegraphics[scale=0.45]{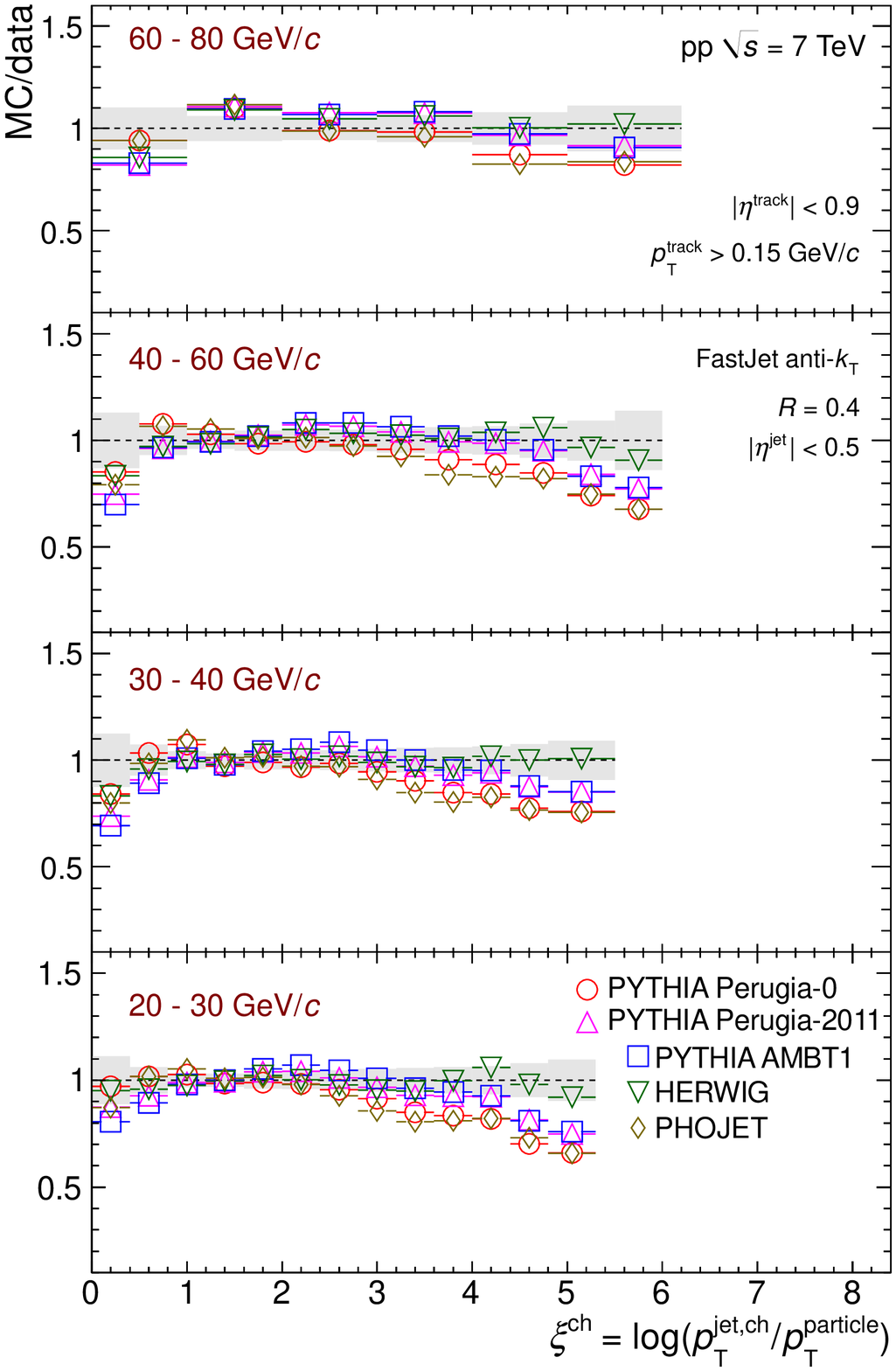}
  \caption{(Color online) Same as Fig.~\ref{fig:FFxi} without UE subtraction.}  
  \label{fig:FFxi-woUEsub}
\end{figure*}

The left panels of Figs.~\ref{fig:FFpt-woUEsub},~\ref{fig:FFz-woUEsub}, and~\ref{fig:FFxi-woUEsub} 
present the measured $p_{\rm T}$ spectra $F^{p_{\rm T}}$
and scaled $p_{\rm T}$ spectra $F^{z}$ and $F^{\xi}$ of charged particles in 
leading charged jets reconstructed with a resolution parameter $R$ = 0.4.
The UE is not subtracted for both data and MC.
The ratios of the calculated MC distributions to measured
distributions are shown in the right panels of Figs.~\ref{fig:FFpt-woUEsub},
~\ref{fig:FFz-woUEsub}, and~\ref{fig:FFxi-woUEsub}.

\clearpage
\section{ALICE Collaboration}
\label{app:collab}



\begingroup
\small
\begin{flushleft}
B.~Abelev\Irefn{org71}\And
J.~Adam\Irefn{org37}\And
D.~Adamov\'{a}\Irefn{org79}\And
M.M.~Aggarwal\Irefn{org83}\And
G.~Aglieri~Rinella\Irefn{org34}\And
M.~Agnello\Irefn{org107}\textsuperscript{,}\Irefn{org90}\And
A.~Agostinelli\Irefn{org26}\And
N.~Agrawal\Irefn{org44}\And
Z.~Ahammed\Irefn{org126}\And
N.~Ahmad\Irefn{org18}
% ...
\renewcommand\labelenumi{\textsuperscript{\theenumi}~}

\section*{Affiliation notes}
\renewcommand\theenumi{\roman{enumi}}
\begin{Authlist}
\item \Adef{0}Deceased
\item \Adef{idp1126816}{Also at: St. Petersburg State Polytechnical University}
\item \Adef{idp3061824}{Also at: Department of Applied Physics, Aligarh Muslim University, Aligarh, India}
\item \Adef{idp3741600}{Also at: M.V. Lomonosov Moscow State University, D.V. Skobeltsyn Institute of Nuclear Physics, Moscow, Russia}
\item \Adef{idp3998400}{Also at: University of Belgrade, Faculty of Physics and "Vin\v{c}a" Institute of Nuclear Sciences, Belgrade, Serbia}
\item \Adef{idp4317104}{Permanent Address: Permanent Address: Konkuk University, Seoul, Korea}
\item \Adef{idp4880768}{Also at: Institute of Theoretical Physics, University of Wroclaw, Wroclaw, Poland}
\item \Adef{idp5813632}{Also at: University of Kansas, Lawrence, KS, United States}
\end{Authlist}

\section*{Collaboration Institutes}
\renewcommand\theenumi{\arabic{enumi}~}
\begin{Authlist}

\item \Idef{org1}A.I. Alikhanyan National Science Laboratory (Yerevan Physics Institute) Foundation, Yerevan, Armenia
\item \Idef{org2}Benem\'{e}rita Universidad Aut\'{o}noma de Puebla, Puebla, Mexico
\item \Idef{org3}Bogolyubov Institute for Theoretical Physics, Kiev, Ukraine
\item \Idef{org4}Bose Institute, Department of Physics and Centre for Astroparticle Physics and Space Science (CAPSS), Kolkata, India
\item \Idef{org5}Budker Institute for Nuclear Physics, Novosibirsk, Russia
\item \Idef{org6}California Polytechnic State University, San Luis Obispo, CA, United States
\item \Idef{org7}Central China Normal University, Wuhan, China
\item \Idef{org8}Centre de Calcul de l'IN2P3, Villeurbanne, France
\item \Idef{org9}Centro de Aplicaciones Tecnol\'{o}gicas y Desarrollo Nuclear (CEADEN), Havana, Cuba
\item \Idef{org10}Centro de Investigaciones Energ\'{e}ticas Medioambientales y Tecnol\'{o}gicas (CIEMAT), Madrid, Spain
\item \Idef{org11}Centro de Investigaci\'{o}n y de Estudios Avanzados (CINVESTAV), Mexico City and M\'{e}rida, Mexico
\item \Idef{org12}Centro Fermi - Museo Storico della Fisica e Centro Studi e Ricerche ``Enrico Fermi'', Rome, Italy
\item \Idef{org13}Chicago State University, Chicago, USA
\item \Idef{org14}Commissariat \`{a} l'Energie Atomique, IRFU, Saclay, France
\item \Idef{org15}COMSATS Institute of Information Technology (CIIT), Islamabad, Pakistan
\item \Idef{org16}Departamento de F\'{\i}sica de Part\'{\i}culas and IGFAE, Universidad de Santiago de Compostela, Santiago de Compostela, Spain
\item \Idef{org17}Department of Physics and Technology, University of Bergen, Bergen, Norway
\item \Idef{org18}Department of Physics, Aligarh Muslim University, Aligarh, India
\item \Idef{org19}Department of Physics, Ohio State University, Columbus, OH, United States
\item \Idef{org20}Department of Physics, Sejong University, Seoul, South Korea
\item \Idef{org21}Department of Physics, University of Oslo, Oslo, Norway
\item \Idef{org22}Dipartimento di Fisica dell'Universit\`{a} 'La Sapienza' and Sezione INFN Rome, Italy
\item \Idef{org23}Dipartimento di Fisica dell'Universit\`{a} and Sezione INFN, Cagliari, Italy
\item \Idef{org24}Dipartimento di Fisica dell'Universit\`{a} and Sezione INFN, Trieste, Italy
\item \Idef{org25}Dipartimento di Fisica dell'Universit\`{a} and Sezione INFN, Turin, Italy
\item \Idef{org26}Dipartimento di Fisica e Astronomia dell'Universit\`{a} and Sezione INFN, Bologna, Italy
\item \Idef{org27}Dipartimento di Fisica e Astronomia dell'Universit\`{a} and Sezione INFN, Catania, Italy
\item \Idef{org28}Dipartimento di Fisica e Astronomia dell'Universit\`{a} and Sezione INFN, Padova, Italy
\item \Idef{org29}Dipartimento di Fisica `E.R.~Caianiello' dell'Universit\`{a} and Gruppo Collegato INFN, Salerno, Italy
\item \Idef{org30}Dipartimento di Scienze e Innovazione Tecnologica dell'Universit\`{a} del  Piemonte Orientale and Gruppo Collegato INFN, Alessandria, Italy
\item \Idef{org31}Dipartimento Interateneo di Fisica `M.~Merlin' and Sezione INFN, Bari, Italy
\item \Idef{org32}Division of Experimental High Energy Physics, University of Lund, Lund, Sweden
\item \Idef{org33}Eberhard Karls Universit\"{a}t T\"{u}bingen, T\"{u}bingen, Germany
\item \Idef{org34}European Organization for Nuclear Research (CERN), Geneva, Switzerland
\item \Idef{org35}Faculty of Engineering, Bergen University College, Bergen, Norway
\item \Idef{org36}Faculty of Mathematics, Physics and Informatics, Comenius University, Bratislava, Slovakia
\item \Idef{org37}Faculty of Nuclear Sciences and Physical Engineering, Czech Technical University in Prague, Prague, Czech Republic
\item \Idef{org38}Faculty of Science, P.J.~\v{S}af\'{a}rik University, Ko\v{s}ice, Slovakia
\item \Idef{org39}Frankfurt Institute for Advanced Studies, Johann Wolfgang Goethe-Universit\"{a}t Frankfurt, Frankfurt, Germany
\item \Idef{org40}Gangneung-Wonju National University, Gangneung, South Korea
\item \Idef{org41}Gauhati University, Department of Physics, Guwahati, India
\item \Idef{org42}Helsinki Institute of Physics (HIP), Helsinki, Finland
\item \Idef{org43}Hiroshima University, Hiroshima, Japan
\item \Idef{org44}Indian Institute of Technology Bombay (IIT), Mumbai, India
\item \Idef{org45}Indian Institute of Technology Indore, Indore (IITI), India
\item \Idef{org46}Inha University, Incheon, South Korea
\item \Idef{org47}Institut de Physique Nucl\'eaire d'Orsay (IPNO), Universit\'e Paris-Sud, CNRS-IN2P3, Orsay, France
\item \Idef{org48}Institut f\"{u}r Informatik, Johann Wolfgang Goethe-Universit\"{a}t Frankfurt, Frankfurt, Germany
\item \Idef{org49}Institut f\"{u}r Kernphysik, Johann Wolfgang Goethe-Universit\"{a}t Frankfurt, Frankfurt, Germany
\item \Idef{org50}Institut f\"{u}r Kernphysik, Westf\"{a}lische Wilhelms-Universit\"{a}t M\"{u}nster, M\"{u}nster, Germany
\item \Idef{org51}Institut Pluridisciplinaire Hubert Curien (IPHC), Universit\'{e} de Strasbourg, CNRS-IN2P3, Strasbourg, France
\item \Idef{org52}Institute for Nuclear Research, Academy of Sciences, Moscow, Russia
\item \Idef{org53}Institute for Subatomic Physics of Utrecht University, Utrecht, Netherlands
\item \Idef{org54}Institute for Theoretical and Experimental Physics, Moscow, Russia
\item \Idef{org55}Institute of Experimental Physics, Slovak Academy of Sciences, Ko\v{s}ice, Slovakia
\item \Idef{org56}Institute of Physics, Academy of Sciences of the Czech Republic, Prague, Czech Republic
\item \Idef{org57}Institute of Physics, Bhubaneswar, India
\item \Idef{org58}Institute of Space Science (ISS), Bucharest, Romania
\item \Idef{org59}Instituto de Ciencias Nucleares, Universidad Nacional Aut\'{o}noma de M\'{e}xico, Mexico City, Mexico
\item \Idef{org60}Instituto de F\'{\i}sica, Universidad Nacional Aut\'{o}noma de M\'{e}xico, Mexico City, Mexico
\item \Idef{org61}iThemba LABS, National Research Foundation, Somerset West, South Africa
\item \Idef{org62}Joint Institute for Nuclear Research (JINR), Dubna, Russia
\item \Idef{org63}Konkuk University, Seoul, South Korea
\item \Idef{org64}Korea Institute of Science and Technology Information, Daejeon, South Korea
\item \Idef{org65}KTO Karatay University, Konya, Turkey
\item \Idef{org66}Laboratoire de Physique Corpusculaire (LPC), Clermont Universit\'{e}, Universit\'{e} Blaise Pascal, CNRS--IN2P3, Clermont-Ferrand, France
\item \Idef{org67}Laboratoire de Physique Subatomique et de Cosmologie, Universit\'{e} Grenoble-Alpes, CNRS-IN2P3, Grenoble, France
\item \Idef{org68}Laboratori Nazionali di Frascati, INFN, Frascati, Italy
\item \Idef{org69}Laboratori Nazionali di Legnaro, INFN, Legnaro, Italy
\item \Idef{org70}Lawrence Berkeley National Laboratory, Berkeley, CA, United States
\item \Idef{org71}Lawrence Livermore National Laboratory, Livermore, CA, United States
\item \Idef{org72}Moscow Engineering Physics Institute, Moscow, Russia
\item \Idef{org73}National Centre for Nuclear Studies, Warsaw, Poland
\item \Idef{org74}National Institute for Physics and Nuclear Engineering, Bucharest, Romania
\item \Idef{org75}National Institute of Science Education and Research, Bhubaneswar, India
\item \Idef{org76}Niels Bohr Institute, University of Copenhagen, Copenhagen, Denmark
\item \Idef{org77}Nikhef, National Institute for Subatomic Physics, Amsterdam, Netherlands
\item \Idef{org78}Nuclear Physics Group, STFC Daresbury Laboratory, Daresbury, United Kingdom
\item \Idef{org79}Nuclear Physics Institute, Academy of Sciences of the Czech Republic, \v{R}e\v{z} u Prahy, Czech Republic
\item \Idef{org80}Oak Ridge National Laboratory, Oak Ridge, TN, United States
\item \Idef{org81}Petersburg Nuclear Physics Institute, Gatchina, Russia
\item \Idef{org82}Physics Department, Creighton University, Omaha, NE, United States
\item \Idef{org83}Physics Department, Panjab University, Chandigarh, India
\item \Idef{org84}Physics Department, University of Athens, Athens, Greece
\item \Idef{org85}Physics Department, University of Cape Town, Cape Town, South Africa
\item \Idef{org86}Physics Department, University of Jammu, Jammu, India
\item \Idef{org87}Physics Department, University of Rajasthan, Jaipur, India
\item \Idef{org88}Physik Department, Technische Universit\"{a}t M\"{u}nchen, Munich, Germany
\item \Idef{org89}Physikalisches Institut, Ruprecht-Karls-Universit\"{a}t Heidelberg, Heidelberg, Germany
\item \Idef{org90}Politecnico di Torino, Turin, Italy
\item \Idef{org91}Purdue University, West Lafayette, IN, United States
\item \Idef{org92}Pusan National University, Pusan, South Korea
\item \Idef{org93}Research Division and ExtreMe Matter Institute EMMI, GSI Helmholtzzentrum f\"ur Schwerionenforschung, Darmstadt, Germany
\item \Idef{org94}Rudjer Bo\v{s}kovi\'{c} Institute, Zagreb, Croatia
\item \Idef{org95}Russian Federal Nuclear Center (VNIIEF), Sarov, Russia
\item \Idef{org96}Russian Research Centre Kurchatov Institute, Moscow, Russia
\item \Idef{org97}Saha Institute of Nuclear Physics, Kolkata, India
\item \Idef{org98}School of Physics and Astronomy, University of Birmingham, Birmingham, United Kingdom
\item \Idef{org99}Secci\'{o}n F\'{\i}sica, Departamento de Ciencias, Pontificia Universidad Cat\'{o}lica del Per\'{u}, Lima, Peru
\item \Idef{org100}Sezione INFN, Bari, Italy
\item \Idef{org101}Sezione INFN, Bologna, Italy
\item \Idef{org102}Sezione INFN, Cagliari, Italy
\item \Idef{org103}Sezione INFN, Catania, Italy
\item \Idef{org104}Sezione INFN, Padova, Italy
\item \Idef{org105}Sezione INFN, Rome, Italy
\item \Idef{org106}Sezione INFN, Trieste, Italy
\item \Idef{org107}Sezione INFN, Turin, Italy
\item \Idef{org108}SSC IHEP of NRC Kurchatov institute, Protvino, Russia
\item \Idef{org109}SUBATECH, Ecole des Mines de Nantes, Universit\'{e} de Nantes, CNRS-IN2P3, Nantes, France
\item \Idef{org110}Suranaree University of Technology, Nakhon Ratchasima, Thailand
\item \Idef{org111}Technical University of Split FESB, Split, Croatia
\item \Idef{org112}The Henryk Niewodniczanski Institute of Nuclear Physics, Polish Academy of Sciences, Cracow, Poland
\item \Idef{org113}The University of Texas at Austin, Physics Department, Austin, TX, USA
\item \Idef{org114}Universidad Aut\'{o}noma de Sinaloa, Culiac\'{a}n, Mexico
\item \Idef{org115}Universidade de S\~{a}o Paulo (USP), S\~{a}o Paulo, Brazil
\item \Idef{org116}Universidade Estadual de Campinas (UNICAMP), Campinas, Brazil
\item \Idef{org117}University of Houston, Houston, TX, United States
\item \Idef{org118}University of Jyv\"{a}skyl\"{a}, Jyv\"{a}skyl\"{a}, Finland
\item \Idef{org119}University of Liverpool, Liverpool, United Kingdom
\item \Idef{org120}University of Tennessee, Knoxville, TN, United States
\item \Idef{org121}University of Tokyo, Tokyo, Japan
\item \Idef{org122}University of Tsukuba, Tsukuba, Japan
\item \Idef{org123}University of Zagreb, Zagreb, Croatia
\item \Idef{org124}Universit\'{e} de Lyon, Universit\'{e} Lyon 1, CNRS/IN2P3, IPN-Lyon, Villeurbanne, France
\item \Idef{org125}V.~Fock Institute for Physics, St. Petersburg State University, St. Petersburg, Russia
\item \Idef{org126}Variable Energy Cyclotron Centre, Kolkata, India
\item \Idef{org127}Vestfold University College, Tonsberg, Norway
\item \Idef{org128}Warsaw University of Technology, Warsaw, Poland
\item \Idef{org129}Wayne State University, Detroit, MI, United States
\item \Idef{org130}Wigner Research Centre for Physics, Hungarian Academy of Sciences, Budapest, Hungary
\item \Idef{org131}Yale University, New Haven, CT, United States
\item \Idef{org132}Yonsei University, Seoul, South Korea
\item \Idef{org133}Zentrum f\"{u}r Technologietransfer und Telekommunikation (ZTT), Fachhochschule Worms, Worms, Germany
\end{Authlist}
\endgroup

\end{document}